\documentclass[oneside, openany]{aa}  
\usepackage{graphicx}
\usepackage{txfonts}
\usepackage{xcolor}

\usepackage[hidelinks]{hyperref}
\hypersetup{
    colorlinks=true,
    allcolors=blue,}

\begin{document} 
   \title{Probabilistic and progressive deblended far-infrared and sub-millimetre point source catalogues}
   
   \subtitle{I. Methodology and first application in the COSMOS field}

   \author{Lingyu Wang
          \inst{1, 2}\fnmsep\thanks{l.wang@sron.nl}
          \and
          Antonio La Marca\inst{1, 2}
          \and
          Fangyou Gao\inst{2, 3, 4}
          \and
          William J. Pearson\inst{5}
          \and
          Berta Margalef-Bentabol\inst{1}
          \and
          Matthieu B\'{e}thermin\inst{6, 7}
          \and
          Longji Bing\inst{8}
          \and
          James Donnellan\inst{8}
          \and
          Peter D. Hurley\inst{8}
          \and
          Seb J. Oliver\inst{8}
          \and
          Catherine L. Hale\inst{9, 10}
          \and
          Matt J. Jarvis\inst{10, 11}
          \and
Lucia Marchetti\inst{12, 13, 14}
          \and
Mattia Vaccari\inst{13,14, 15}
          \and
          Imogen H. Whittam\inst{10, 11} 
          }

   \institute{SRON Netherlands Institute for Space Research, Landleven 12, 9747 AD Groningen,The Netherlands\\
              \email{l.wang@sron.nl}
         \and
             Kapteyn Astronomical Institute, University of Groningen, Postbus 800, 9700 AV Groningen,The Netherlands
        \and
Department of Astronomy, Nanjing University, Nanjing 210093, China
\and
Key Laboratory of Modern Astronomy and Astrophysics (Nanjing University), Ministry of Education, Nanjing 210093, China
\and
        National Centre for Nuclear Research, Pasteura 7, 02-093 Warszawa, Poland
        \and
         Aix-Marseille Univ., CNRS, CNES, LAM, Marseille, France
         \and
Universit\'{e} de Strasbourg, CNRS, Observatoire astronomique de Strasbourg, UMR 7550, 67000 Strasbourg, France
        \and
         Astronomy Centre, Department of Physics and Astronomy, University of Sussex, Falmer, Brighton BN1 9QH, UK
         \and
 Institute for Astronomy, School of Physics and Astronomy, University of Edinburgh, Royal Observatory Edinburgh, Blackford Hill, Edinburgh EH9 3HJ
  \and
  Sub-Department of Astrophysics, University of Oxford, Denys Wilkinson Building, Keble Road, Oxford, OX1 3RH, UK
         \and
         Department of Physics and Astronomy, University of the Western Cape, Robert Sobukwe Road, 7535 Bellville, Cape Town, South Africa
         \and
Department of Astronomy, University of Cape Town, 7701 Rondebosch, Cape Town, South Africa
         \and
         Inter-University Institute for Data Intensive Astronomy, Department of Astronomy, University of Cape Town, 7701 Rondebosch, Cape Town, South Africa
\and
INAF - Istituto di Radioastronomia, via Gobetti 101, 40129 Bologna, Italy 
\and
Inter-University Institute for Data Intensive Astronomy, Department of Physics and Astronomy, University of the Western Cape, 7535 Bellville, Cape Town, South Africa
         \\
             }

   \date{Received; accepted}

  \abstract
   {Single-dish far-infrared (far-IR) and sub-millimetre (sub-mm) point source catalogues and their connections with galaxy catalogues at other wavelengths are of paramount importance to studying galaxy evolution. However, due to the large mismatch in spatial resolution, cross-matching galaxies detected at different wavelengths is not straightforward.}
   {This work aims to develop the next-generation deblended far-IR and sub-mm catalogues in deep extragalactic survey fields, by extracting photometry in these low-resolution images at the positions of known sources. We present the first application of our methodology in the COSMOS field.}
   {Our progressive deblending used the Monte Carlo Markov Chain (MCMC)-based Bayesian probabilistic framework known as XID+.  The deblending process started from the {\it Spitzer}/MIPS 24 $\mu$m data, using an initial prior list composed of sources selected from the COSMOS2020 catalogue and radio catalogues from the VLA and the MeerKAT surveys, based on spectral energy distribution (SED) modelling which predicts fluxes of the known sources at the deblending wavelength. To speed up flux prediction, we made use of a neural network-based emulator. After deblending the 24 $\mu$m data, we proceeded to the {\it Herschel} PACS (100 \& 160 $\mu$m) and SPIRE wavebands (250, 350 \& 500 $\mu$m). Each time we constructed a tailor-made prior list based on the predicted fluxes of the known sources, taking into account the deblended photometry from the previous steps.}
   {Using simulated far-IR and sub-mm sky, we detailed the performance of our deblending pipeline. After validation with simulations, we then deblended the real observations from 24 to 500 $\mu$m and compared with blindly extracted  catalogues and previous versions of deblended catalogues. As an additional test, we deblended the SCUBA-2 850 $\mu$m map and compared our deblended fluxes with ALMA measurements, which demonstrates a higher level of flux accuracy compared to previous results. We publicly release our XID+ deblended point source catalogues, including best estimates and posterior probability distribution functions. These deblended long-wavelength data, which are cross-matched with multi-band photometry by construction, are crucial for studies such as deriving the fraction of dust-obscured star formation and better separation of quiescent galaxies from dusty star-forming galaxies.}
   {}


\titlerunning{Probabilistic and progressive deblended far-IR and sub-mm point source catalogues}
\authorrunning{Wang et al.}

   \maketitle

\section{Introduction}

Around half of all the luminous power from star formation and supermassive black hole (SMBH) accretion is obscured by dust \citep{1996A&A...308L...5P, 1998ApJ...508..123F, 2001ARA&A..39..249H, 2008ApJ...678L.101D}. The interstellar dust heated by ultraviolet (UV)/optical light re-emits the absorbed photons approximately as a modified blackbody with a peak at far-infrared (IR) and sub-millimetre (sub-mm) wavelengths \citep{2012A&A...539A.155M, 2012MNRAS.425.3094C, 2014PhR...541...45C,  2016MNRAS.461.1898W}. Therefore, far-IR and sub-mm observations of dusty star-forming galaxies (DSFG) are of key importance to achieving an unbiased census of the cosmic star-formation history \citep{2014ARA&A..52..415M,  2019A&A...624A..98W, 2020A&A...643A...8G, 2021ApJ...909..165Z}. It is also well known that massive galaxies at higher redshifts ($z>1$) are increasingly dominated by DSFGs due to the correlation between stellar mass and the level of dust obscuration for star-forming galaxies \citep{2009ApJ...698L.116P,  2011ApJ...742...96W, 2012ApJ...754L..29W, 2017ApJ...850..208W, 2023MNRAS.518.6142A} and the increasing fraction of the overall star-forming galaxy population with increasing redshift \citep{2016ApJ...827L..25M, 2023A&A...677A.184W}. Thus, the DSFG population also plays a crucial role in understanding massive galaxy formation and evolution \citep{ 2017MNRAS.467.1360B,  2021A&A...654A.117G,  2023ApJ...953...11L, 2020MNRAS.494.3828D}. 

The overarching goal of this work is to develop the next-generation deblended far-IR and sub-mm galaxy photometric catalogues in the premier deep extragalactic survey fields, building on our experience and expertise in analysing single-dish long-wavelength data with coarse spatial resolution ($\sim10$ times or more worse than optical and near-IR observations). There are several reasons why deblending is critical. Deep long-wavelength observations often result in confusion-limited maps. Due to the confusion limit which reaches as high as $\sim20$ mJy (5$\sigma$) in the {\it Herschel}/SPIRE bands \citep{2010A&A...518L...5N}, blind detections are limited to only a small number of bright sources \citep{2011A&A...533A.119E, 2015A&A...573A.113B} accounting for $<\sim$ 15\% of the cosmic IR background at SPIRE wavelengths \citep{2010A&A...518L..21O}. Because of the large beam size, blindly detected sources suffer from large positional uncertainty and flux boosting which become increasingly severe with decreasing flux \citep{ 2012MNRAS.419..377S, 2014MNRAS.444.2870W}. In addition, a single blind detection may not correspond to one source but instead a blend of several sources within a single spatial beam, i.e. the so-called multiplicity issue \citep{2015ApJ...812...43B, 2016MNRAS.460.1119S, 2018MNRAS.480.4974H, 2021A&A...648A...8W}. 
Therefore, it can be extremely challenging to match detections from single-dish far-IR/sub-mm maps with their counterparts in the optical/near-IR. To probe fainter sources below the confusion limit as well as to overcome the difficulty in multi-wavelength source cross-matching, deblending is needed to correctly distribute fluxes among the contributing sources.

In the past decade or so, several techniques have been developed that can deblend sources in low-resolution far-IR/sub-mm maps, with varying degrees of success. These techniques are typically based on the positions of sources detected in high-resolution imaging at other wavelengths, for example at the {\it Spitzer}/MIPS 24 $\mu$m and the radio 1.4 GHz  \citep{2010A&A...518L..28M, 2010A&A...516A..43B, 
 2010MNRAS.409...48R, 2012MNRAS.419.2758R, 2011MNRAS.411..505C, 2013ApJ...778..131L} wavelengths, thanks to the strong correlation between the 24-$\mu$m/radio emission and the far-IR/sub-mm dust emission \citep{2009ApJ...692..556R, 2010MNRAS.402..245I, 2017A&A...602A...4D}. Most of these techniques use a maximum-likelihood optimisation approach partly due to its computational ease. However, they suffer from two major issues. The first is that variance and covariance of source fluxes cannot be properly estimated, which is problematic as neighbouring sources are expected to be highly correlated (particularly when the source surface density is high). The second is overfitting when many of the sources are intrinsically faint which can lead to a systematic flux underestimation. The list-driven algorithm DESPHOT \citep{2010MNRAS.409...48R, 2012MNRAS.419.2758R, 2014MNRAS.444.2870W} specifically developed for the {\it Herschel} Multi-tiered Extragalactic Survey (HerMES; \citealt{2012MNRAS.424.1614O})  tried to overcome this by using the non-negative weighted least absolute shrinkage and selection operator (LASSO) algorithm (\citealt{tibshirani1996regression, zou2006adaptive, braak2010identity}), a shrinkage and selection method that introduces an additional penalty term in order to decrease the number of sources needed to account for the emission in the map. However, when multiple sources are very close to each other, this method can have the opposite problem by incorrectly assigning all the flux to one source. 

In this paper, we aim to construct state-of-the-art deblended far-IR and sub-mm point source catalogue in the Cosmic Evolution Survey (COSMOS; \citealt{2007ApJS..172....1S}) field, using a progressive and probabilistic approach. There are several deblended catalogues in this field already. For example, \citet{2018ApJ...864...56J} presented a super-deblended catalogue from 24 $\mu$m to mm wavelengths. Their deblending method is progressive (i.e., moving from short to long wavelengths) and relies on a $K_s$-selected prior catalogue, with additional sources selected in the radio. \citet{2017MNRAS.464..885H} presented a deblended {\it Herschel}/SPIRE catalogue using prior sources selected at 24 $\mu$m and a Bayesian Markov Chain Monte Carlo (MCMC)-based probabilistic deblending and source extraction tool called XID+. In developing the next generation state-of-the-art deblended catalogues, we need to combine the strengths of the previous approaches, for example, the progressive deblending in \citet{2018ApJ...864...56J} to exploit the information encoded in the multi-wavelength data and the probabilistic deblending in \citet{2017MNRAS.464..885H} to fully explore the posterior, the variance and covariance between sources. At the same time, we want to improve on several aspects of the previous deblending works. For example, we need to take into account the correlation between different far-IR/sub-mm bands rather than fitting each band independently. The prior catalogue itself, which has a significant impact on the quality of the deblended far-IR/sub-mm fluxes, has also changed substantially. For example, in the COSMOS field, there is now a newer and deeper multi-wavelength photometric catalogue with improved photometric redshift (photo-$z$) and stellar mass estimates. While we do not make use of the photo-$z$ and stellar mass estimations, the improved photometric data and better spectral energy distribution (SED) coverage lead to a more complete and reliable prior list for deblending the long-wavelength data. 

The structure of this paper is as follows. In Section \ref{prior}, we explain how we build the initial prior catalogue for deblending the long-wavelength data based on the latest photometric catalogue in COSMOS and additional radio source catalogues. First, we give an overview of the various datasets and our selection criteria. Then, we present a deep learning-based SED modelling and fitting emulator which is used to predict fluxes at the wavelengths to be deblended. This step is critical as it allows us to select the most relevant sources to be included in the prior. In Section \ref{deblend}, we describe the key properties of the far-IR and sub-mm data to be deblended from 24 $\mu$m to 500 $\mu$m and our probabilistic deblending tool XID+. In Section \ref{validation}, we focus on validation of our deblending methodology using realistic simulations of the far-IR and sub-mm sky. Issues such as flux accuracy, flux precision and accuracy of flux uncertainty are discussed in detail. In Section \ref{real}, we deblend the real observations from 24 $\mu$m to 500 $\mu$m. Our final deblended far-IR and sub-mm point source catalogue is then compared with blindly extracted catalogues as well as the super-deblended catalogue from \citet{2018ApJ...864...56J}. As an additional test, we also deblend the SCUBA-2 850 $\mu$m map and compare with ALMA measurements. In Section \ref{science}, we show two example science applications of our deblended long-wavelength catalogue by examining the galaxy star formation main sequence (SFMS) and the far-IR to radio correlation (FIRC). In Section \ref{conclude}, we present our conclusions and future directions. Throughout the paper, we adopt the \citet{2003PASP..115..763C} initial mass function (IMF) and the Wilkinson Microwave Anisotropy Probe year 7 (WMAP7) cosmology \citep{2011ApJS..192...18K, 2011ApJS..192...16L} with $\Omega_M = 0.272$, $\Omega_{\Lambda} = 0.728$, and $H_0 = 70.4$ km s$^{-1}$ Mpc$^{-1}$.

\section{Construction of the initial prior catalogue}\label{prior}

In this section, first we introduce the multi-wavelength datasets used to construct the initial prior catalogue for deblending the {\it Spitzer}/MIPS 24 $\mu$m data, i.e., the COSMOS2020 catalogue and the joint radio catalogue from combining the VLA and MeerKAT data. The COSMOS2020 catalogue is the main dataset which can be regarded as a proxy for a stellar mass-based selection, while the radio source catalogues provide additional high-redshift (high-$z$) sources which may be missing from the COSMOS2020 catalogue \citep{2018ApJ...853..172L, 2018ApJ...864...56J}. Then, we describe the SED fitting procedure used to predict the 24 $\mu$m flux, taking into account systematic uncertainties due to varying assumptions in SED fitting. This is another major improvement over previous deblending efforts. In addition, to speed up the flux prediction step, we use a Deep Learning Neural Network (DLNN) as an emulator which is trained on the outputs from the SED fitting step.  

\subsection{The input multi-wavelength datasets}

\subsubsection{The COSMOS2020 catalogue}

The main component in building our prior catalogue for deblending the far-IR and sub-mm data is the latest photometric catalogue from COSMOS. This catalogue named COSMOS2020 \citep{2022ApJS..258...11W} is an updated version of the previous catalogue COSMOS2015 \citep{2016ApJS..224...24L}. It includes multi-band photometry from far-UV (FUV) and near-UV (NUV) band of the Galaxy Evolution Explorer (GALEX) satellite to {\it Spitzer}/Infrared Array Camera (IRAC) over the wavelength range 0.1–10 $\mu$m for about 1.7 million sources. Compared to COSMOS2015, the latest release includes new imaging data, such as the U-band data from the Canada-France-Hawaii Telescope (CFHT) large area U-band deep survey (CLAUDS; \citealt{2019MNRAS.489.5202S}), {\it grizy}-band data from the Hyper Suprime-Cam (HSC) Subaru Strategic Program (SSP) PDR2 \citep{2019PASJ...71..114A}, {\it YJHKs}-band data from the fourth data release (DR4) of the UltraVISTA survey \citep{2012A&A...544A.156M, 2023yCat.2373....0M}, and {\it Spitzer}/IRAC data from the Cosmic Dawn Survey \citep{2022A&A...658A.126E}. Source detection was performed on a very deep multi-band chi-squared {\it izYJHKs} detection image. There are two versions of COSMOS2020: a Classic one produced by a pipeline similar to COSMOS2015 (with an exception for the IRAC bands), and the Farmer version (limited to the UltraVISTA coverage) produced by the software Tractor \citep{2016ascl.soft04008L}. Photometric redshift (Photo-$z$) was derived using either the LePhare \citep{2002MNRAS.329..355A,  2006A&A...457..841I} or EAZY code \citep{2008ApJ...686.1503B}. Therefore, there are in total four possible combinations in choosing a version of the photometric measurements and a version of the photo-$z$. In this paper, we chose the Farmer catalogue with photo-$z$ derived using LePhare as this combination performs the best in terms of comparisons to spectroscopic redshifts\footnote{We do not make use of photo-$z$ in our deblending pipeline. Only source type information provided by LePhare is used.} \citep{2022ApJS..258...11W}.  Extinctions due to the Milky Way and photometric offsets are corrected in COSMOS2020. 


The photometric catalogue, downloaded from the release site\footnote{\url{https://cosmos2020.calet.org/catalogues/}}, contains 964\,506 sources. Following recommendations from the COSMOS team, we set the flag \textsc{FLAG\_COMBINED} equal to zero to select areas that have coverage from HSC, UltraVISTA, and {\it Spitzer}/IRAC and are not affected by bright stars or large artifacts, which reduced the catalogue to 746\,976 sources. We also made use of the star/galaxy separation provided in the Le Phare photo-$z$ and physical parameters. We selected sources which have the star/galaxy separation flag lp\_type 
 value equal to 0 (galaxy) or 2 (X-ray source), which further reduced the catalogue to a total of 711\,106 sources. According to the readme file\footnote{\url{https://cosmos2020.calet.org/catalogues/MASKS/MASKS_README.txt}}, the area selected by setting \textsc{FLAG\_COMBINED} equal to zero is 1.278 deg$^2$, which is much smaller than the full COSMOS field of $\sim2$ deg$^2$. In the future, with new ground- and space-based observations such as the planned observations from the European Space Agency survey mission {\it Euclid} \citep{2011arXiv1110.3193L} and the {\it James Webb} Space Telescope (JWST; \citealt{2006SSRv..123..485G}), it will be possible extend this work over a larger area in COSMOS. 

\subsubsection{The joint radio catalogue}


In order to add additional high-$z$ sources which may be missing from  COSMOS2020, we made use of the source catalogue from the VLA-COSMOS 3 GHz Large Project \citep{2017A&A...602A...1S}, which is based on 384 hours of observations with the Karl G. Jansky Very Large Array (VLA) and uniformly covers the entire COSMOS field down to a median rms of 2.3 $\mu$Jy/beam at an angular resolution of 0.75\arcsec. The catalogue contains 10\,830 blindly extracted sources down to 5$\sigma$, with a total fraction of spurious sources $<$ 2.7\%. Positional accuracy is estimated to be around 0.01\arcsec for bright sources with S/N $>20$.


We also made use of the MeerKAT International GHz Tiered Extragalactic Exploration (MIGHTEE)  survey \citep{2016mks..confE...6J, 2022MNRAS.509.2150H}, which is an ongoing flagship Large Survey Project on MeerKAT \citep{2016mks..confE...1J} and precursor to the Square Kilometre Array. When completed, MIGHTEE will cover 20 deg$^2$ across four extragalactic deep fields to a depth of $\sim$1 $\mu$Jy/beam rms at a central bandwidth frequency of $\sim1.3$ GHz. In this paper, we used the COSMOS catalogue extracted from the Data Release 1 total intensity map with an angular resolution of $\sim8\arcsec$ (Hale, Heywood, Jarvis et al., in prep). Images at a higher resolution ($5\arcsec$) are also available but are shallower by a factor of $\sim3$. The thermal noise in the MIGHTEE COSMOS image reaches 1.7$\mu$Jy/beam. For comparison, the VLA-COSMOS 1.4 GHz Large Project map \citep{2010ApJS..188..384S} has a depth of $\sigma=12\mu$Jy / beam in the deepest central 50'$\times$50' area. Consequently, the MIGHTEE continuum images are limited by confusion noise rather than thermal noise. By comparing to sources detected in the shallower but much-higher resolution VLA-COSMOS 3 GHz Large Project, the mean offsets in (RA, Dec) of the MIGHTEE sources have been determined to be (-0.25\arcsec$\pm$0.01\arcsec, -0.08\arcsec$\pm$0.01\arcsec). Cross-matching MIGHTEE sources with the VLA 1.4 GHz source catalogue \citep{2010ApJS..188..384S} and comparing measurements of the integrated flux density also demonstrate a high level of photometric consistency (with a median MIGHTEE to VLA flux ratio of 0.95).

To find radio sources which are missing in COSMOS2020, we cross-matched the VLA 3 GHz and the MIGHTEE 1.3 GHz  catalogues with COSMOS2020 (after applying the flags as described in Section 2.1.1). If a VLA or MIGHTEE source does not have a match in COSMOS2020 within 1\arcsec, we treated it as an additional  source to be included in the prior catalogue. In total, we found 2\,791 such sources as most of the radio sources are already included in  COSMOS2020. In our previous work \citep{2021A&A...648A...8W}, we used the radio flux as an informative flux prior to deblend the far-FIR/sub-mm data. In this work, we decided to  just use the detection information to keep the potential bias in the deblended flux at a minimum. In Section \ref{science}, we examine whether MIGHTEE sources not included in COSMOS2020 follow the FIRC, which also serves as an independent test of the quality of our deblended photometry.

\subsection{Flux prediction using SED fitting}

The number of sources in COSMOS2020  and the joint radio catalogue (totalling over 700\,000 sources) is too large to be directly deblended in the far-IR and sub-mm maps which have much worse spatial resolutions than optical and near-IR maps. We thus need to develop a way of selecting the most relevant sources to be included in the prior catalogue. As discussed in \citet{2018ApJ...853..172L}, ideally we want to keep the average number of sources to be deblended under a single spatial beam to be $\lesssim1$ in all bands. 

\begin{figure} \centering
    \includegraphics[width=8.5cm, height=8.cm]{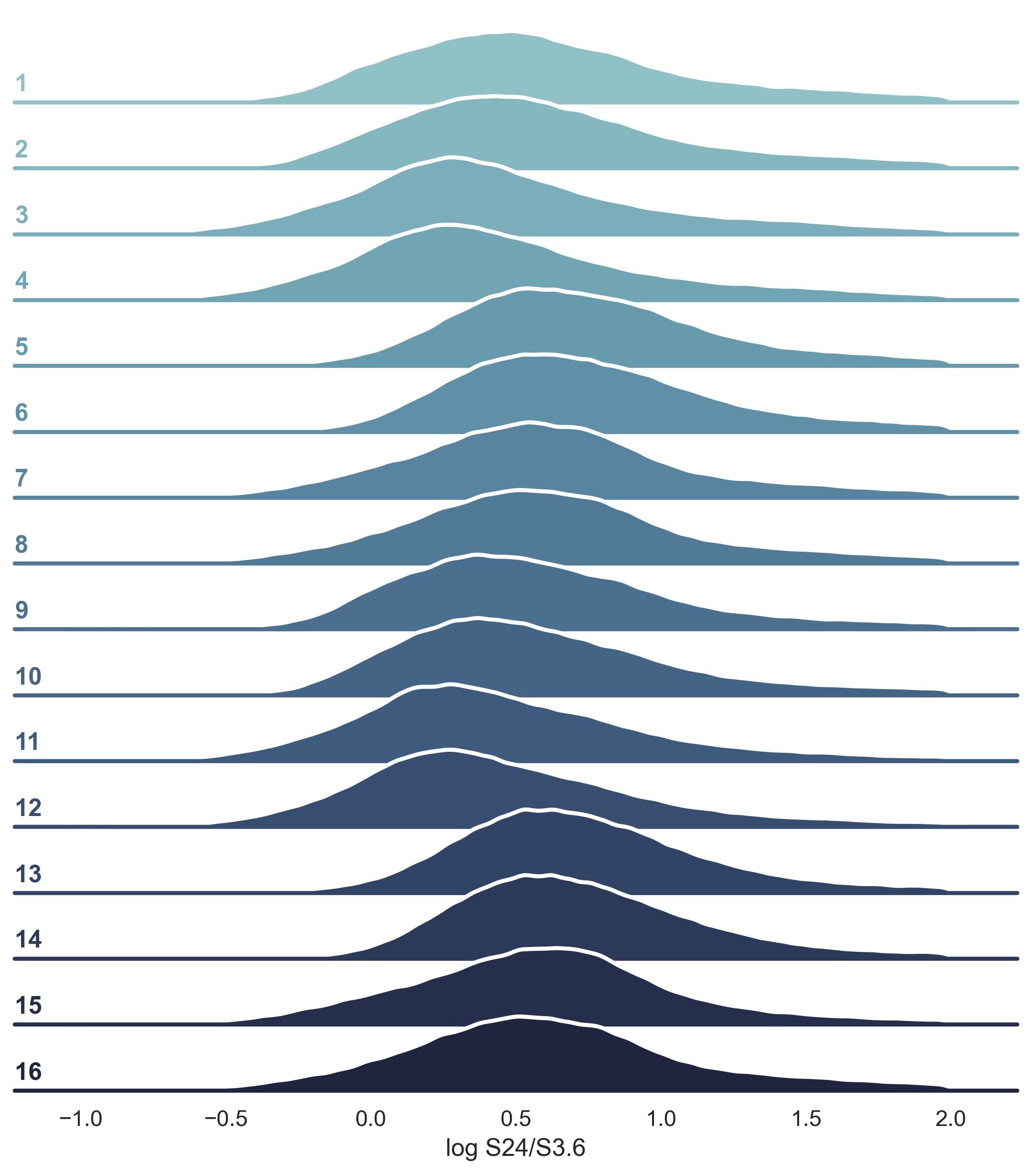}
    \caption{Distributions of the ratio of the predicted MIPS 24 $\mu$m flux  to the observed IRAC 3.6 $\mu$m flux from the 16 CIGALE SED fitting runs, showing systematic effects arising from varying assumptions (such as the adopted SFH, AGN templates, dust attenuation and emission models) in the SED modelling and fitting process.}
      \label{ridge}
\end{figure}

\subsubsection{The CIGALE SED library}

Our probabilistic and progressive deblending starts from the {\it Spitzer}/MIPS 24 $\mu$m and then moves to the {\it Herschel} PACS and SPIRE bands. This approach has the benefit of staying as close as possible to the measured data without having to extrapolate too far down in wavelength, and at the same time extracting as much information as possible from the data. First we select a subset of sources from COSMOS2020 which are most likely to contribute to the emission in the 24 $\mu$m map. 

We made use of the SED modelling and fitting tool Code Investigating GALaxy Emission (CIGALE; \citealt{2009A&A...507.1793N}) with an improved fitting procedure by \citet{2011ApJ...740...22S} to predict the 24 $\mu$m flux in the observed frame based on the multi-band photometry in COSMOS2020. To make the process of SED fitting and flux prediction for the entire catalogue faster, we first used CIGALE to predict the observed 24 $\mu$m fluxes for a random sample of 200,000 sources. To explore systematic uncertainties introduced by varying assumptions in SED fitting, we carried out a total of 16 CIGALE runs with all possible combinations of two star-formation history models (a delayed $\tau$ model plus a late starburst or double exponential forms), two dust attenuation models (modified  \citealt{2000ApJ...539..718C} or modified \citealt{2000ApJ...533..682C} attenuation law), two dust emission models (IR SED templates from the  \citealt{2014ApJ...780..172D} model or the  \citealt{2014ApJ...784...83D} model), and two active galactic nuclei (AGN) models (smooth AGN templates from the \citealt{2006MNRAS.366..767F} model or clumpy AGN templates from the SKIRTOR model in \citealt{2012MNRAS.420.2756S}). The SED model components and parameters (representing a range of common configurations in the literature) are listed in Table \ref{tab:parameter}. Fig.~\ref{ridge} shows the distributions of the ratio between the predicted 24 $\mu$m flux to the observed IRAC 3.6 $\mu$m flux from the 16 CIGALE SED fitting runs. Overall, the systematic differences between the different runs are smaller than the spread in each run, demonstrating that the SEDs are well sampled enough to allow for predictions which are not very sensitive to inherent assumptions in the SED fitting.

\subsubsection{The deep learning-based SED emulator}\label{222}

\begin{figure}
   \centering
   \includegraphics[width=9cm, height=7.5cm]{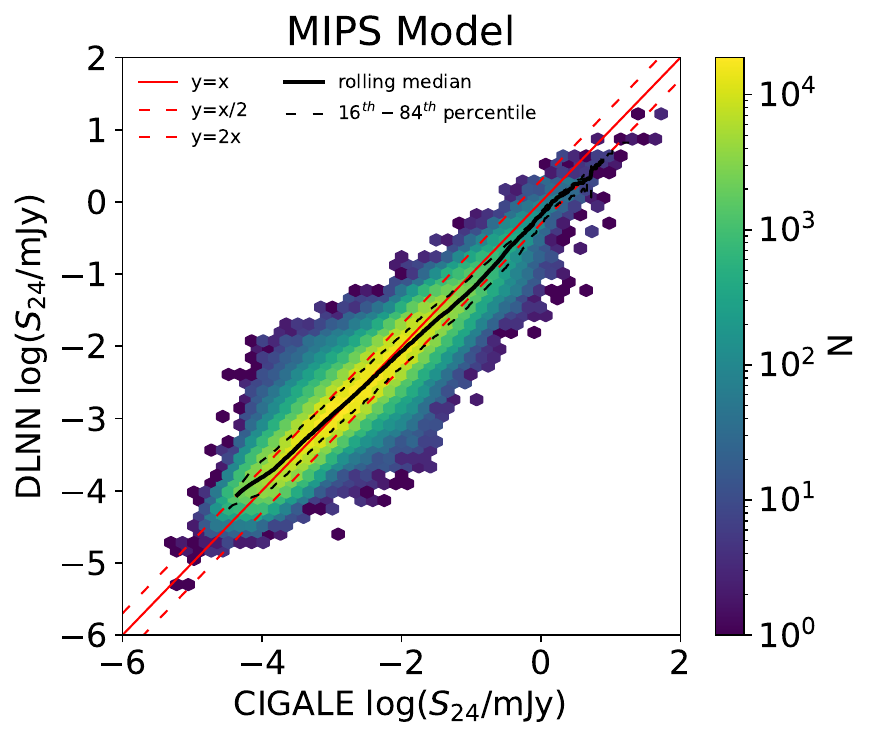}
      \caption{Predicted MIPS 24 $\mu$m flux from the DLNN on the test set compared to the predicted values from the 16 CIGALE runs. Colour coding is based on number of sources. The running median is close to the one-to-one line. The 16th and 84th percentile ranges are mostly within a factor of two from the one-to-one line.}
         \label{MIPS24}
\end{figure}



The inputs (i.e. the multi-band photometry in  COSMOS2020) and outputs (i.e. the predicted flux in a  specific far-IR/sub-mm band to be deblended) of the 16 CIGALE runs form an SED library which is then used to train a DLNN developed in \citet{2023MNRAS.520.3529E}. The DLNN model consists of four linear fully connected layers with  different numbers of neurons and non-linear activation functions, as listed in Table \ref{tab:DLNN}. The purpose of using a DLNN as an SED emulator was to speed up the process of predicting the observed 24 $\mu$m flux density. To ensure reliability of the predicted fluxes, we selected sources which are detected at a signal-to-noise ratio (S/N) $>0.1$ in at least 40\% of the available bands. All flux measurements were converted to log-space. We augmented the COSMOS2020 multi-band photometric data by calculating 16 colours\footnote{A popular way of improving the performance of machine learning methods is through combining existing features with new features, which can be obtained from the differences and/or ratios of the existing features.}, namely $u-r$, $g-r$, $r-i$, $i-z$, $z-y$, $J-K_s$, $H-Y$, $J-H$, $Y-K_s$, $r-K_s$, $u-K_s$, $r-$ IRAC channel 1 (3.6 $\mu$m), $u-$ IRAC channel 4 (8.0 $\mu$m), IRAC channel 1 $-$ channel 3 (5.8 $\mu$m), IRAC channel 1 $-$ channel 4, and IRAC channel 2 (4.5 $\mu$m) $-$ channel 4. In addition, we applied a normalisation process to scale data to a range of 0 - 1 using the transformation below,
\begin{equation}
        x' = \frac{x-min}{max-min},
\end{equation}
where $x$ refers to a given feature of a given source and the $min$ and $max$ refer to the minimum and maximum value for all sources. Missing data were assigned a value of $-1$.

\begin{table}[ht]
    \caption{DLNN architecture used as an SED emulator.}
    {\small
    \begin{center}
        \begin{tabular}{lll}
        \hline \hline
        Layer & Properties \\
        \hline
        Input & $N_{in}$ neurons$^a$ & \\
        Linear, fully connected & 2000 neurons & ReLU activation \\
        Dropout & Dropout rate 0.35 & \\
        Linear, fully connected & 1000 neurons & ReLU activation \\
        Linear, fully connected & 500 neurons & ReLU activation \\
        Dropout & Dropout rate 0.35 & \\
        Output, fully connected & $N_{out}$ neurons$^b$ & ReLU activation  \\
        \hline
        \end{tabular}
    \end{center}
    \label{tab:DLNN}}
    \tablefoot{The input and output layers have a number of neurons that depends on the particular task (i.e., predicting fluxes in the MIPS, PACS or SPIRE bands).}  
\footnotesize{$^a$ To predict the 24 $\mu$m flux, the number of neurons in the input layer $N_{in}$ corresponds to the number of the multi-band photometry (satisfying the S/N requirements) in COSMOS2020 and the 16 colours as explained in Section \ref{222}. To predict the PACS and SPIRE fluxes, $N_{in}$ also includes the deblended far-IR/sub-mm fluxes in the previous step(s). \\
$^b$ To predict the 24 $\mu$m flux, the output is a single number. To predict the PACS and SPIRE fluxes, the output consists of two and three numbers, respectively.}\\
\end{table}

We used the TensorFlow framework \citep{2016arXiv160508695A} to build and train the DLNN model.  We used as activation function for each layer a Rectified Linear Unit \citep[ReLU;][]{nair2010rectified} which returns $max(x,0)$ when passed $x$.  To prevent overfitting, we included two dropout layers \citep{srivastava2014dropout} with a dropout rate of 0.35. One after the first layer and the other after the third layer. To train the DLNN we adopted a mean squared error loss function (a typical performance metric for regression problems) which is optimised using the Adam algorithm \citep{kingma2014adam}. The learning rate was set to 0.001. The sample was split into a 40\% training set, a 10\% validation set, and a 50\% {\bf test set}. The DLNN is trained using the training sample, while the validation sample is employed to derive an independent estimate of the loss function. The training and validation samples were passed to the DLNN in batches of 4\,000 objects.
Therefore, the parameters of the DLNN were updated after every 4\,000 objects.  The training stops when the loss function of the validation sample converged, to avoid overfitting the training set. The third dataset, the test sample, is used only to evaluate the network performance. 

In Fig.~\ref{MIPS24}, we compare the predicted value from the trained DLNN ($S_{24}^{DLNN}$) on the test set with the CIGALE predicted 24 $\mu$m fluxes ($S_{24}^{CIGALE}$). Note that as there are 16 CIGALE runs, each galaxy in the test set is plotted 16 times. The running median is close to the one-to-one line across the entire dynamic range with the 16th and 84th percentile ranges are mostly within a factor of two from the 1:1. At the very faint end ($<10^{-4}$ mJy), $S_{24}^{DLNN}$ starts to deviate significantly from $S_{24}^{CIGALE}$. This is not an issue, as we apply a flux cut at $10^{-2}$ mJy or 10 $\mu$Jy to decide whether to keep a source in the prior list (see Section 2.3 and Section 4). We  measure the median bias
\begin{equation}
\widetilde{\Delta S} = \rm{median}\left(S_{24}^{DLNN} - S_{24}^{CIGALE}\right),   
\end{equation}
the normalised median absolute deviation (equivalent to the standard deviation for a normal distribution)
\begin{equation}
    {\rm NMAD} = 1.48\cdot \rm{median( | (S_{24}^{DLNN} - S_{24}^{CIGALE}) - \widetilde{\Delta S} | ) },
\end{equation}
and the median fractional difference
\begin{equation}
    {\rm MFD} = {\rm median}\left(\frac{S_{24}^{DLNN} - S_{24}^{CIGALE} }{S_{24}^{CIGALE}} \right).
\end{equation}
We report these metrics in Table \ref{tab:DLNN_preformance}. The perfect algorithm would have zero bias, MFD and NMAD.



\begin{figure}
\centering
\includegraphics[width=9.8cm]{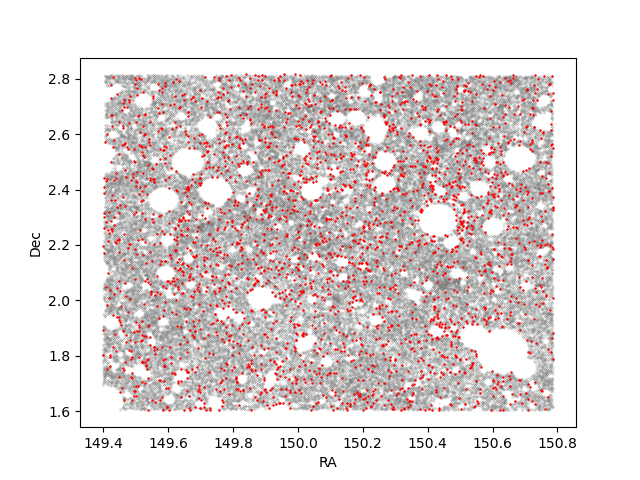}
\caption{Sky distribution of the sources in our initial prior catalogue for deblending the MIPS 24 $\mu$m data. The grey data points are the sources selected from COSMOS2020. The red data points are the additional radio sources from the VLA-COSMOS 3GHz and MIGHTEE 1.3 GHz surveys which are not included in COSMOS2020.}
\label{Figpos}
\end{figure}

\begin{table}[ht]
    \caption{DLNN performance.}
    \begin{center}
   \begin{tabular}{llll}
        \hline
        DLNN Model & $\widetilde{\Delta S}$ & NMAD  & MFD \\
        \hline
        MIPS & 0.0000 & 0.0011 & $-0.0028$ \\
        PACS100 & $-0.0001$ & 0.0096 & $-0.0110$ \\
        SPIRE250 (opt. - PACS) & $-0.0006$ &  0.0066 & $-0.0230$ \\
        SPIRE250 (opt. - MIPS) & $-0.0001$ & 0.0240 & $-0.0063$\\
        \hline
        \end{tabular}
    \end{center}
    \label{tab:DLNN_preformance}
    \tablefoot{
Both the median bias $\widetilde{\Delta S}$ and the normalised median absolute deviation NMAD are in unit of mJy. The median fractional difference MFD is dimensionless. For the PACS DLNN, we report the metrics measured for the 100 $\mu m$ band. For the two SPIRE DLNNs (one with optical to MIPS data and one with optical to PACS data), we report the 
    results for the 250 $\mu m$ band.

    }
\end{table}


\subsection{The initial prior catalogue for deblending the MIPS 24 $\mu$m data}

After training the DLNN, we then applied it to the whole dataset of 711\,106 sources selected from  COSMOS2020 to predict their observed 24 $\mu$m fluxes. We constructed an initial prior catalogue for deblending the MIPS 24 $\mu$m image as follows:
\begin{enumerate}
\item We included 121\,227 sources from COSMOS2020 with lp\_type = 0 (galaxies) and predicted $S24>$ 10 $\mu$Jy. Note that we use the notations $S_{24}$ and $S24$ (and similarly for fluxes at other wavelengths) interchangeably. This flux cut is chosen in light of the noise properties of the 24 $\mu$m map and the validation results based on simulations (see Section \ref{sim_mips}).
\item We included 5\,141 sources from COSMOS2020 with lp\_type = 0 but no predicted $S24$, in order to catch sources for which we could not predict their 24 $\mu$m fluxes due to insufficient SED coverage and/or low S/N.
\item We included 2\,019 sources from COSMOS2020 with lp\_type = 2 (X-ray sources) for which the SED modelling and fitting procedure may not work as well as for other source types. 
\item We included 2\,791 radio sources from the VLA-COSMOS 3 GHz and MIGHTEE 1.3 GHz surveys which are not present in COSMOS2020. This step is to include additional high-$z$ far-IR/sub-mm emitting sources.
\end{enumerate}
To summarise, we found a total of 131\,178 sources (corresponding to 0.29 sources per beam) in the initial prior catalogue for deblending the MIPS 24 $\mu$m map. Fig.~\ref{Figpos} shows the sky distribution of these sources, with the missing radio sources in red. In particular, the extra radio sources seem to cluster around the masked region in COSMOS2020. 

\begin{figure}
   \centering
   \includegraphics[width=9.cm, height=11.5cm]{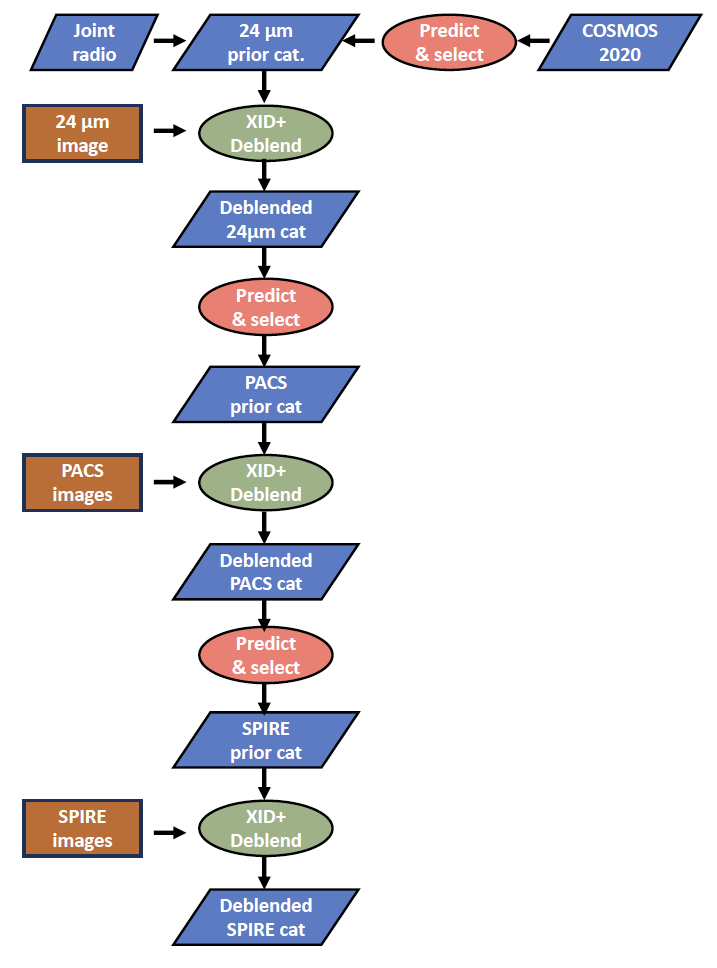}
      \caption{Flowchart of our progressive deblending from 24 $\mu$m to the sub-mm wavelengths. Our initial prior catalogue for deblending the 24 $\mu$m map is constructed using COSMOS2020 and additional radio sources from the VLA-COSMOS 3 GHz and MIGHTEE 1.3 GHz catalogues. At each step of the deblending, we use a DLNN trained on an SED library to predict the fluxes of the sources in the prior catalogue at the relevant wavelength. Based on the predicted flux, we decide whether to keep the source in the prior list.}
         \label{flow}
\end{figure}

\section{Deblending the far-IR and sub-mm data}\label{deblend}

We adopt a progressive deblending approach from the 24 $\mu$m data to the  PACS and SPIRE data.  After the 24 $\mu$m map is de-blended, we add the deblended 24 $\mu$m flux to the sources in the prior list and continue with the prediction of the PACS 100 and 160 $\mu$m fluxes, using the aforementioned DLNN trained with the SED library from the 16 CIGALE runs. Similarly, after the PACS maps are deblended and the deblended PACS fluxes are added to the sources in the prior list, we continue with the prediction of the SPIRE 250, 350 and 500 $\mu$m fluxes. A schematic representation of our progressive deblending is shown in Fig.~\ref{flow}. Below we first give an account of the far-IR and sub-mm data to be deblended. Then we describe the characteristics of  XID+ which is a prior-based probabilistic deblending tool and how we use it in this work. 


\subsection{The long wavelength data}

For the MIPS 24 $\mu$m imaging data, we used the GO3 image from the COSMOS-{\it Spitzer} programme \citep{2007ApJS..172...86S}. The noise level corresponds to a 1$\sigma$ standard deviation of 14 to 18 $\mu$Jy in the point source noise \citep{2009ApJ...703..222L}. The $1\sigma$ confusion noise is 11.2 $\mu$Jy \citep{2004ApJS..154...93D}. The map is in unit of MJy/sr. To use XID+, we converted to mJy/beam by multiplying a factor of 1.156, assuming an idealised Gaussian beam with a full width at half maximum (FWHM) of 5.7\arcsec. However, this assumption ignores the side lobes of the real beam. We thus applied a correction factor of 1.369 to account for the lost flux \citep{2006MNRAS.371.1891R}. 


The PACS maps came from the PACS Evolutionary Probe (PEP; \citealt{2011A&A...532A..90L}) survey\footnote{ \url{https://www.mpe.mpg.de/ir/Research/PEP/DR1}}. The FWHM of the point spread function (PSF) is 7.2\arcsec and 12.0\arcsec at 100 and 160 $\mu$m, respectively. The maps, in unit of Jy/pixel, were convert to mJy/beam using a multiplicative factor of 40.8 and 28.3 at 100 and 160 $\mu$m, assuming Gaussian beam. We applied a correction factor of 1.68 and 1.64 at 100 and 160 $\mu$m to account for aperture correction and flux loss due to the high-pass filtering, as detailed in the data release\footnote{ \url{https://www.mpe.mpg.de/resources/PEP/DR1_tarballs/readme_PEP_global.pdf}}. The instrument noise level corresponds to a 1$\sigma$ standard deviation of 1.4 and 3.5 mJy at 100 and 160 $\mu$m. The 1$\sigma$ confusion noise is  0.15 and 0.68 mJy at 100 and 160 $\mu$m \citep{2011A&A...532A..49B, 2013A&A...553A.132M}. 

The SPIRE maps were obtained from the fourth  release of the {\it Herschel} Database in Marseille hosting data from the {\it Herschel} Multi-tiered Extragalactic Survey (HerMES; \citealt{2012MNRAS.424.1614O})\footnote{\url{https://hedam.lam.fr/HerMES/}}. These maps are already in units of mJy/beam. The PSF FWHM is equal to 18.2, 24.9, and 36.3\arcsec at 250, 350, and 500 $\mu$m, respectively. The 1$\sigma$ instrument noise is 1.7, 2.7, and 2.9 mJy at 250, 350, and 500 $\mu$m \citep{2012MNRAS.424.1614O}. In comparison, the 1$\sigma$ confusion noise is 6.8, 6.3, and 5.8 mJy at 250, 350, and 500 $\mu$m \citep{2010A&A...518L...5N}. Therefore, the SPIRE maps in COSMOS are dominated by confusion.  

In Table \ref{tab:summary}, we show a summary of the long-wavelength imaging data to be deblended in the COSMOS field, including the nominal wavelength, the FWHM of the PSF, the image pixel size, the flux cut applied to the predicted flux in a given band, the number of sources to be fitted, the source density per beam area, the 1-$\sigma$ instrument noise and confusion noise.

\begin{table*}
\caption{Summary of the long-wavelength imaging data.} 
\centering                      
\begin{tabular}{l l l l l l l l l}        
\hline\hline                 
Wavelength & Telescope/Instrument & PSF FWHM & pixel size & $S_{cut}$& $N_{fit}$ & $\rho_{fit}$& $\sigma_{instrument}$ & $\sigma_{confusion}$\\    
\hline                        
   24 $\mu$m & {\it Spitzer}/MIPS & 5.7\arcsec &  1.2\arcsec & 10 $\mu$Jy &  131\,178 &0.29& 18.8 $\mu$Jy& 11.2 $\mu$Jy\\      
   100 $\mu$m & {\it Herschel}/PACS & 7.2\arcsec    & 1.2\arcsec & 1.0 mJy& 35\,819& 0.13 &1.4 mJy& 0.15 mJy\\
   160 $\mu$m & {\it Herschel}/PACS & 12.0\arcsec    & 2.4\arcsec  & - &35\,819&0.35 & 3.5 mJy& 0.68 mJy\\
   250 $\mu$m & {\it Herschel}/SPIRE & 18.2\arcsec    & 6.0\arcsec  & 7.0 mJy & 14\,869 & 0.34 & 1.7 mJy& 6.8 mJy\\
   350 $\mu$m & {\it Herschel}/SPIRE & 24.9\arcsec    & 8.3\arcsec  & - & 14\,869 & 0.63 & 2.7 mJy& 6.3 mJy\\
   500 $\mu$m & {\it Herschel}/SPIRE & 36.3\arcsec    & 12.0\arcsec  & - & 14\,869 & 1.34  & 2.9 mJy& 5.8 mJy\\
\hline         
\hline
\label{tab:summary}
\end{tabular}
  \tablefoot{
The columns (from left to right) are the nominal wavelength, telescope/instrument, FWHM of the PSF, image pixel size, flux cut applied to the predicted flux, number of sources to be fitted, source density per beam area, the 1$\sigma$ instrument noise and the 1$\sigma$ confusion noise.
  }
\end{table*}

\subsection{Probabilistic deblending with XID+}

\begin{figure}
   \centering
   \includegraphics[width=9.2cm, height=7.5cm]{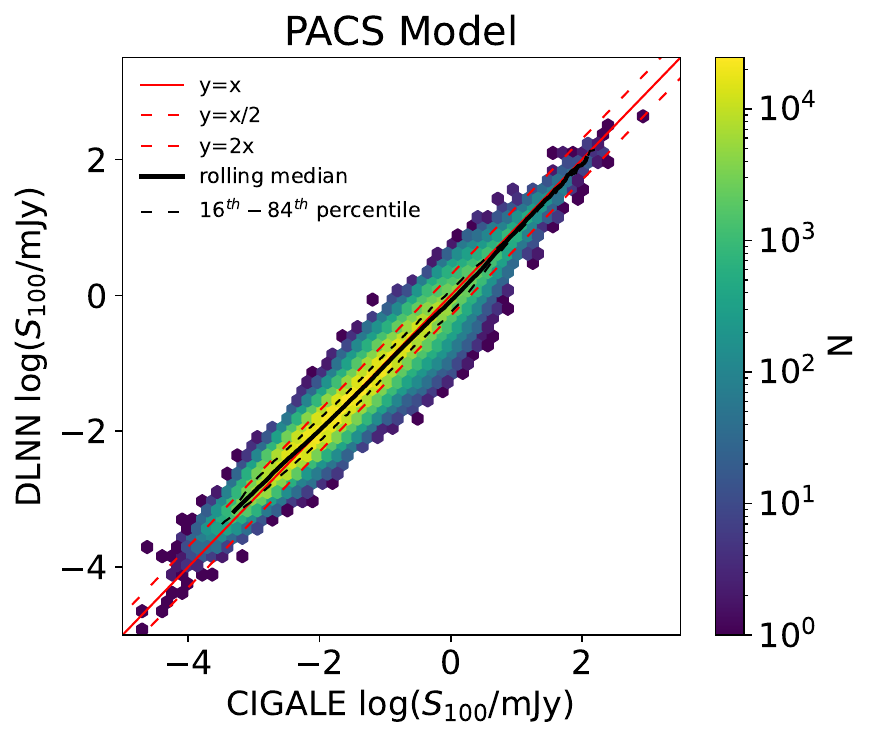}
   \caption{DLNN predicted PACS 100 $\mu$m fluxes on the test set compared to the values from the 16 CIGALE runs. The running median is close to the 1:1 ratio. The 16th and 84th percentile ranges are mostly within a factor of two from the 1:1.}
   \label{NN_pacs}
\end{figure}

\begin{figure}
   \centering
\includegraphics[width=9.2cm, height=7.5cm]
{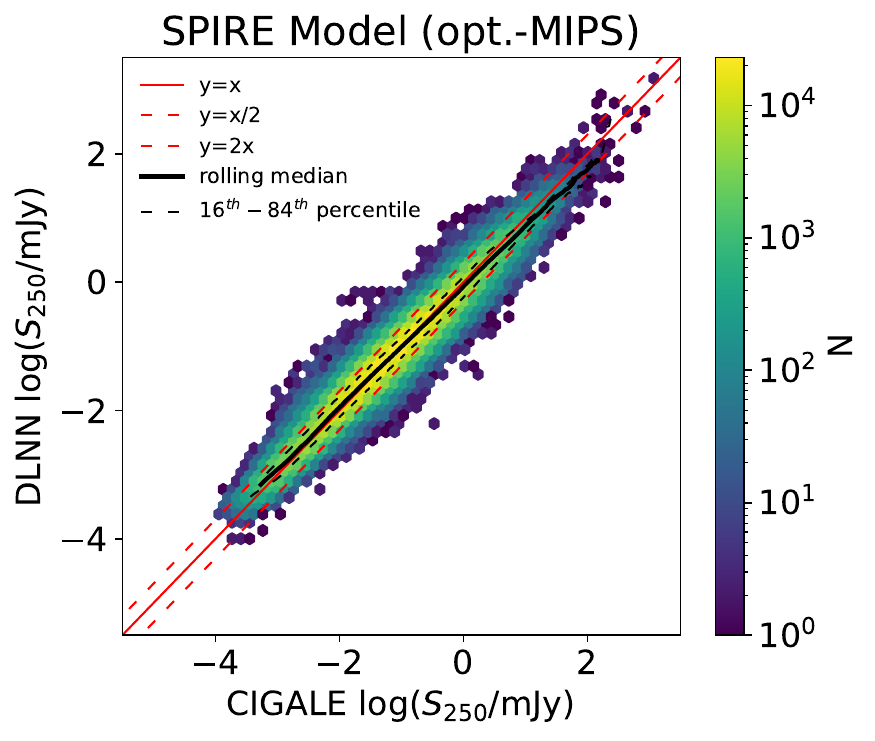}
\includegraphics[width=9.2cm, height=7.5cm]{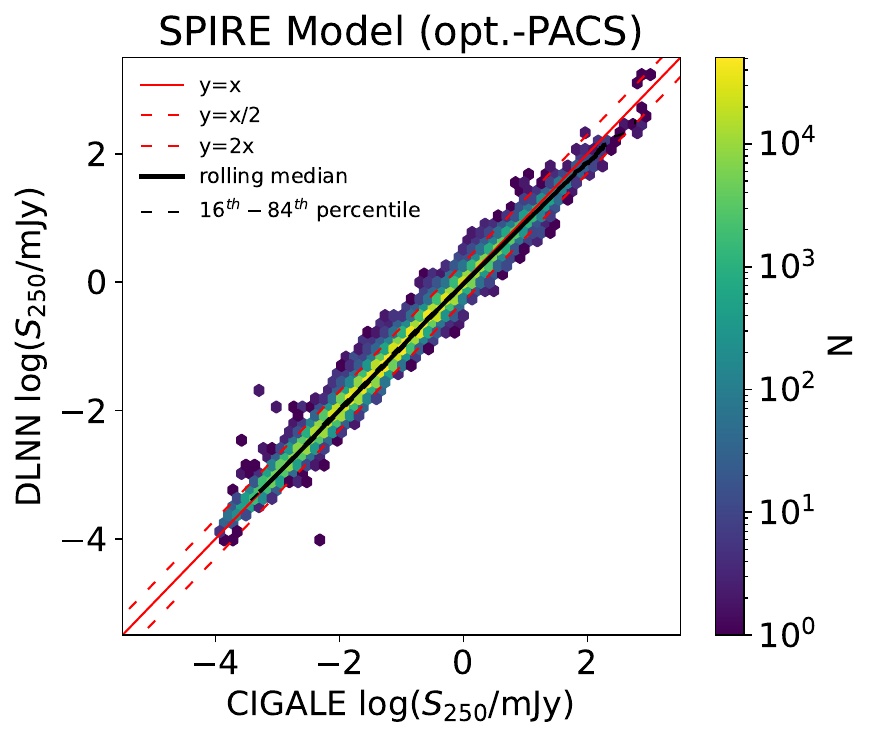}
   \caption{DLNN predicted SPIRE 250 $\mu$m flux on the test set compared to the values from the 16 CIGALE runs. The top panel corresponds to predictions from the DLNN using photometry from COSMOS2020 and the deblended 24 $\mu$m flux. The bottom panel corresponds to predictions from the DLNN using all available photometry out to the PACS wavelengths.}
   \label{NN_spire}
\end{figure}

\begin{figure*}
    \centering
\includegraphics[width=.99\textwidth]{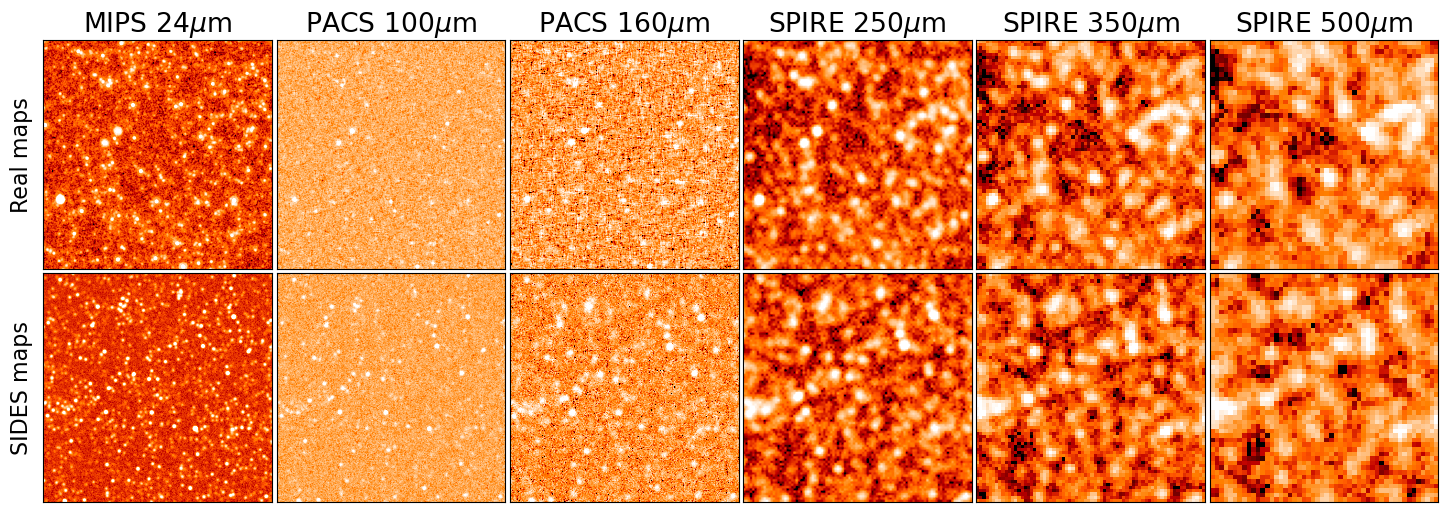}
    \caption{Same $10\arcmin\times10\arcmin$ patch of the sky at different wavelengths from real observations (top) and the SIDES simulation including Gaussian random noise (bottom). The simulation can statistically reproduce the observed far-IR and sub-mm sky reasonably well. One can also notice that the PACS 100 and 160 $\mu$m maps are much more dominated by instrument noise compared to the maps at other wavelengths.}
    \label{fig:cutouts}
\end{figure*}

\citet{2017MNRAS.464..885H} developed a prior-based source extraction tool, XID+\footnote{\url{https://github.com/H-E-L-P/XID_plus}}, to extract flux information at the positions of known sources. XID+ uses a probabilistic Bayesian framework that provides a natural way to include prior information, and uses the Bayesian inference tool Stan \citep{stan2015stan, stan2018pystan} to obtain the full posterior probability distribution on flux estimates. The basic model of our data (i.e., far-IR and sub-mm maps) can be built by the sum of three components,
\begin{equation}
\rm{d} = \Sigma_{i=1}^{S} \rm{\bf{P}} \rm{f_i} + N(0, \Sigma_{instrumental}) + N(B, \Sigma^{residual}_{confusion}).    
\label{eq_xid}
\end{equation}
The first component represents the contribution of the known sources which can be derived by multiplying the pointing matrix (calculated using a Gaussian PSF) and the flux vector (composed of the flux density from each source). The second term represents the instrumental noise which is provided by the uncertainty map of the observations. The third component is a global estimate for the background term with Gaussian fluctuations accounting for the unknown sources\footnote{The unknown sources could include sources which are fainter than our applied flux cut and sources which are not detected in the COSMOS2020 catalogue nor the joint radio catalogues.}. This model can then be compared with the observed data.

A key feature of XID+ (in contrast with maximum-likelihood based photometry methods) is that in addition to estimates of the source flux, it also determines the full posterior probability distributions. The original XID+ used a flat flux prior. Later works added informative flux prior \citep{2017A&A...603A.102P, 2018A&A...615A.146P, 2021A&A...648A...8W} which has the advantage of being able to go down to fainter flux levels. 
In this paper, we use the informative flux prior only in deciding whether a given source should be kept in the prior catalogue. When actually running XID+, we use a flat flux prior in order not to introduce bias in the deblended flux.  Finally, we fit the two PACS bands and the three SPIRE bands simultaneously to take into account the correlations between the bands that are close together.



For deblending the PACS maps, we took the initial prior catalogue (used for deblending the 24 $\mu$m map) and added the deblended 24 $\mu$m fluxes. We then fed these data to the CIGALE-trained DLNN to predict simultaneously the 100 and 160 $\mu$m flux $S_{100}$ and $S_{160}$. In Fig.~\ref{NN_pacs}, we show the comparison between the predicted PACS 100 $\mu$m fluxes by the DLNN ($S_{100}^{DLNN}$) and the predicted values from the 16 CIGALE SED fitting runs ($S_{100}^{CIGALE}$) for the test set. Performance metrics such as the median bias, MFD and NMAD of $(S_{100}^{DLNN} - S_{100}^{CIGALE})$  are reported in Table \ref{tab:DLNN_preformance}. We constructed the prior catalogue for deblending the PACS maps as follows:
\begin{enumerate}
\item We included sources with predicted $S_{100}>1.0$ mJy. This flux cut is chosen in light of the noise properties and the validation results based on simulations (see Section \ref{sim_herschel}).
\item We included sources with predicted $S_{100}<1.0$ mJy or without predicted $S_{100}$, but present in the sample of 2\,019 X-ray sources.
\item We included the 2\,791 radio sources not present in COSMOS2020.
\end{enumerate}
In total, there were 35\,819 sources in the prior catalogue for deblending the PACS maps, corresponding to 0.13 sources per beam at 100 $\mu$m and 0.35 sources per beam at 160 $\mu$m.

To build the prior catalogue for SPIRE, we took the initial prior catalogue and added the deblended 24 $\mu$m and PACS fluxes. Then we fed these data to the DLNN to predict the 250, 350, and 500 $\mu$m flux $S_{250}$, $S_{350}$, and $S_{500}$. 
When the deblended PACS fluxes are available, we use the DLNN trained using also the 100 and 160 $\mu$m CIGALE fluxes. Otherwise, we predict SPIRE fluxes using the DLNN trained with data up to 24 $\mu$m.  In Fig.~\ref{NN_spire}, we show the comparison between the predicted SPIRE 250 $\mu$m fluxes by the two DLNNs ($S_{250}^{DLNN}$) and the predicted values from the 16 CIGALE SED fitting runs ($S_{250}^{CIGALE}$) for the test set. Performance metrics such as the median bias, MFD and NMAD of $(S_{250}^{DLNN} - S_{250}^{CIGALE})$ for the two DLNNs (one with optical to MIPS data and one with optical to PACS data) are shown in Table \ref{tab:DLNN_preformance}. We constructed the prior catalogue for deblending the SPIRE maps as follows:
\begin{enumerate}
\item We included sources with predicted $S_{250}>$ 7.0 mJy. This flux cut is chosen in light of the noise properties and the validation results based on simulations (see Section \ref{sim_herschel}).
\item We included sources with predicted $S_{250}<$ 7.0 mJy or without predicted $S_{250}$ but present in the sample of 2\,019 X-ray sources.
\item We included the 2\,791 radio sources not present in COSMOS2020.
\end{enumerate}
In total, there were 14\,869 sources in the prior catalogue for deblending the SPIRE maps, corresponding to a source density of 0.34 sources per beam at 250 $\mu$m, 0.63 sources per beam at 350 $\mu$m, and 1.34 sources per beam at 500 $\mu$m.

\begin{figure}
   \centering
   \includegraphics[width=9.cm, height=7cm]{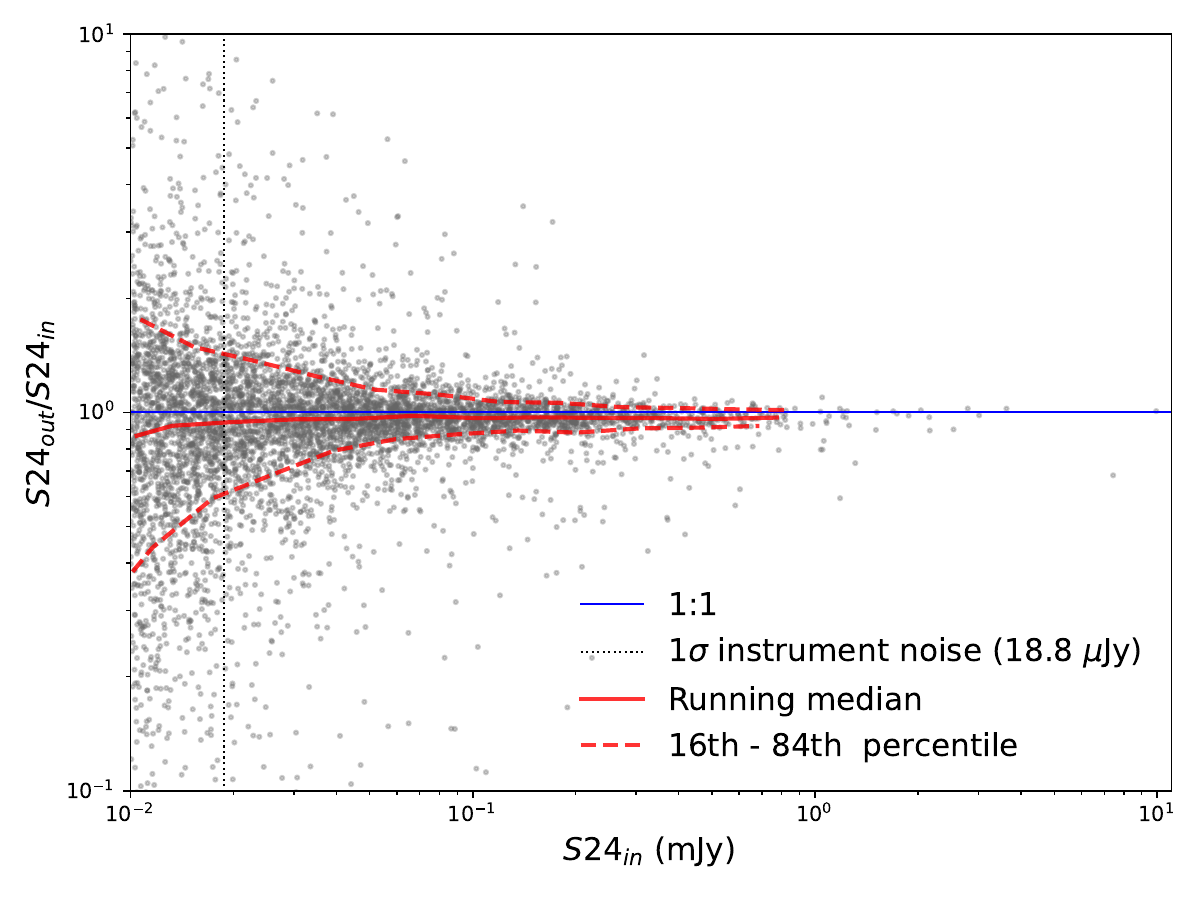}
      \caption{Ratio of the output XID+ flux to the input true flux vs. the true flux at 24 $\mu$m, down to the flux cut of 10 $\mu$Jy. The blue line is the one-to-one correspondence. The vertical dotted line corresponds to the 1$\sigma$ instrument noise. The solid red line is the running median and the dashed red lines are the 16$^{th}$ and 84$^{th}$ percentiles.}
         \label{Fig24}
\end{figure}

\begin{figure}
   \centering
\includegraphics[width=9.cm, height=7.cm]{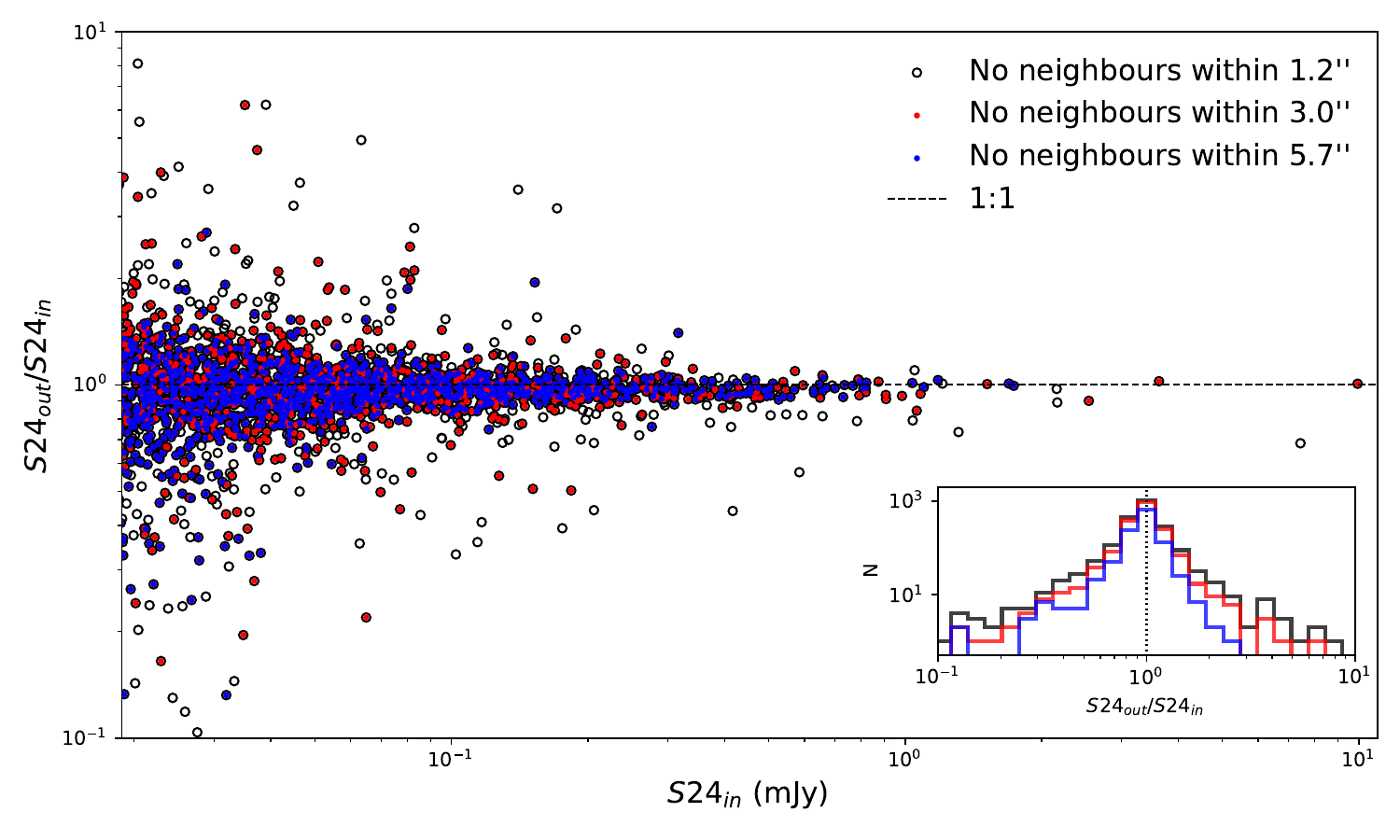}
      \caption{Similar to Fig.~\ref{Fig24}, but only for sources which have true flux $S24_{in}>18.8$ $\mu$Jy (i.e., above the $1\sigma$ instrument noise level) and do not have any neighbours in the prior list within a certain radius (black circles - no neighbours within 1.2"; red dots - no neighbours within 3.0"; blue dots- no neighbours within 5.7"). The inset shows the histograms of the ratio of the output XID+ flux to the input true flux for the three groups.}
         \label{Fig24_accu}
\end{figure}

\section{Validation}\label{validation}

In this section, we validate our deblending methodology using simulations. Specifically, we make use of the continuum observables in Simulated Infrared Dusty Extragalactic Sky (SIDES; \citealt{2017A&A...607A..89B, 2022ascl.soft04016B}) which is a 2 deg$^2$ simulation of the extragalactic sky, including clustering, based on dark matter simulations and empirical prescriptions such as abundance matching between the galaxy stellar mass function (SMF) and the dark matter halo mass function, the fraction of star-forming galaxies as a function of stellar mass and redshift, the SFMS and the fraction of starbursts as a function of redshift. This empirical model is designed to reproduce a range of observed statistical properties of far-IR and sub-mm galaxies, such as the distribution of the surface brightness in the pixels (P(D)), the number counts and the power spectra of the cosmic IR background anisotropies \citep{2023A&A...670A..16G}. To generate simulated far-IR and sub-mm maps, we adopted idealised Gaussian beams which are commonly used as representations of the PSF at these wavelengths. The FWHM of the beams and the pixel sizes are set to the same values as in the observations (Table~\ref{tab:summary}). We also added Gaussian random noise based on the uncertainty map from the real observations. A realistic level of the confusion noise is already included as the SIDES simulation reproduces the observed number counts. The top row of Fig.~\ref{fig:cutouts} shows cutouts of the same $10\arcmin\times 10\arcmin$  patch of the sky in COSMOS, observed at different wavelengths by MIPS, PACS, and SPIRE. In the bottom row, we show cutouts of the same size from the simulated MIPS, PACS, and SPIRE maps, demonstrating that the simulated maps can statistically reproduce the observed far-IR and sub-mm sky reasonably well.

\subsection{Deblending results from the simulated MIPS map}\label{sim_mips}

To make the prior catalogue, we selected a total of 154\,748 sources with a true 24 $\mu$m flux $S24>10\mu$Jy, corresponding to 0.22 sources per beam. This flux cut is equal to the $1\sigma$ confusion noise at 24 $\mu$m. For validation, only the first 100 tiles\footnote{ In XID+, the map to be fitted is segmented into tiles using the Hierarchical Equal
Area isoLatitude Pixelization of a sphere (HEALPix), as it is computationally unfeasible to process the whole map simultaneously. The tiles have equal areas determined by the HEALPix level.} (out of 2470 in total) were run. Fig.~\ref{Fig24} shows the relation between the true  24 $\mu$m flux ($S24_{in}$) and the XID+ deblended flux ($S24_{out}$). The running median of the flux ratio ($S24_{out}/S24_{in}$) is very close to the 1:1 line (with deviation $<5$\%) above the 1$\sigma$ instrument noise of 18.8 $\mu$Jy, demonstrating a high level of flux accuracy. The 16th - 84th percentile ranges widens towards fainter sources, as expected. We can also notice that even at the bright end, XID+ can significantly underestimate or overestimate the fluxes of a few sources. This is due to very nearby sources, as illustrated in Fig.~\ref{Fig24_accu}. The black circles represent the sources without any neighbours within 1.2\arcsec (i.e., the pixel size). The red dots are the sources without any neighbours within 3.0\arcsec. We can  see that the number of sources for which the XID+ flux is very different from the true flux is greatly reduced. So, as long as the sources are separated by at least two pixels, XID+ can correctly deblend them in most cases. Finally, the blue dots represent the sources without any neighbours within 5.7\arcsec (i.e., the PSF FWHM). The deblended fluxes for these sources show further improvement, i.e., better agreement with the true fluxes. 

\begin{figure}
   \centering
   \includegraphics[width=9.cm, height=6.7cm]{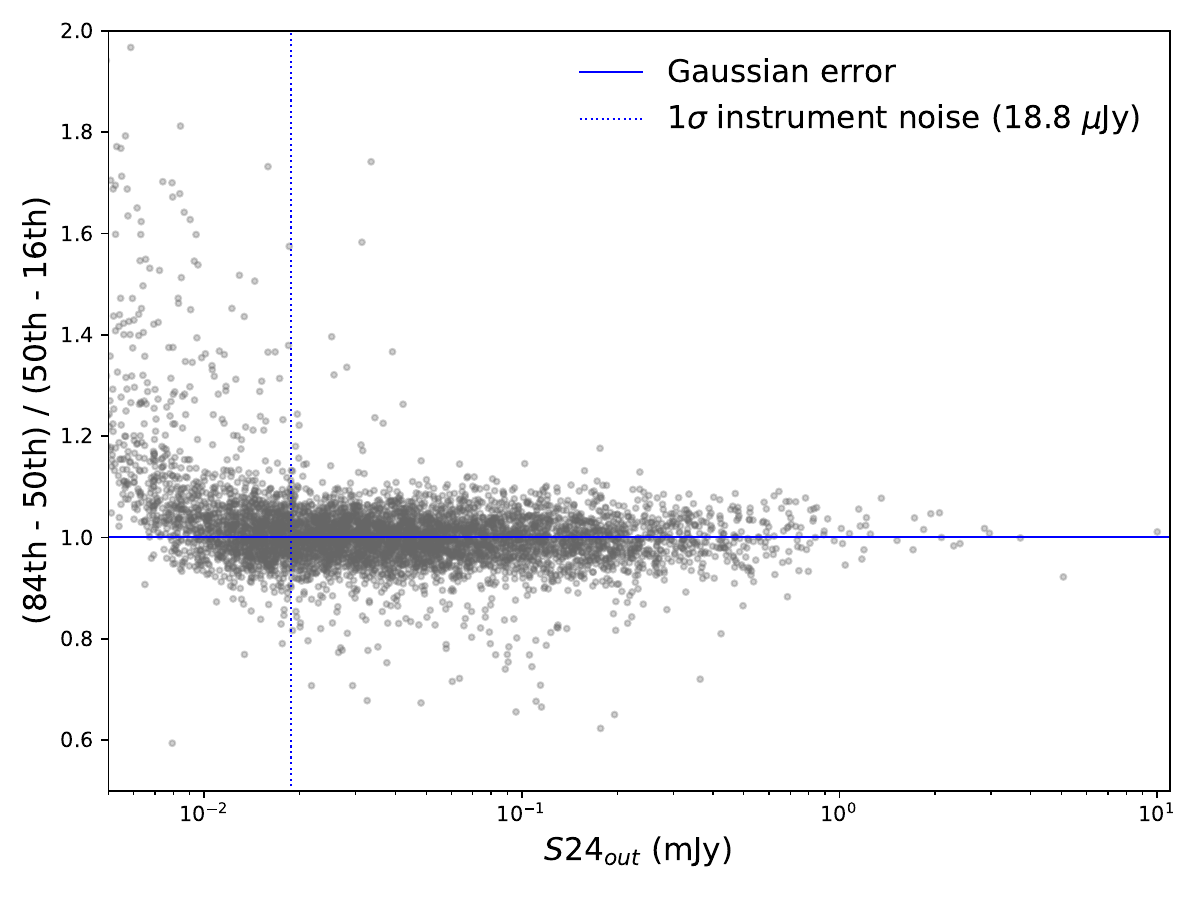} 
      \caption{$\frac{84th - 50th}{50th - 16th}$ percentile ratio (i.e. $\sigma^+/\sigma^-$) of the flux posterior distribution as a function of the output XID+ deblended flux. If the flux uncertainties are Gaussian, this ratio would be equal to one (the horizontal line), which is roughly the case at fluxes above the $1\sigma$ instrument noise of 18.8 $\mu$Jy.}
         \label{Fig24_sigma}
\end{figure}
\begin{figure}
   \centering
   \includegraphics[width=9.cm, height=7cm]{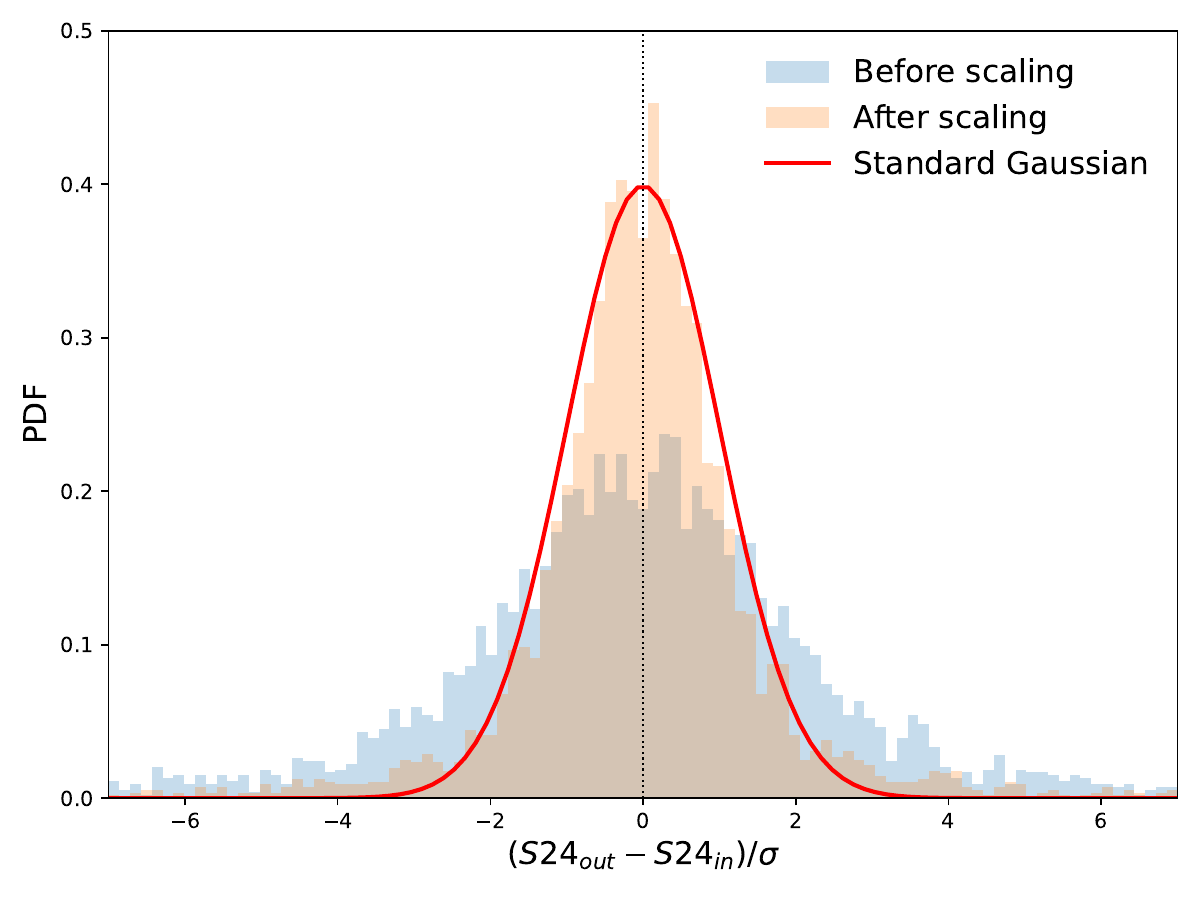}
      \caption{Distribution of the flux difference (between the output XID+ flux and the true flux) normalised by the XID+ flux uncertainty estimate, before and after multiplying the $1\sigma$ uncertainty by a factor of two. The red line corresponds to a standard Gaussian.}
         \label{Fig24_norm}
\end{figure}

\citet{2017MNRAS.464..885H} showed that for faint sources comparable to or below the noise level, their flux posterior distributions will be non-Gaussian, due to the limit imposed by noise. To illustrate this, we calculated from the posterior $\sigma^+$ (from the difference between the 84th percentile and the median) and $\sigma^-$ (from the difference between the median and 16th percentile). Fig.~\ref{Fig24_sigma} shows the ratio of $\sigma^+$ to $\sigma^-$ as a function of the median (i.e. what we assign as the best estimate of the XID+ deblended flux). The flux uncertainties become roughly Gaussian at fluxes above the 1$\sigma$ instrument noise. 
We also checked the accuracy of the flux uncertainty. First, we derived the $1\sigma$ flux uncertainty estimate from the maximum of $\sigma^+$ and $\sigma^-$. Following \citet{2018ApJ...853..172L} and \citet{2018ApJ...864...56J}, we plot in Fig.~\ref{Fig24_norm} the probability distribution of the difference between the XID+ flux $S24_{out}$ and the true flux $S24_{in}$ divided by the $1\sigma$ flux uncertainty, for sources with $S24_{out}>18.8$ $\mu$Jy which is the $1\sigma$ instrument noise. The distribution (the blue histogram) is wider than the standard Gaussian with a mean of zero and standard deviation of one, indicating that the flux uncertainties are underestimated\footnote{Based on a frequentist interpretation, we expect the estimated fluxes to fall within $1\sigma$ roughly 68\% of the time and within $2\sigma$ roughly 95 of the time.}. If we multiply the $1\sigma$ error by a factor of two, the distribution (the brown histogram) then closely follows the standard Gaussian. In fact, the same multiplicative scaling factor is also needed in the {\it Herschel} SPIRE bands. For a detailed explanation of why we recommend applying a scale factor of two to the $1\sigma$ error in the MIPS and SPIRE bands, please refer to Appendix \ref{appendix_scaling}.



\begin{figure}
   \centering 
\includegraphics[width=9.cm, height=6.7cm]{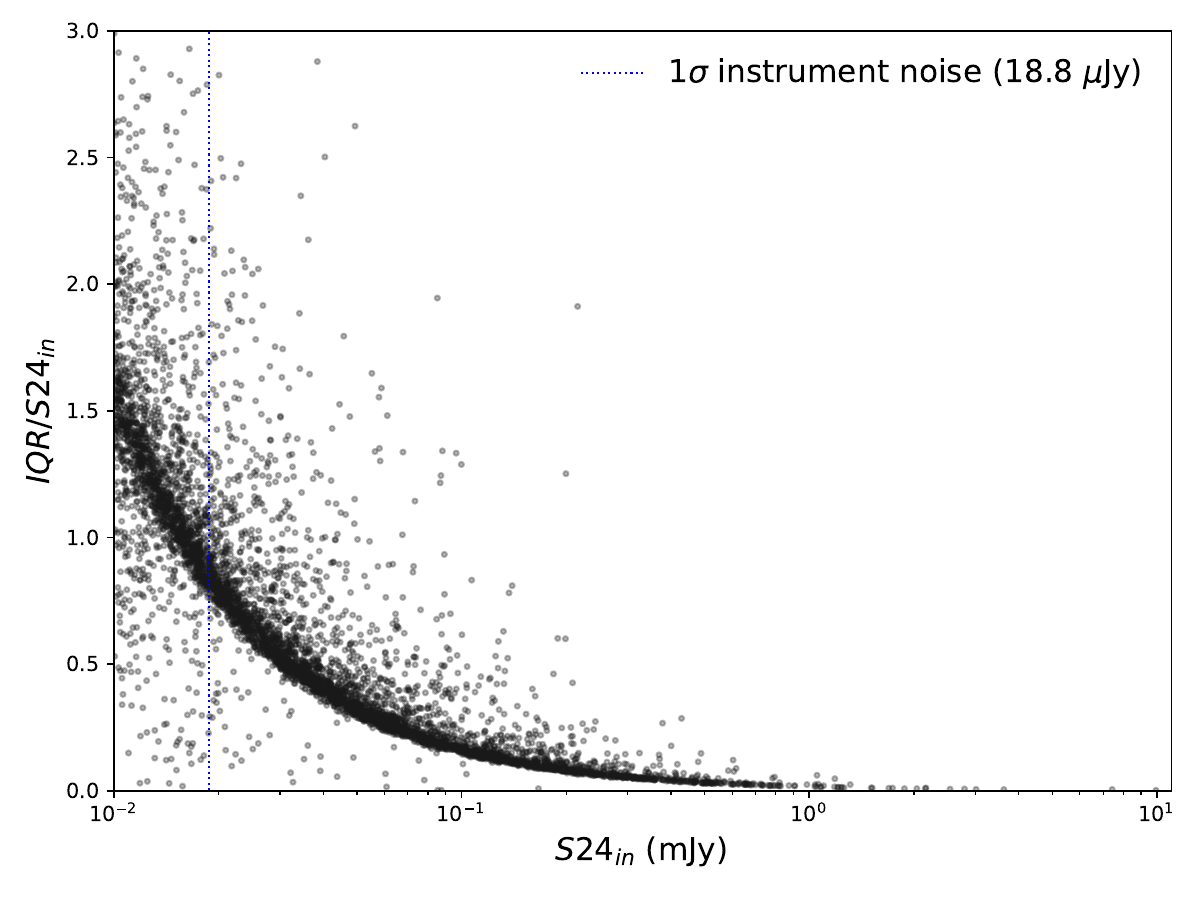}
      \caption{Ratio of the IQR to the true flux vs. the true flux, after scaling by a factor of two. The vertical line is the $1\sigma$ instrument noise.}
         \label{Fig24_unc}
\end{figure}

\begin{figure}
   \centering
   \includegraphics[width=9.cm, height=7cm]{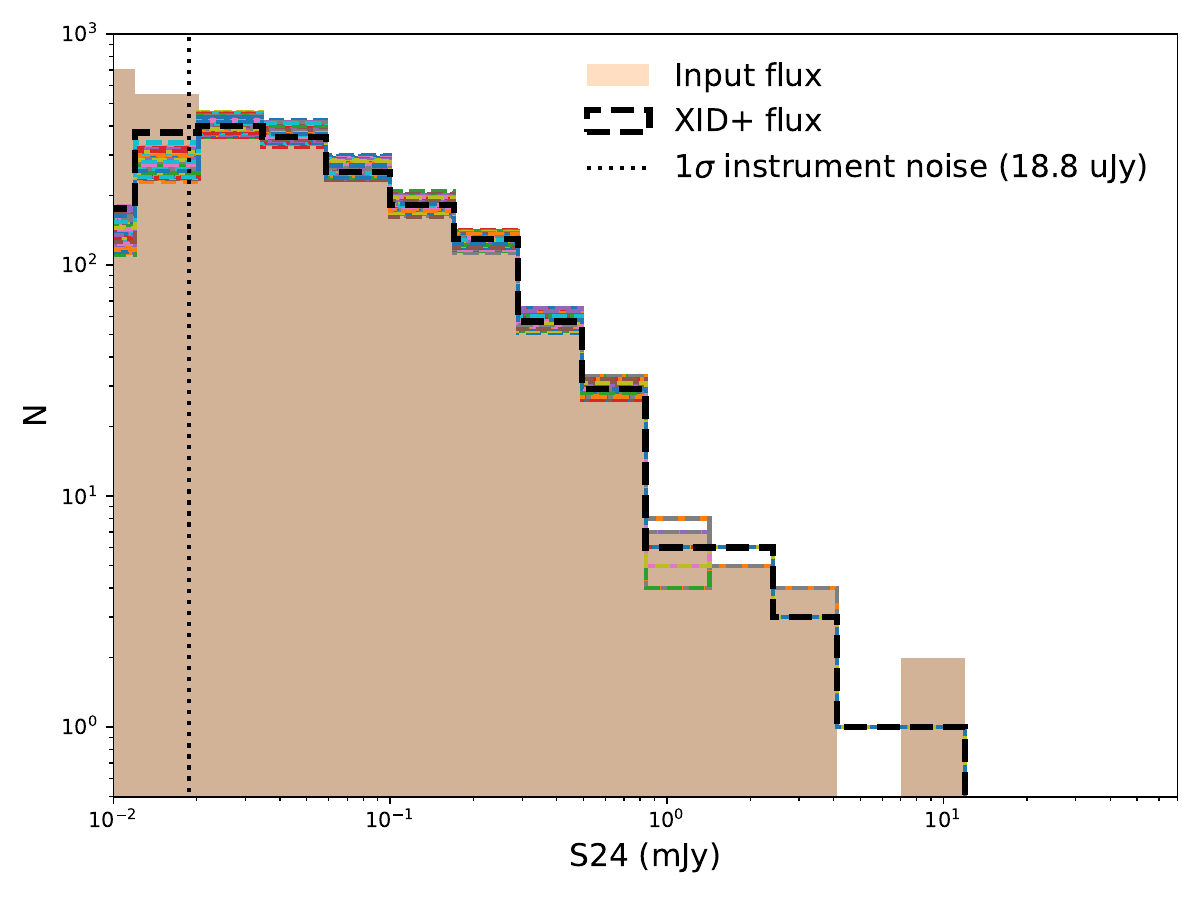}
      \caption{Comparison of the true 24 $\mu$m number counts (filled histogram) with the counts extracted from the output XID+ flux (black line). The dashed histograms of different colours represent the 3000 samplings from the posterior PDFs. The vertical dotted line is the $1\sigma$ instrument noise.}
         \label{Fig24_NC}
\end{figure}

To investigate how well constrained the XID+ deblended flux is (i.e. flux precision), we show in Fig.~\ref{Fig24_unc}  the interquartile range (IQR) divided by the true flux and how it varies as a function of the true flux. The IQR is the difference between the 84th and 16th percentile of the posterior. To account for the systematic underestimation in the flux uncertainty, we multiply the IQR by a factor of two. The ratio of the IQR to the true flux generally decreases (indicating higher precision) for intrinsically brighter sources. When the sources are very faint (similar to the $1\sigma$ instrument noise), their IQR values are comparable to (or even larger than) the true fluxes. 
Finally, we examined how well we can recover the true number counts. In Fig.~\ref{Fig24_NC}, we show the true counts as the brown histogram which rise steeply with decreasing flux. The black dashed line represents the output counts derived from the best estimate of the deblended flux. The dashed lines with different colours correspond to the extracted counts using the 3000 samplings from the XID+ posterior probability distribution functions (PDFs). To account for the systematic flux uncertainty underestimation, we rescaled the 3000 samplings from the posterior for each source as follows,
\begin{equation}
S'_i = S_{50th} + 2 \times (S_i - S_{50th}), {\rm with}\  i = 1, ..., 3000,
\label{eq:scale}
\end{equation}
where $S_{50th}$ is the median (i.e., the best estimate) of the source flux.
The output counts are close to the true counts above the $1\sigma$ instrument noise. Below the $1\sigma$ instrument noise (where both flux accuracy and flux precision are poorer), the output counts fall increasingly under the true counts. 



\subsection{Deblending results from the simulated {\it Herschel} PACS \& SPIRE maps}\label{sim_herschel}

\begin{figure}
   \centering
    \includegraphics[width=9cm, height=7.cm]{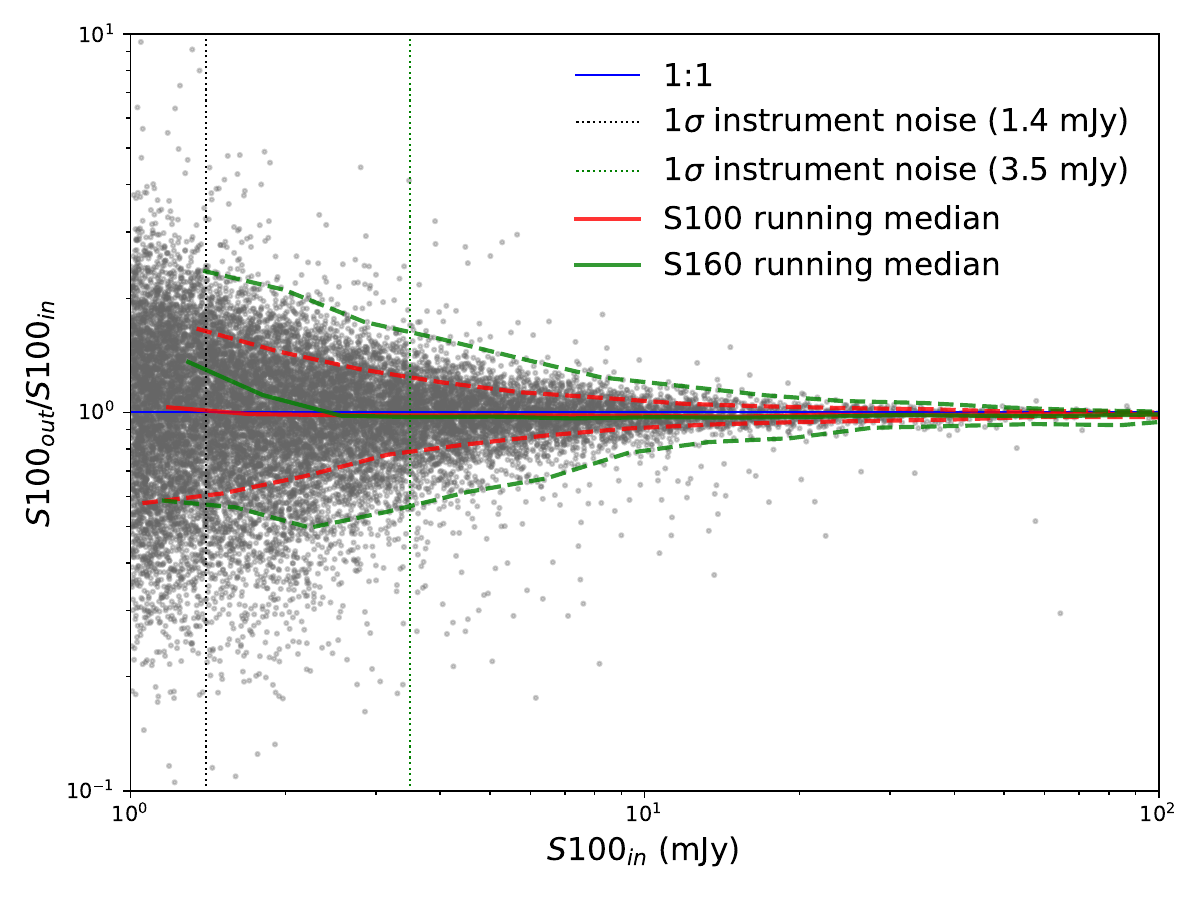}
      \caption{Ratio of the output XID+ flux to the input flux vs the input flux
at 100 $\mu$m, down to the flux cut of 1 mJy. The blue line is the one-to-
one correspondence. The black and green vertical dotted line corresponds to the 1$\sigma$ instrument noise level of 1.4 and 3.5 mJy at 100 and 160 $\mu$m respectively. The red/green solid line is the running
median at 100/160 $\mu$m. The red/green dashed lines are the 16th and 84th percentiles at 100/160 $\mu$m.}
         \label{sim_100_compare}
\end{figure}

The PACS maps are dominated by instrument noise. Therefore, we selected sources with a true 100 $\mu$m flux $>1$ mJy (close to the $1\sigma$ instrument noise) which gave us a total of 43\,615 sources (0.10 sources per beam). Fig.~\ref{sim_100_compare} compares the output to input flux ratio at 100 $\mu$m as a function of the true 100 $\mu$m flux. The red solid line is the running median and the red dashed lines correspond to the 16th - 84th percentile ranges. We also overplot the running median and the 16th - 84th percentiles at 160 $\mu$m as the green solid line and green dashed lines, respectively. The black and green vertical dotted lines represent the instrument noise level at 100 and 160 $\mu$m, which is 1.4 and 3.5 mJy, respectively. Again, we find a very good agreement between the output flux and the true flux above $1\sigma$ instrument noise in both PACS bands.

\begin{figure}
    \centering
\includegraphics[width=9cm, height=7.cm]{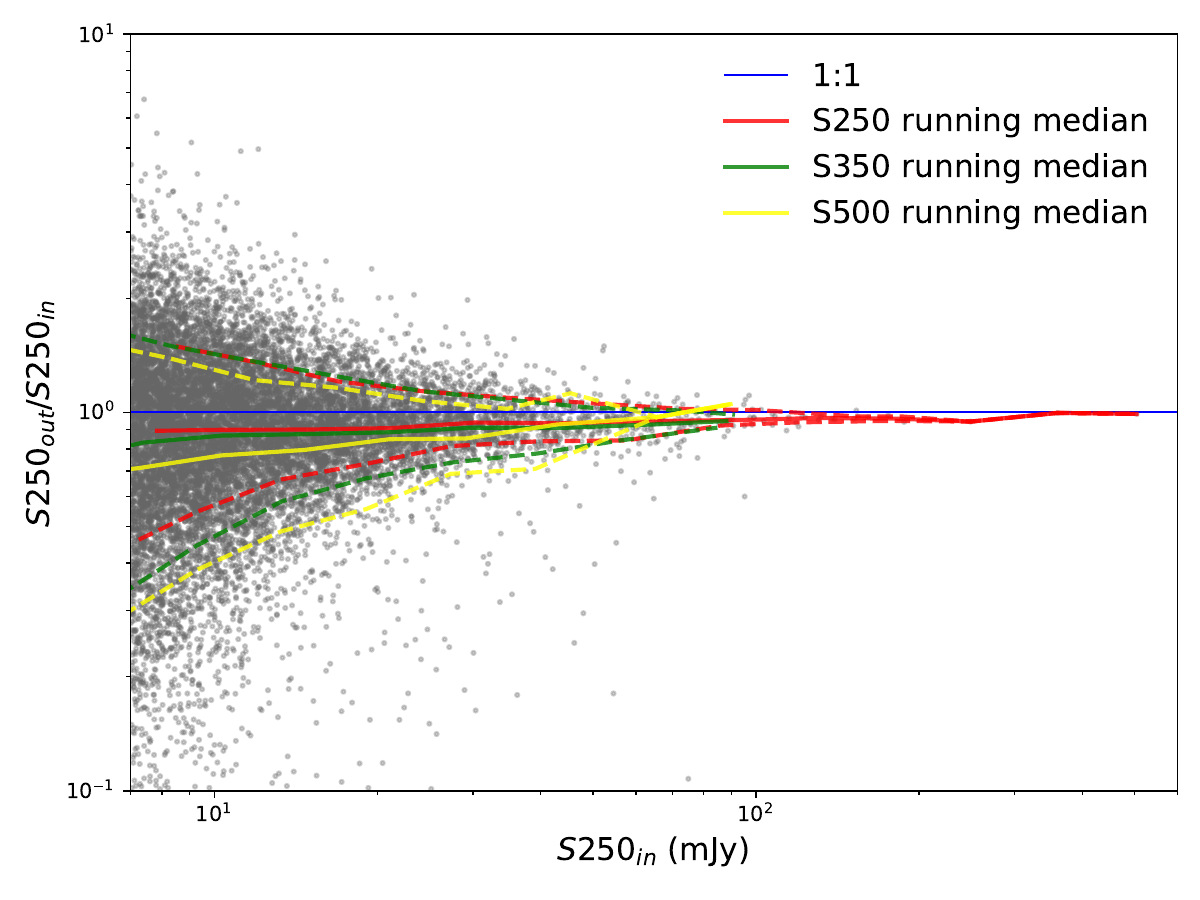}
    \caption{Ratio of the output XID+ flux to the input true flux vs the true flux at 250 $\mu$m, down to the flux cut of 7 mJy. The blue line is the 1:1 ratio. The solid and dashed red/green/yellow lines are the 250/350/500 $\mu$m running median and the 16th - 84th percentiles, respectively.}
    \label{fig:spire_frac_err}
\end{figure}

\begin{figure}
    \centering
\includegraphics[width=.49\textwidth]{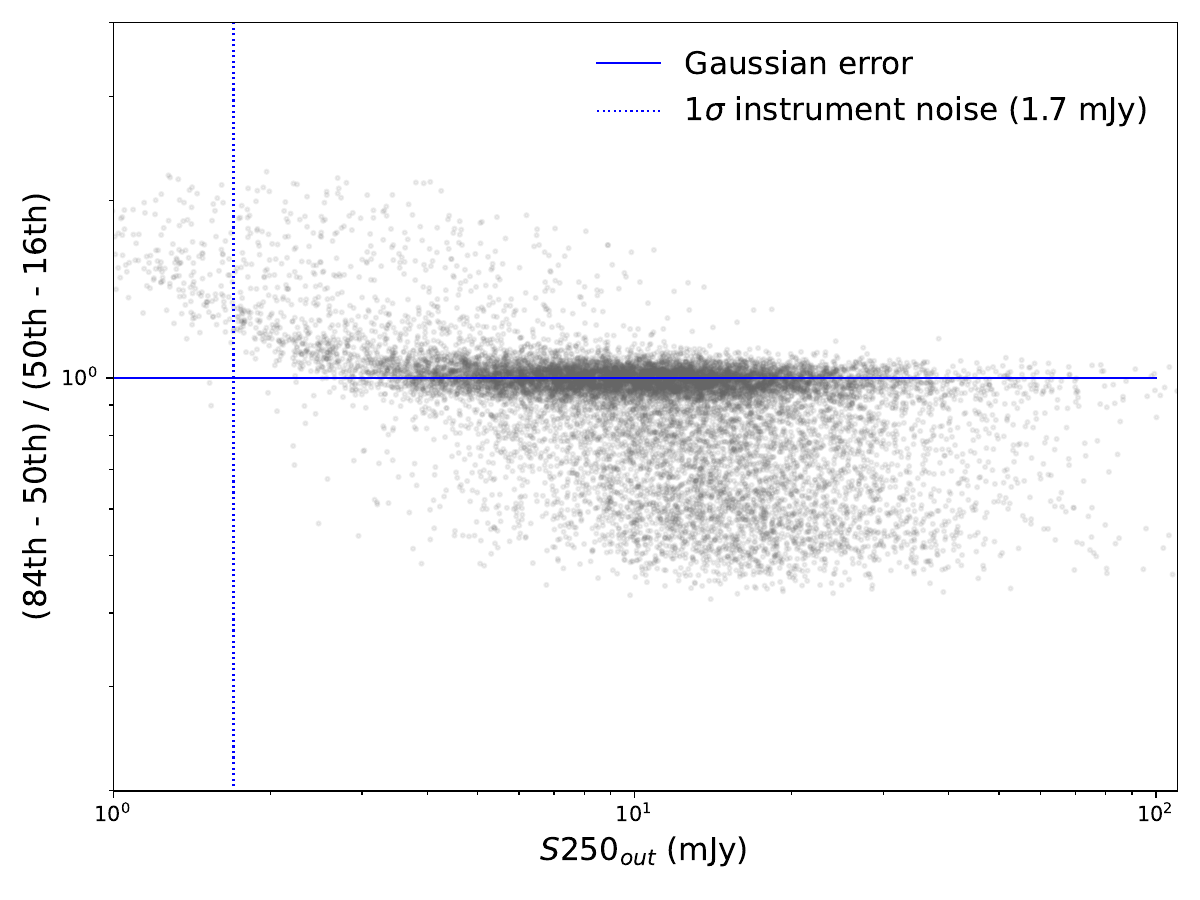}
    \caption{$\frac{84th - 50th}{50th - 16th}$ percentile ratio  (i.e. $\sigma^+/\sigma^-$) of the flux posterior as a function of the output XID+ deblended flux). If the flux uncertainties are Gaussian, this ratio would be equal to one (the horizontal blue line).}
    \label{fig:spire_unc}
\end{figure}

\begin{figure}
    \centering
\includegraphics[width=.49\textwidth]{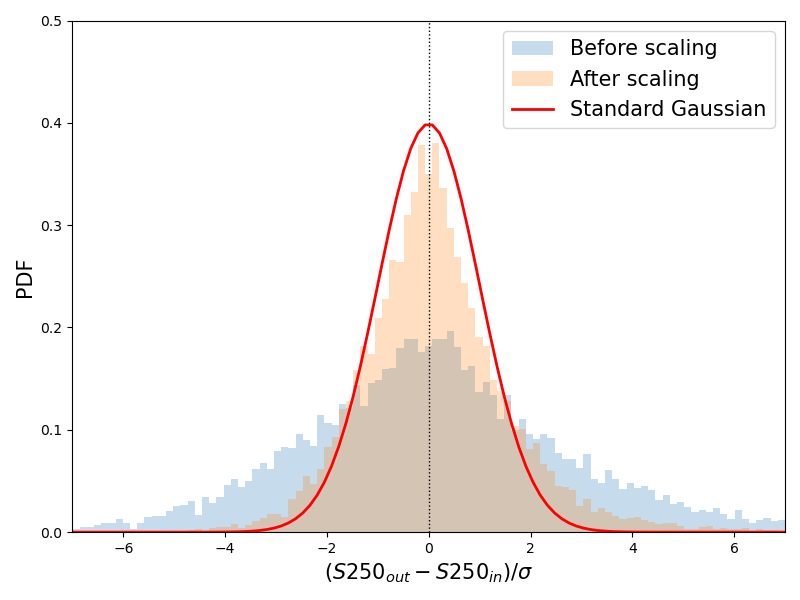}
    \caption{Distribution of flux difference (between the output XID+ flux and the input true flux) normalised by the XID+ flux uncertainty estimate, before and after scaling by a factor of two. The red line corresponds to a standard Gaussian distribution.}
    \label{fig:spire_norm}
\end{figure}

\begin{figure}
    \centering
\includegraphics[width=.49\textwidth]{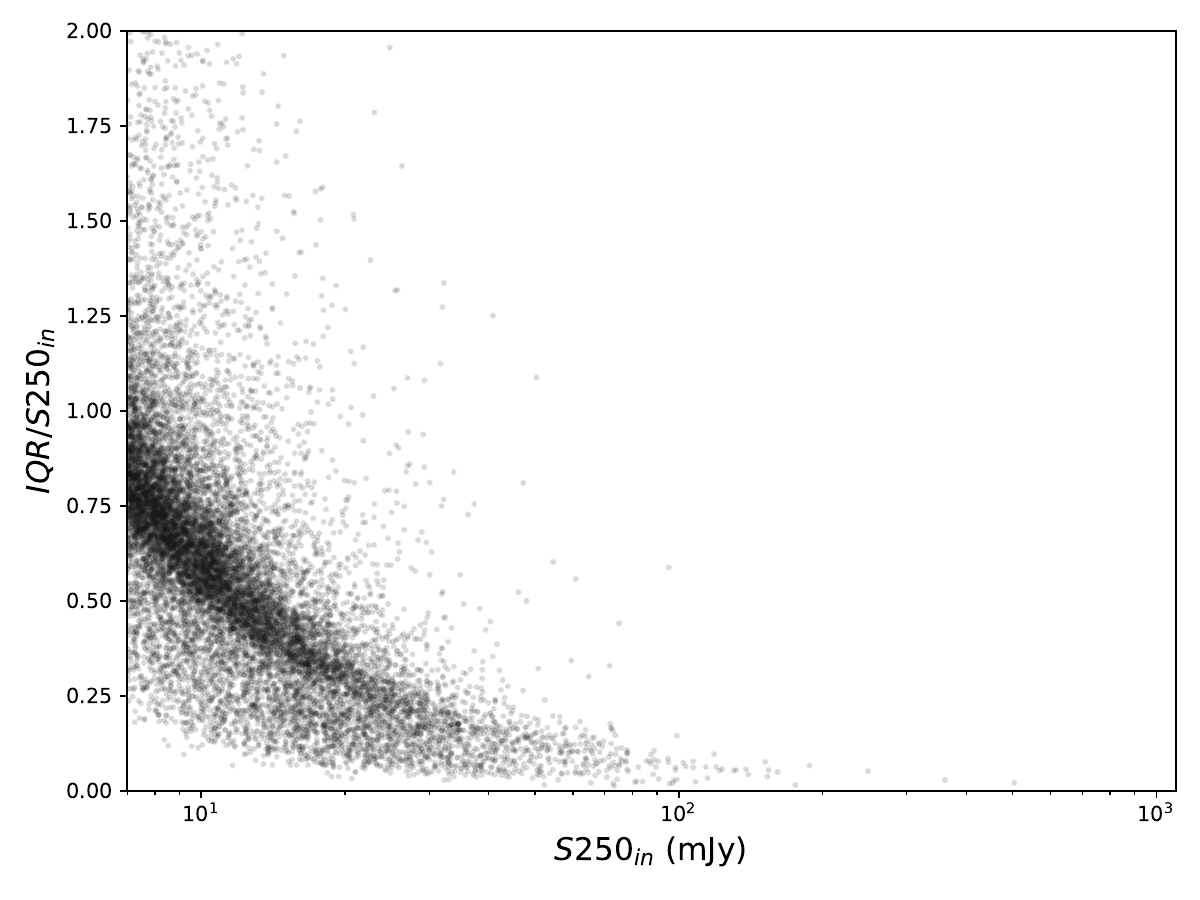}
    \caption{Ratio of the IQR to the input flux as a function of the true 250 $\mu$m flux, after applying the scaling factor.}
    \label{fig:spire_iqr}
\end{figure}

\begin{figure}
    \centering
\includegraphics[width=.49\textwidth]{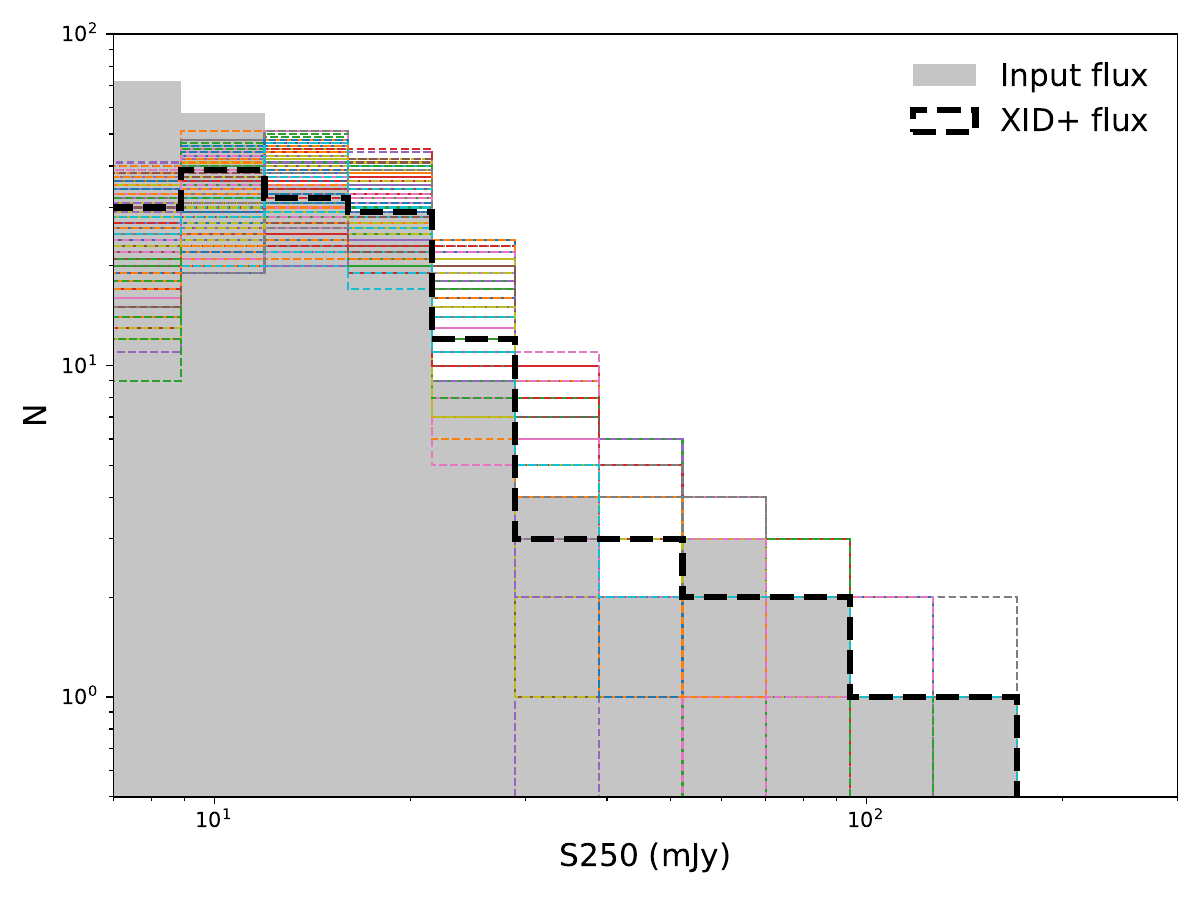}
    \caption{Comparison of the input 250 $\mu$m number counts (filled histogram) with the counts extracted from the output XID+ flux (black line). The dashed histograms of different colours represent the 3000 samplings from the posterior.}
    \label{fig:spire_NC}
\end{figure}

\begin{figure}[b]
    \centering
\includegraphics[width=0.49\textwidth]{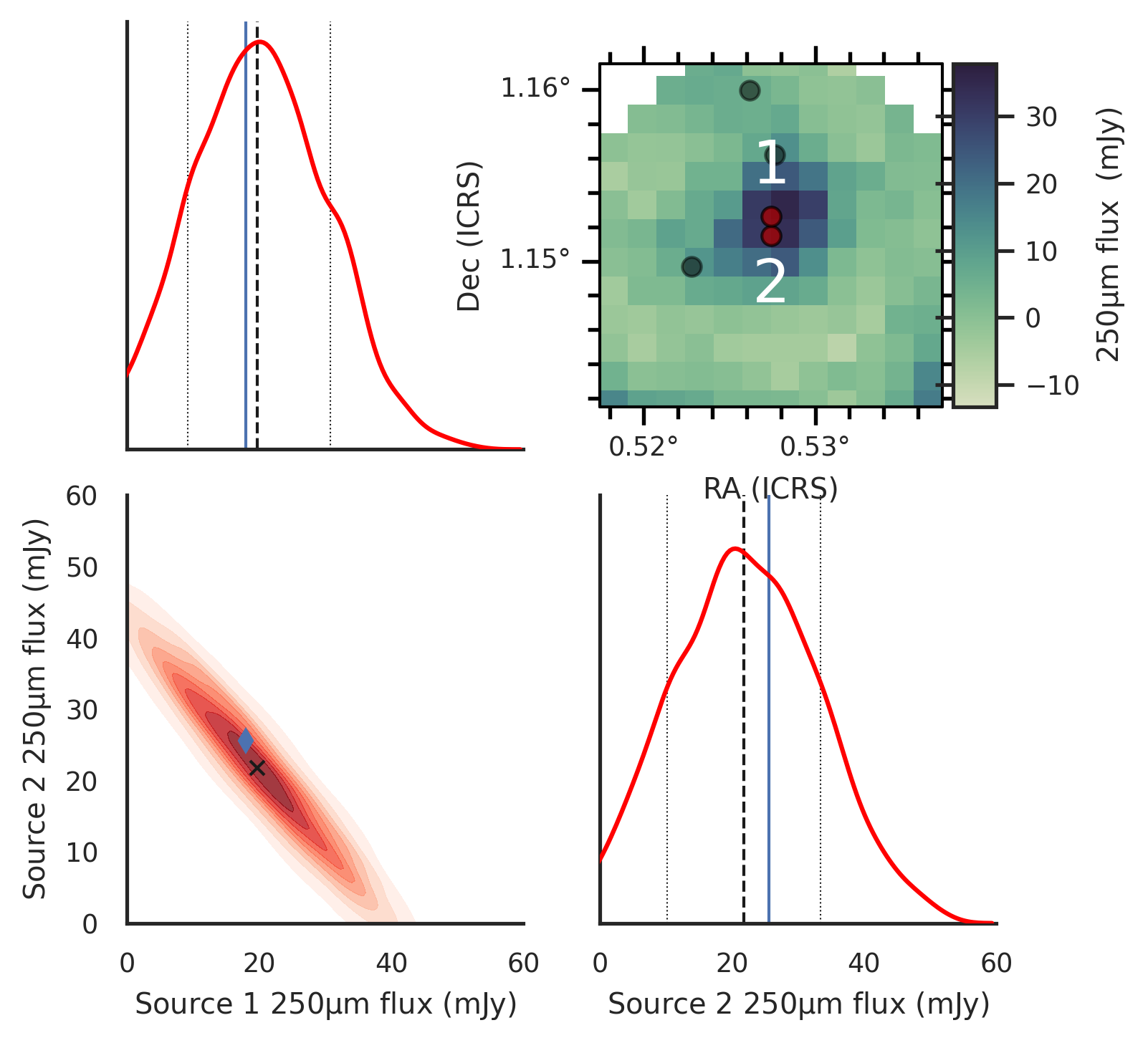}
    \caption{Joint and marginalised posterior plot of two correlated sources ($\sim 4\arcsec$ apart) in the simulated 250 $\mu$m maps. Note that the joint and marginalised posteriors have been scaled by a factor of two using Eq. \ref{eq:scale}. The dashed black lines and the black cross indicate the fluxes extracted by XID+. The solid blue lines and the blue diamond indicate the true fluxes. The vertical dotted lines correspond to the 16th - 84th percentiles in the posterior.
    }
    \label{fig:posterior_plot_sides}
\end{figure}

\begin{figure}
   \centering
   \includegraphics[width=9.5cm, height=7cm]{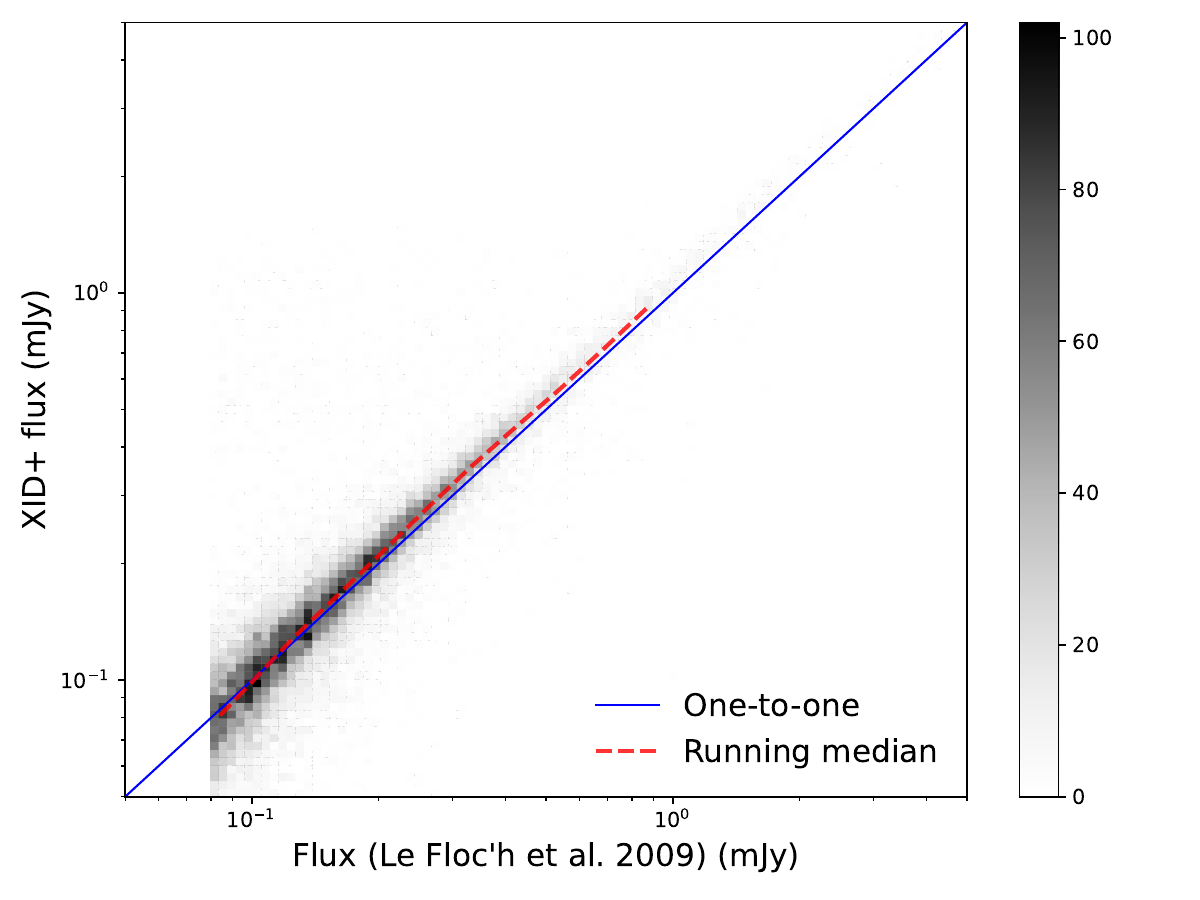}
     \includegraphics[width=9.5cm, height=7cm]{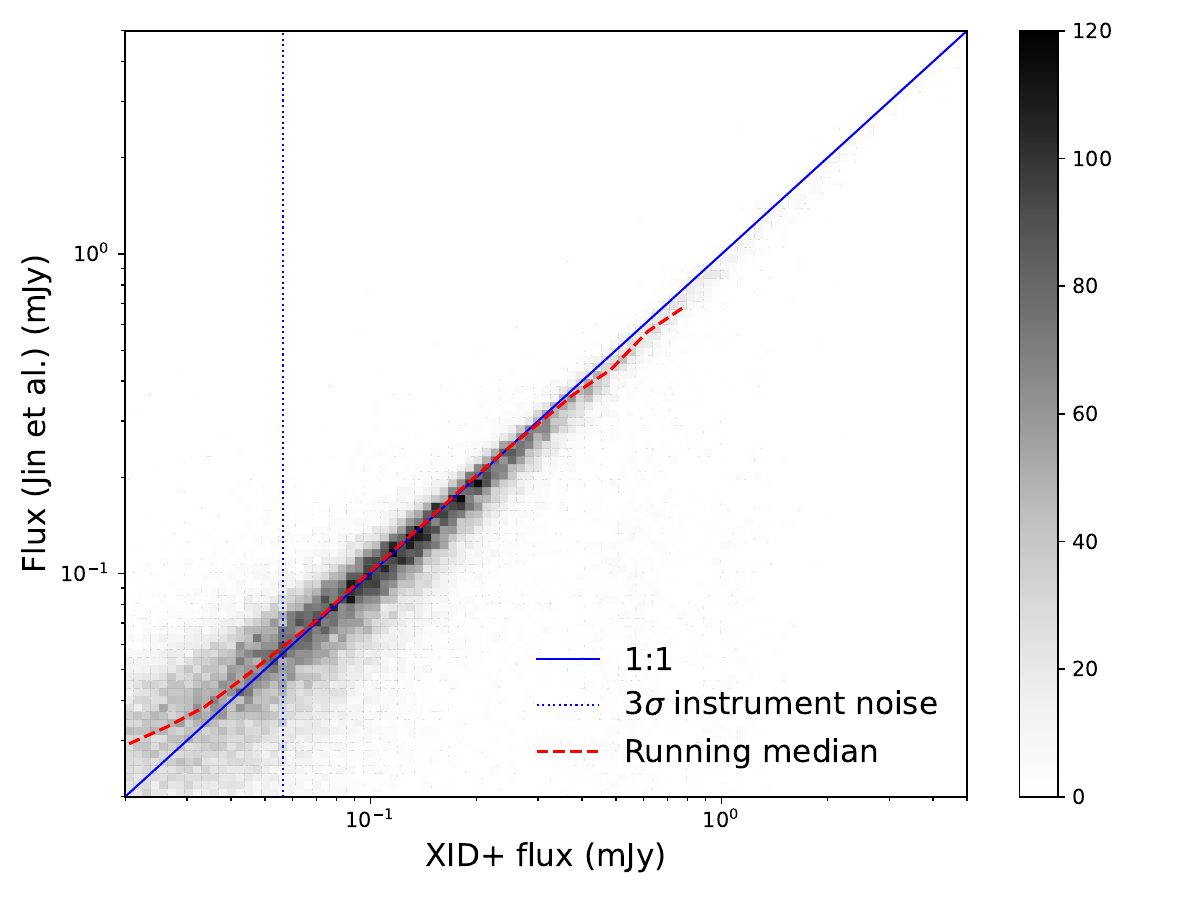}
      \caption{{\it Spitzer} MIPS 24 $\mu$m flux comparison. Top panel: A 2D histogram plot comparing the XID+ deblended fluxes with the blind catalogue from \citet{2009ApJ...703..222L}. There is a good agreement down to the limit of the blind catalogue (80 $\mu$Jy). Bottom panel: Comparison with the super deblended catalogue from \citet{2018ApJ...864...56J}. Our deblended fluxes agree quite well with the super deblended fluxes, but with increasing scatter towards fainter sources.}
         \label{Fig24_blind_super}
\end{figure}

\begin{figure}
   \centering
   \includegraphics[width=4.4cm, height=8.5cm]{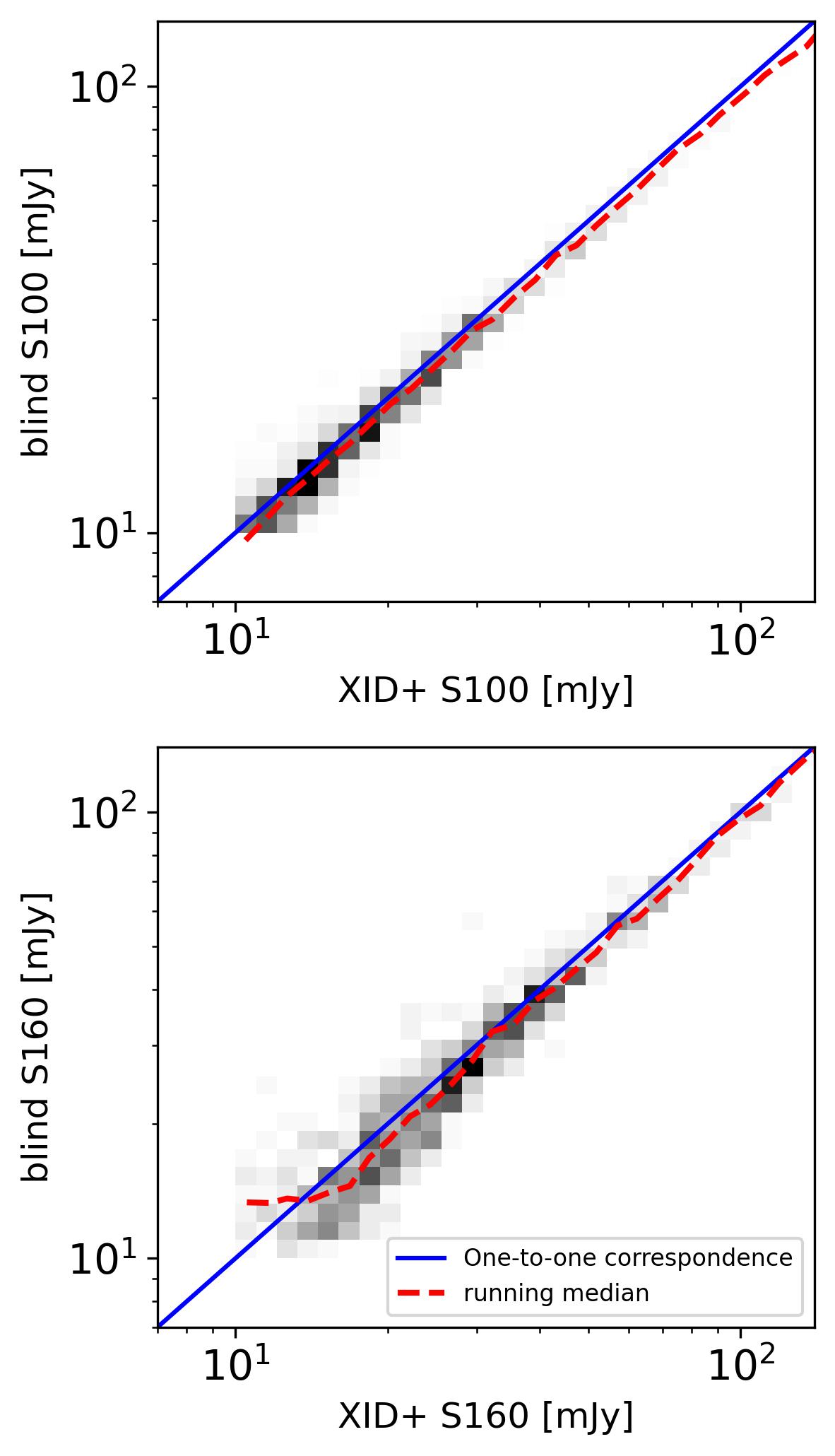}
\includegraphics[width=4.4cm, height=8.5cm]{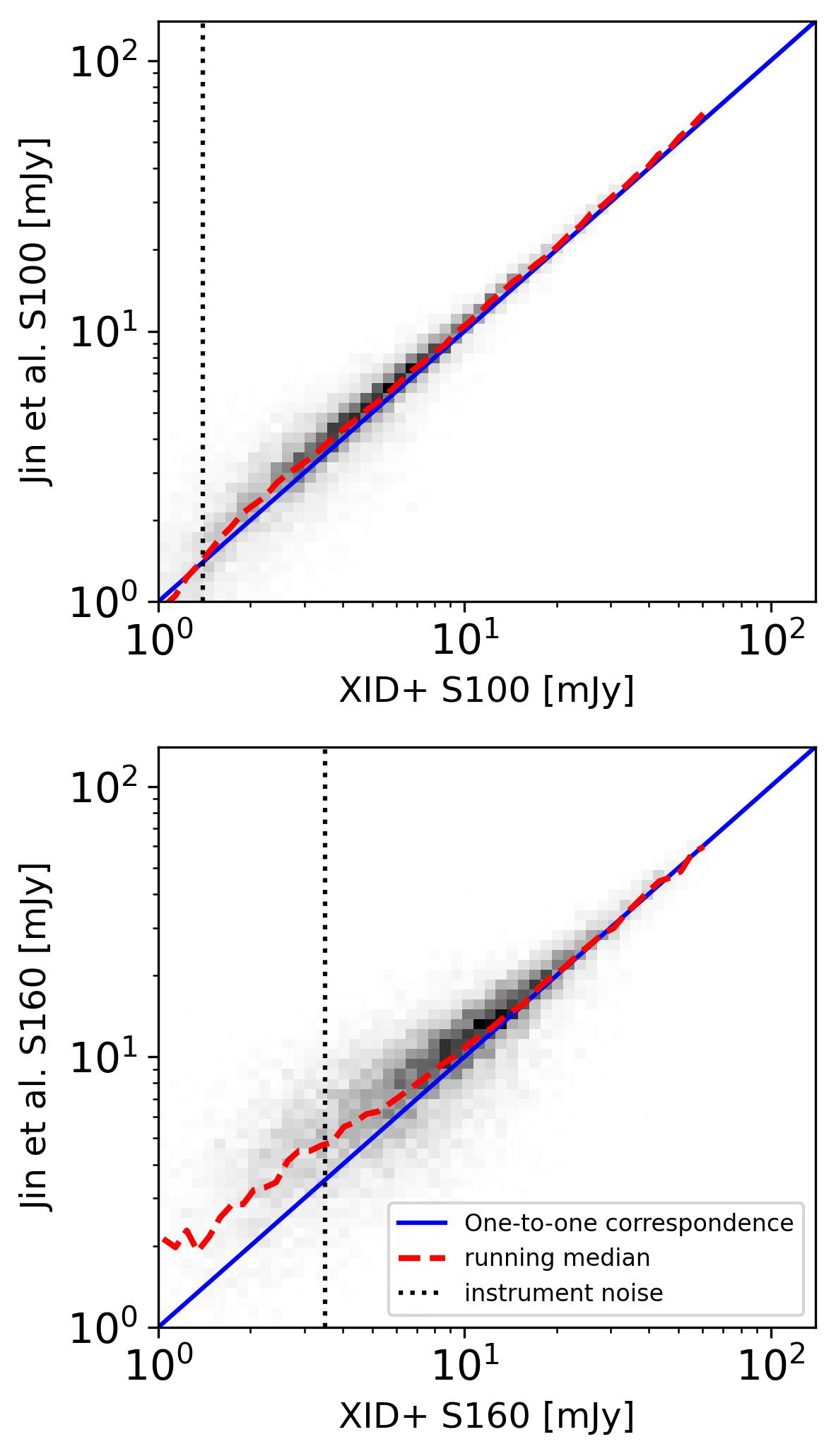}
      \caption{{\it Herschel} PACS 100 \& 160 $\mu$m flux comparison. Left column: 2D histogram plots comparing the XID+ deblended fluxes with the blind catalogue. Right column: Comparison with the super-deblended catalogue from \citet{2018ApJ...864...56J}. The vertical dotted line corresponds to the $1\sigma$ instrument noise level, which is 1.4 mJy at 100 $\mu$m  and 3.5 mJy at 160 $\mu$m.}
         \label{pacs_compare}
\end{figure}

\begin{figure*}[ht]
    \centering
    \includegraphics[height=5cm]{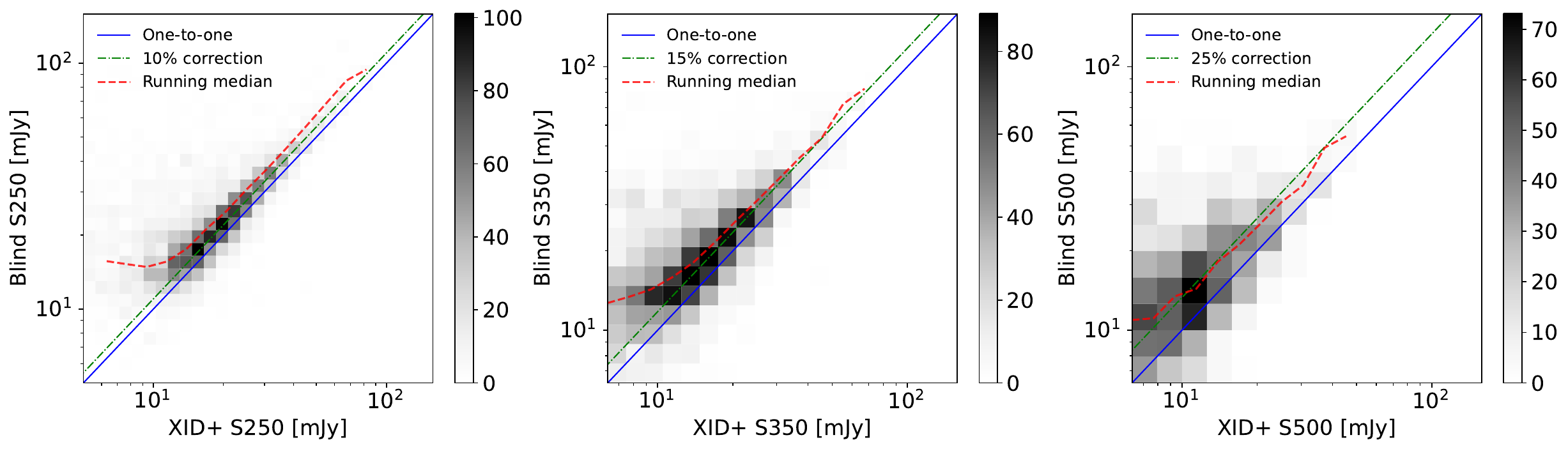}
    \includegraphics[height=5cm]{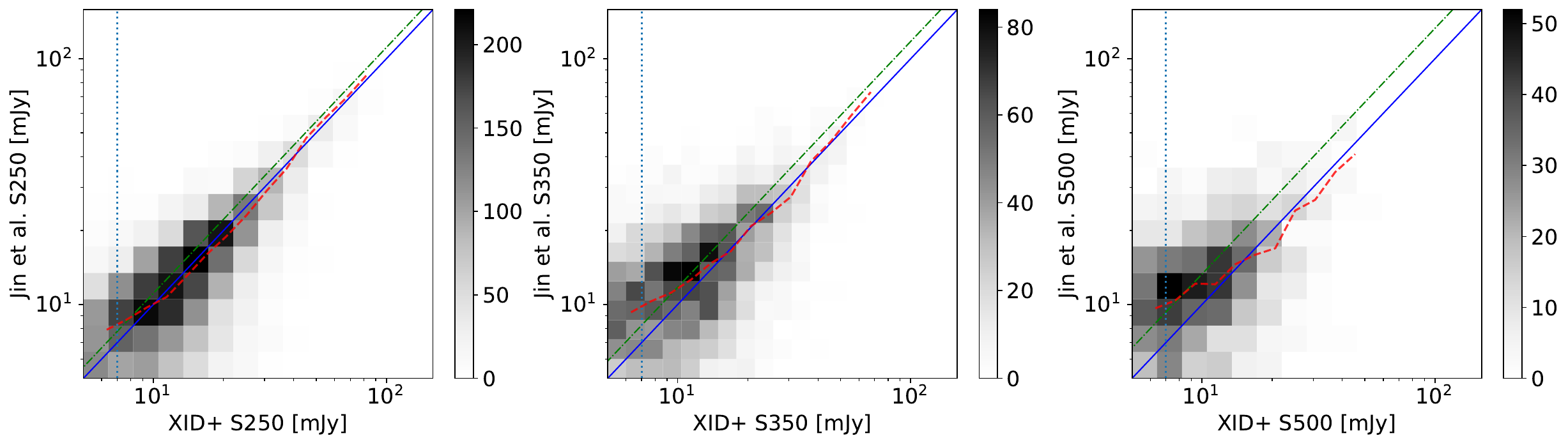}
    \caption{{\it Herschel} SPIRE 250, 350 \& 500 $\mu$m flux comparison. Top panels: 2D histogram plots comparing the XID+ deblended SPIRE fluxes with the blind catalogue. Bottom panels: Comparison of the XID+ deblended SPIRE fluxes with the Jin et al. (2018) super-deblended catalogue. The blue line corresponds to the one-to-one correspondence. The red line is the shifted one-to-one line, after taking into account the maximum level of the systematic underestimation of our XID+ deblended SPIRE fluxes (see Section \ref{sim_herschel}).}
    \label{fig:spire_blind_jin}
\end{figure*}

\begin{figure}
   \centering
   \includegraphics[width=9cm]{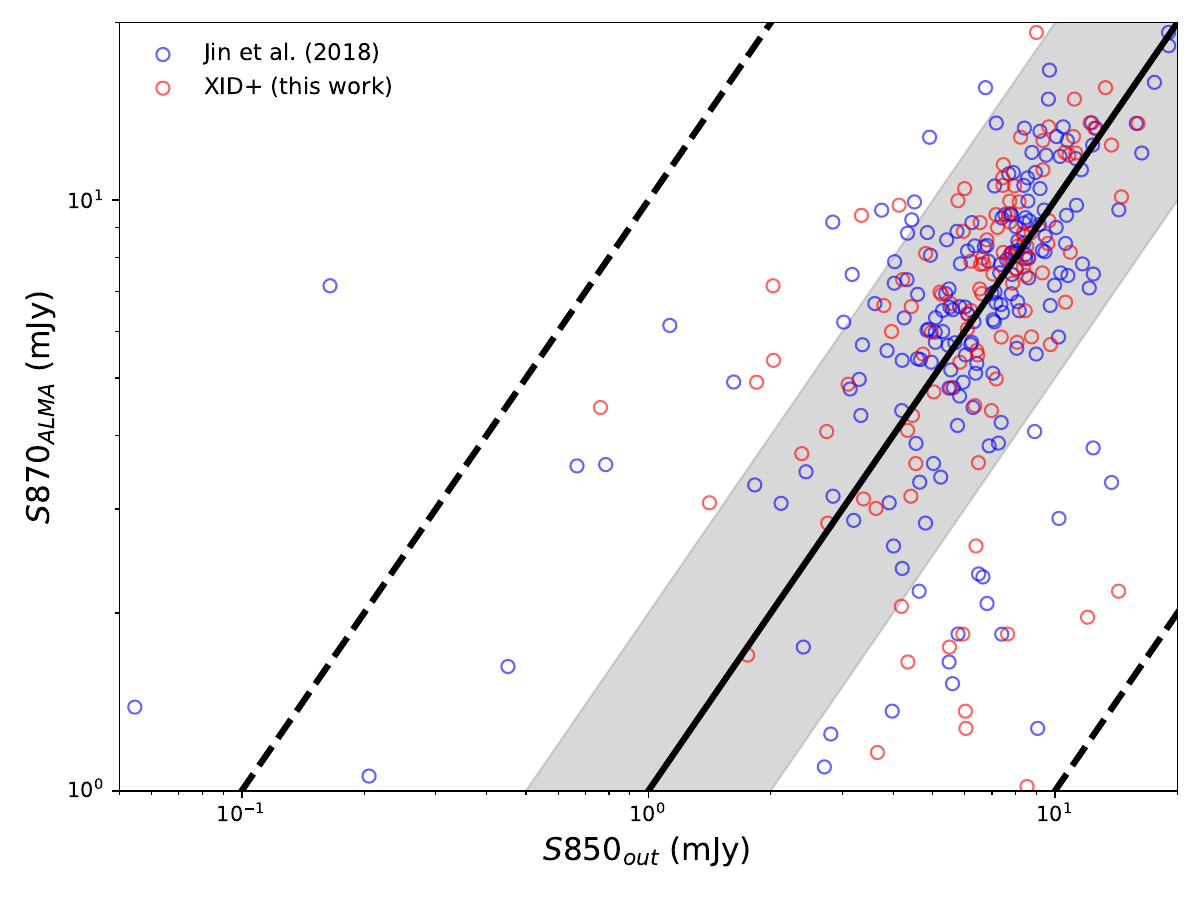}
      \includegraphics[width=9cm]{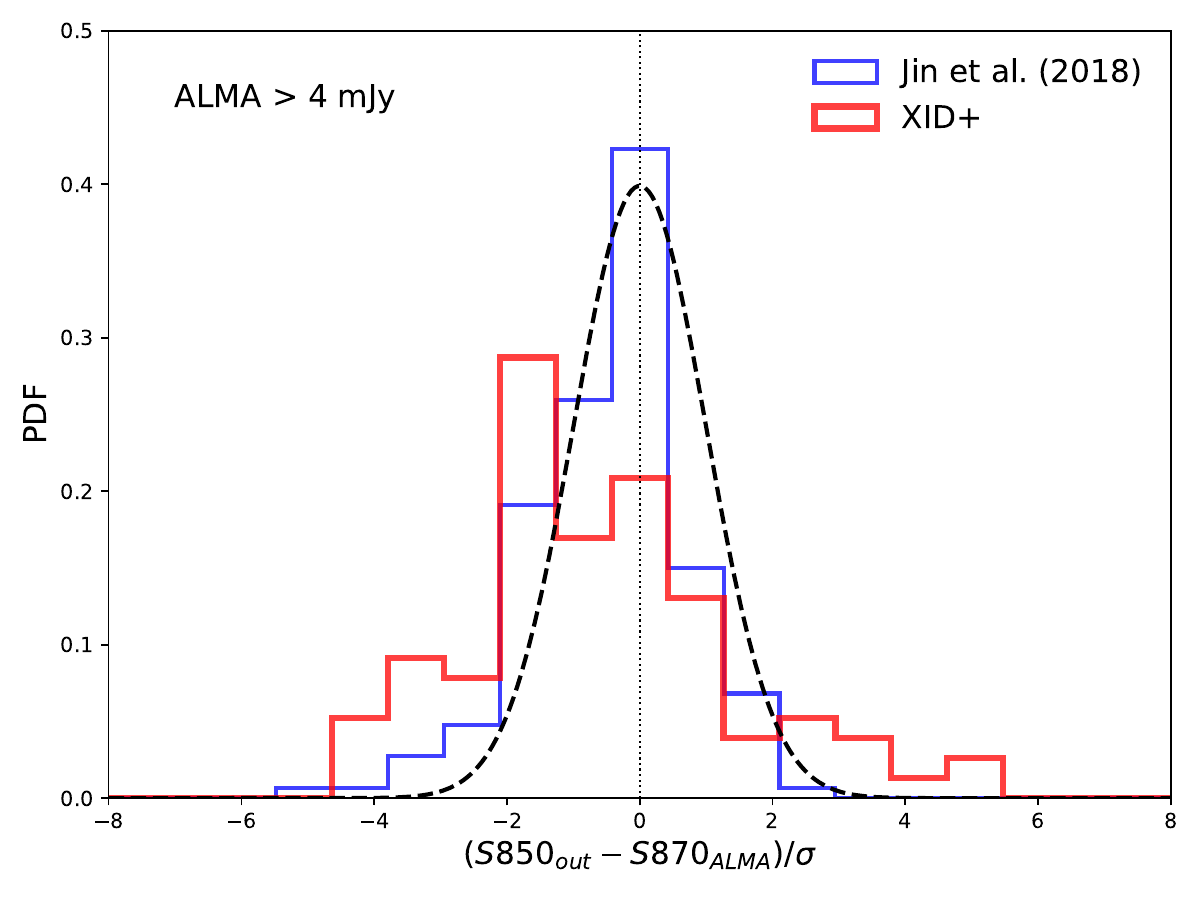}
      \caption{Testing deblended 850 $\mu$m flux densities using AS2COSMOS. Top panel: Comparison between deblended 850 $\mu$m fluxes (red: XID+; blue: \citet{2018ApJ...864...56J}) and ALMA 870 $\mu$m measurements from AS2COSMOS. The solid line corresponds to the 1:1 ratio. The shaded region is within a factor of two from the 1:1 and the dashed lines are within a factor of 10.  Bottom panel: Probability distributions of the difference between the deblended fluxes and ALMA measurements normalised by the $1\sigma$ flux uncertainty, for sources with ALMA measured fluxes $>4$ mJy.}
         \label{AS2COSMOS}
\end{figure}

\begin{figure*}
   \centering
   \includegraphics[width=9cm]{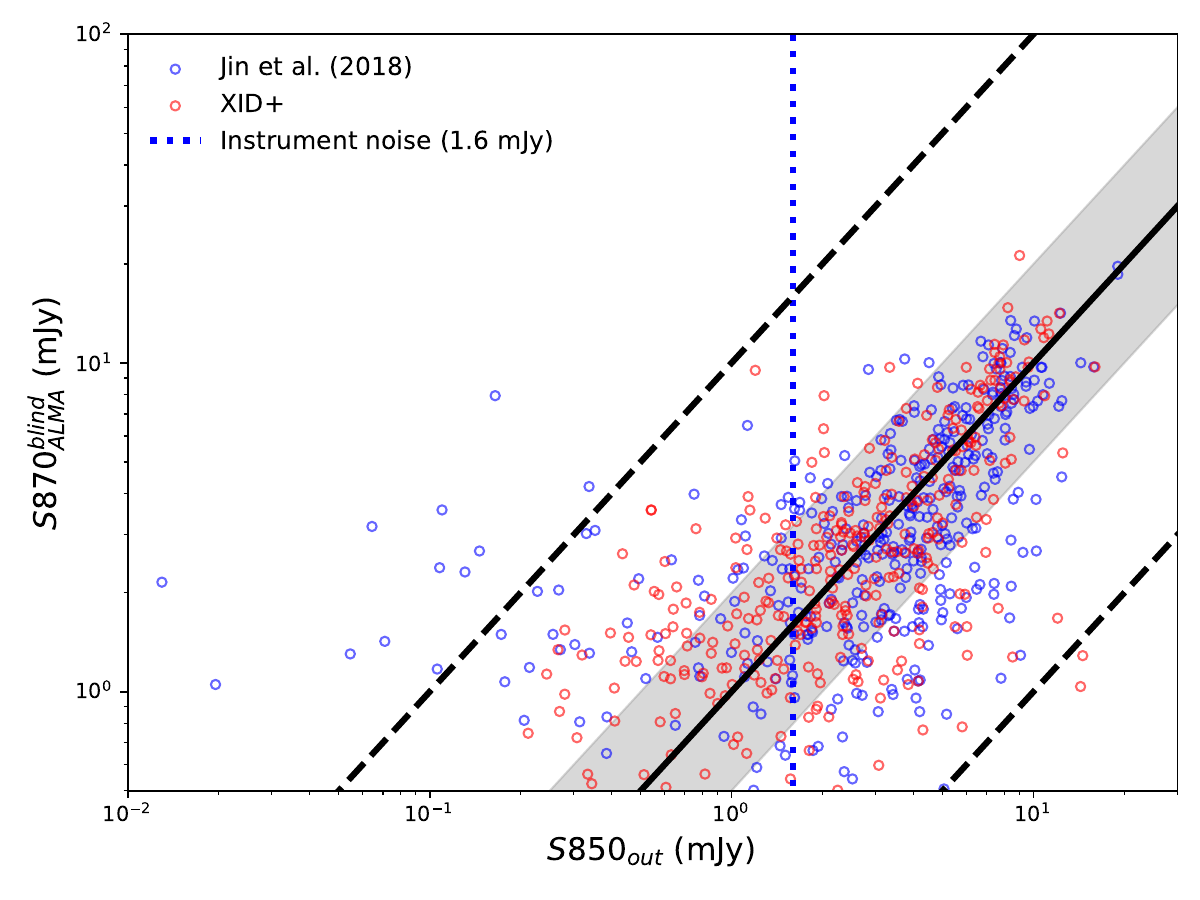}
    \includegraphics[width=9cm]{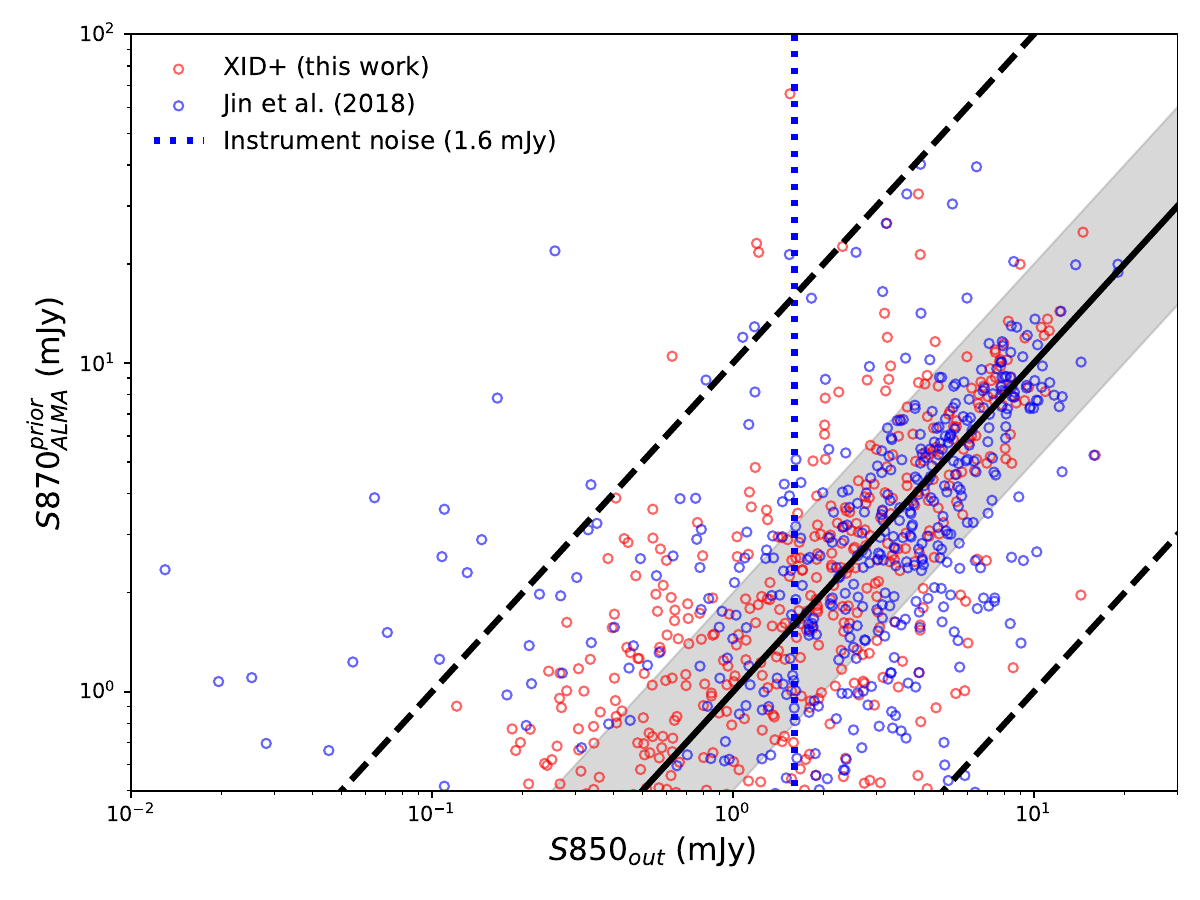}
    \includegraphics[width=9cm]{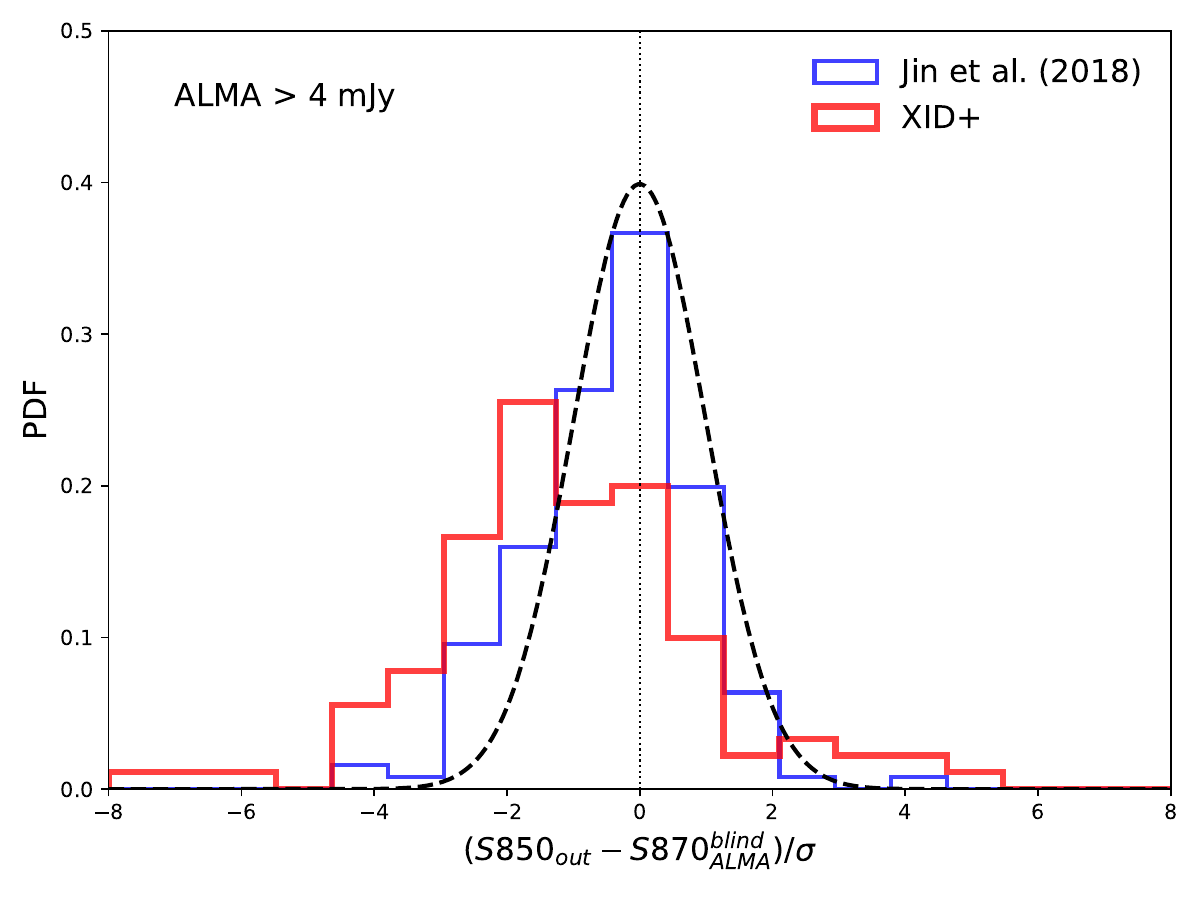}        \includegraphics[width=9cm]{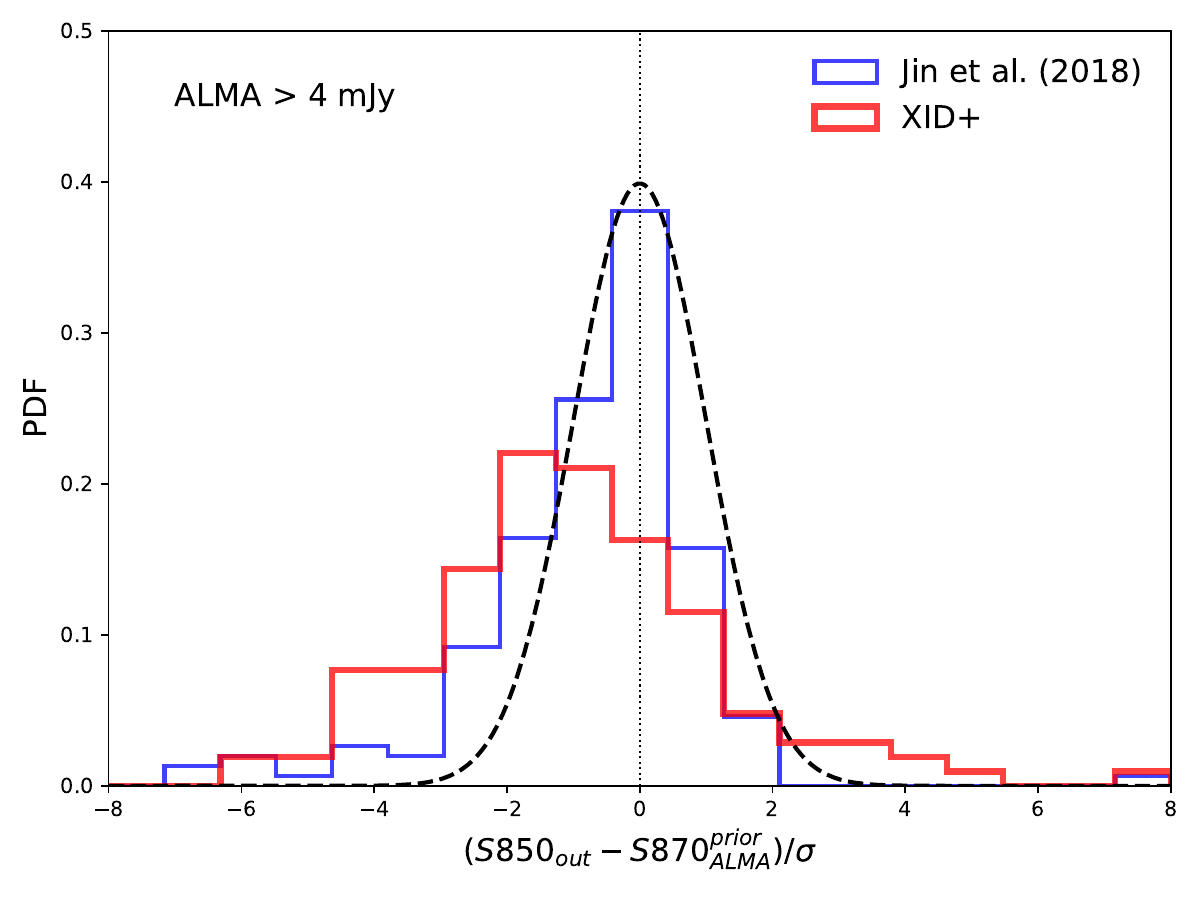}
      \caption{Testing deblended 850 $\mu$m flux densities using A3COSMOS. Top left: Comparison between deblended 850 $\mu$m flux and ALMA 870 $\mu$m photometry from the blind ALMA catalogue (red: XID+ deblended fluxes; blue: super-deblended fluxes from \citealt{2018ApJ...864...56J}). The blue vertical dotted line corresponds to the rms noise level of S2CLS-COSMOS. Top right: Similar to the top left panel but for comparison with the ALMA prior photometry measurements. Bottom left: Normalised distributions of flux difference (between deblended fluxes and ALMA blind photometry) divided by the $1\sigma$ flux uncertainty, for sources with ALMA measured fluxes $>4$ mJy. The vertical dotted line corresponds to no difference between the deblended flux and the ALMA flux. Bottom right: Similar to the bottom left panel but for comparison with the ALMA prior photometry catalogue.}
         \label{A3COSMOS}
\end{figure*}


In contrast to PACS, the SPIRE maps are dominated by confusion noise. Thus, we selected sources with a true 250 $\mu$m flux $>7$ mJy (equal to the $1\sigma$ total noise = $\sqrt{\sigma_{instrument}^2 + \sigma_{confusion}^2}$), which gave us 14\,583 sources (0.54 sources per beam). 
Fig. \ref{fig:spire_frac_err} compares the XID+ deblended flux to the input flux ratio as a function of the true 250 $\mu$m flux. 
The red solid line represents the running median, while the red dashed lines are the 16th-84th percentiles. We over-plot the same quantities for the other two SPIRE bands. The 350 and 500 $\mu$m running medians and percentiles have qualitatively the same behavior.
At the bright end, the output flux is median unbiased relative to the true flux across the three  bands as the running median is close to the 1:1. Towards fainter fluxes, the median flux bias increases from a few percent to $\sim10\%$, $\sim15\%$ and $\sim25\%$ at the faintest flux level at 250, 350 and 500 $\mu$m, respectively. 
Thus, the offset representing a systematic flux underestimation increases towards longer wavelengths, which is expected as the surface density of the sources in the prior catalogue increases by almost a factor of four from 250 to 500 $\mu$m. 

After demonstrating the level of flux accuracy in the SPIRE bands, we explored further issues such as accuracy of flux uncertainty, flux precision and number counts, using the 250 $\mu$m band as an example. Fig. \ref{fig:spire_unc} shows the $\frac{84th - 50th}{50th - 16th}$ percentile ratio of the XID+ flux posterior distribution as a function of the median at 250 $\mu$m. The flux uncertainties of most sources are roughly Gaussian. However, there are some relatively bright sources with a much lower $\sigma^+$ flux uncertainty. On the other hand, there are some sources with a much lower $\sigma^-$ flux uncertainty at the faint end. In Fig. \ref{fig:spire_norm}, we plot the probability distribution of difference between the output flux $S250_{out}$ and the true flux at 250 $\mu$m $S250_{in}$ normalised by the $1\sigma$ flux uncertainty estimate  (from the maximum of $\sigma^+$ and $\sigma^-$) for sources with $S250_{out}>7$ mJy, before and after scaling by a factor of two. Without scaling, the distribution is much wider than the standard Gaussian (the red line). After scaling by a factor of two, the distribution of the normalised flux difference is much closer to the standard Gaussian. Fig. \ref{fig:spire_iqr} shows that ratio of the IQR to the input true 250 $\mu$m flux as a function of the true flux, demonstrating that flux precision is in general higher for brighter sources. 



In Fig.~\ref{fig:spire_NC}, we show the input counts at 250 $\mu$m as the grey histogram. The black dashed line represents the output counts derived from the best estimate of the output flux. The dashed histograms with different colours correspond to the extracted counts using the 3000 samplings from the posterior, scaled using Eq. \ref{eq:scale}. The output counts are close to the true counts above the $10$ mJy. Because of the worse scatter and flux accuracy of the fainter sources, the output counts falls increasingly below the true counts at $<10$ mJy. 
Finally, we illustrate how posterior PDFs of the output flux determined by XID+ can be used to investigate sources which are very close to each other, as their flux densities will be highly correlated.  In Fig.~\ref{fig:posterior_plot_sides}, we show the joint and marginalised posteriors of two close-by sources (separated by $\sim4$\arcsec which is smaller than the pixel size of 6\arcsec) in the simulated  250 $\mu$m map. The joint posterior is highly elongated which could cause some source photometry methods to assign all of the flux to one source. With XID+, one can explore the correlation of the neighbouring sources.



\section{The final deblended point source catalogue}\label{real}

After validating our deblending methodology using the SIDES simulations, we proceed to deblend the real maps from 24 $\mu$m to 500 $\mu$m and compare with blindly extracted catalogues which contain relatively bright sources and the super-deblended catalogue from \citet{2018ApJ...864...56J} which extends to very faint sources (down to $\sim1\sigma$ instrument noise). As an additional test, we deblend the SCUBA-2 850 $\mu$m map and compare with ALMA measurements which should be free of source blending issues. This gives us an independent validation which does not rely on simulations (and any inbuilt assumptions). 

\subsection{Deblending the real {\it Spitzer} MIPS 24 $\mu$m map} 



We first compare our XID+ deblended 24 $\mu$m fluxes with the blindly extracted catalogue from \citep{2009ApJ...703..222L} which contains $\sim$30\,000 sources down to the limit of 80 $\mu$Jy (roughly $4\sigma$ instrument noise). By construction, blind catalogues are limited to relatively bright sources in order to keep the rate of spurious detections at a minimum. We cross-matched our deblended catalogue with the blind catalogue by finding the closest match within 2\arcsec which results in 21\,703 matches.
The top panel in Fig. ~\ref{Fig24_blind_super} shows that there is a very good agreement between our XID+ deblended 24 $\mu$m fluxes and the blindly extracted fluxes down to the limit of the blind catalogue. The median difference is 5.4 $\mu$Jy which is much smaller than the $1\sigma$ uncertainty of the blind catalogue. In terms of fractional difference (i.e., flux difference divided by the blind flux) the median difference is 3.3\%. 

We also compare our XID+ deblended catalogue with the super-deblended catalogue from \citet{2018ApJ...864...56J}. There are several differences between the two deblended catalogues. First, the construction of the prior catalogue is different. \citet{2018ApJ...864...56J} used COSMOS2015 while we used the updated COSMOS2020  benefiting from deeper imaging and improved SED coverage. The initial selection of prior sources in \citet{2018ApJ...864...56J} is based on photo-$z$ and stellar mass estimates\footnote{\citet{2018ApJ...864...56J} used photo-$z$ and stellar mass estimates in the initial selection of prior sources to deblend the 24 $\mu$m map and the radio maps. After that, they used SED fitting and the photo-$z$ information to predict the flux of a given prior source at the wavelength to be deblended (e.g., PACS or SPIRE maps). If the predicted flux is above a critical flux value, then the source is kept in the prior list.}. While a stellar mass-based selection is reasonable because of the SFMS which shows a strong correlation between star-formation rate (SFR) and stellar mass for star-forming galaxies, it also includes passive galaxies which are not significant far-IR/sub-mm emitters. The fraction of passive galaxies increases with increasing stellar mass and decreasing redshift. Furthermore, relying directly on the photo-$z$ and stellar mass estimates would be problematic for some sources for which these properties could not be reliably derived. For example, from COSMOS2015 to COSMOS2020, more than 20\% of the sources ($\sim$30,000 sources) have their photo-$z$ estimates changed by $>10$\%. In our deblending methodology, we only used the observed fluxes and colours to predict the flux in the waveband to be deblended, taking into account systematic uncertainties in the SED modelling. Thus, we can avoid relying on derived properties and automatically fold in differences in the SEDs between star-forming galaxies and passive galaxies. There are also differences in the deblending methodology. \citet{2018ApJ...864...56J} used a maximum-likelihood based GALFIT PSF fitting and rely on Monte Carlo simulations to correct the output flux. The Monte Carlo simulations would break down the correlation between sources as the injected source has no correlation with the real sources to be deblended. Bearing these differences in mind, we can see a very good agreement in the bottom panel of Fig. ~\ref{Fig24_blind_super} between the two deblended catalogues with increasing scatter towards fainter fluxes. Below the $3\sigma$ instrument noise level ($3 \times 18.8 \mu$Jy), our XID+ deblended fluxes are slightly lower than the super-deblended fluxes, which could be due to the systematic underestimation effect (which is $<5$\% above $1\sigma$ instrument noise) as seen in the validations using simulations (Section \ref{sim_mips}).

\subsection{Deblending the real {\it Herschel} PACS \& SPIRE maps}

 
First, we compare our deblended PACS catalogues with the blindly extracted catalogues from the PEP survey which contains 7\,443 sources with 100 $\mu$m flux $>4.4$ mJy. Selecting the closest match within 2\arcsec,  we found a total of 3\,654 and 2\,489 matches at 100 and 160 $\mu$m, respectively. 
The left column of Fig. \ref{pacs_compare} shows a  good agreement on flux down to the limit of the blind catalogue (5 and 10 mJy at 100 and 160 $\mu$m respectively). In the right column of Fig. \ref{pacs_compare} we compare our results with the super-deblended catalogue. At 100 $\mu$m, the flux agreement is excellent throughout the explored range. At 160 $\mu$m, the super-deblended flux is slightly higher than our deblended flux at $<\sim10$ mJy. Based on the validation in Section \ref{sim_herschel}, we expect our deblended flux to be median unbiased relative to the true flux down to the $1\sigma$ instrument noise which is 3.5 mJy at 160 $\mu$m. Therefore, we conclude that the \citet{2018ApJ...864...56J} super-deblended 160 $\mu$m fluxes may be slightly overestimated in this flux range. 



We then compare our XID+ deblended results of the SPIRE maps with the blind catalogue from the {\it Herschel} Extragalactic Legacy Project (HELP; \citealt{2021MNRAS.507..129S}). The sources in the blind catalogue are detected as peaks in the matched filter images which maximises the S/N of individual points sources taking into account instrument noise and confusion noise \citep{2011MNRAS.411..505C}. The blind catalogue in COSMOS contains 9\,382 sources directly detected from the 250 $\mu$m map, with a flux cut $S_{250\mu m}\geq13.5$ mJy corresponding to the 85\% completeness level. Selecting the closest match within 5\arcsec  based on the positional uncertainty expected of blind SPIRE sources
 \citep{2014MNRAS.444.2870W}, we found a total of 3\,591 matches.
The top row in Fig. \ref{fig:spire_blind_jin} shows the comparison of our deblended fluxes with the blind photometry, revealing a generally good agreement particularly at 250 $\mu$m. There is a systematic underestimation in our deblended flux  which increases with increasing wavelength. 
The scatter also increases toward fainter fluxes and longer wavelengths. This is expected as the density of the sources in the prior catalogue increases from 250 to 500 $\mu$m (from 0.34 to 1.34 sources per beam). Based on the validation in Section \ref{sim_herschel}, we expect that there is a maximum median offset of $\sim$10\%, 15\% and 25\% at 250, 350 and 500 $\mu$m, respectively, at the faintest flux levels. We add the red dashed lines to guide the eye in gauging the maximum level of median bias in our deblended flux relative to the blind photometry. The bottom row of Fig. \ref{fig:spire_blind_jin} compares our deblended SPIRE fluxes with the \citet{2018ApJ...864...56J} super-deblended catalogue. At 250 $\mu$m, we see a good agreement between our deblended fluxes and the super-deblended fluxes throughout the explored flux range. At 350 and 500 $\mu$m, the agreement between the two deblended catalogues is slightly worse (with larger scatter). The running median lines cross the maximum level of median bias (based on the validation in Section \ref{sim_herschel}) around 7 mJy. Overall, the good agreement of the two deblended catalogues demonstrate the good quality and robustness of the deblended photometry. 


\subsection{Deblending the real SCUBA 850 $\mu$m maps and comparison with ALMA photometry}



We also apply our XID+ deblending to the SCUBA-2 850 $\mu$m map and then compare with ALMA measurements. The ALMA measurements regarded as the truth allow us to independently verify the quality of our deblended photometry. We used the 850 $\mu$m map from  the SCUBA-2 \citep{2013MNRAS.430.2513H} COSMOS survey (S2COSMOS) \citep{2019ApJ...880...43S}, conducted with the East Asian Observatory’s {\it James Clerk Maxwell} Telescope (JCMT). Our prior catalogue  is the same  as used for deblending the SPIRE maps. For discussions on completeness and precision in using $K-$band/mid-IR/radio detected sources as counterparts of sub-mm sources, we refer the reader to \citet{2013ApJ...768...91H, 2018ApJ...862..101A, 2019ApJ...886...48A}. The median instrumental noise level of the 850 $\mu$m map is 1.2 mJy/beam over the main area of 1.6 deg$^2$ (with the deepest area reaching $\sigma_{\rm{inst}}\sim0.5$  mJy/beam), which is where our prior catalogue is located. We use a Gaussian PSF with a FWHM of 14.9\arcsec. Following \citet{2017MNRAS.465.1789G} and \citet{2019ApJ...880...43S}, we also applied a multiplicative factor of 1.13 to account for the flux loss due to filtering  in map making.

We compare our XID+ deblended fluxes with ALMA Band 7 continuum measurements centred at 870 $\mu$m of $\sim180$ brightest sources from S2COSMOS, undertaken as a pilot  of the ALMA-S2COSMOS (AS2COSMOS) survey \citep{2020MNRAS.495.3409S}. No correction is applied to convert the ALMA fluxes to 850 $\mu$m. Benefiting from the high resolution (with a median synthesised beam of $0.80$\arcsec$\times0.79$\arcsec) and sensitivity of interferometric observations (with a median sensitivity of 0.19 mJy/beam), ALMA measurements are free from issues such as blending of multiple sources and are highly complete at $>6.2$ mJy\footnote{The main limiting factor in the completeness level is the completeness of the parent SCUBA-2 sample, which is estimated to be 87\%.}. Within 1\arcsec, we found 113 matches between our deblended catalogue and the AS2COSMOS catalogue. The top panel of Fig. \ref{AS2COSMOS} compares ALMA fluxes with our deblended fluxes. The bottom panel of Fig. \ref{AS2COSMOS} shows the probability distribution of the flux difference normalised by the uncertainty estimates of our deblended fluxes, for sources with ALMA measured fluxes $>4$ mJy. We also compare with the super-deblended photometry from \citet{2018ApJ...864...56J} which used the images from the SCUBA-2 Cosmology Legacy Survey (S2CLS) COSMOS program \citep{2017MNRAS.465.1789G} (Cowie et al. 2017) with an rms noise of 1.6 mJy/beam. Within 1\arcsec, we got 201 matches between the super-deblended catalogue and the AS2COSMOS catalogue. The larger number of matches is mostly due to the larger area covered by the super-deblended catalogue. Overall, the level of agreement between our deblended fluxes and the ALMA measurements is similar to that between the super-deblended fluxes and the ALMA measurements. In the case of our deblended photometry, 92\% of the matches (with ALMA fluxes $>4$ mJy) agree with ALMA fluxes within a factor of two (i.e., in the shaded region). In the case of the super-deblended photometry, 88\% of the matches (with ALMA fluxes $>4$ mJy) agree with ALMA fluxes within a factor of two. There are a few sources with much higher ALMA fluxes than the corresponding super-deblended fluxes. Two sources even have $>10$ times brighter ALMA fluxes, suggesting that their super-deblended fluxes are significantly underestimated. 


As a final and complementary test, we compare with measurements from the A3COSMOS project \citep{2019ApJS..244...40L}, which processes ALMA archival observations in COSMOS every six months. The newest data release, including all data which were made public before 10/03/2020, covers  622.8 arcmin$^2$. The A3COSMOS sample is much larger than AS2COSMOS, but  has a more heterogeneous selection and mix of data with different depths and resolutions. The A3COSMOS data release includes two catalogues. One is a catalogue of blindly extracted ALMA sources with peak S/N $>5.4$, with a $\sim50$\% spurious rate at S/N $=5.4$ and a $\sim12$\% cumulative spurious rate at S/N $>5.4$. The other is a prior photometry catalogue containing fitting of known optical/near-IR/mid-IR/radio sources with S/N peak $>4.35$ and $\sim50$\% spurious rate at S/N $=4.35$ and a $\sim8$\% cumulative spurious rate for S/N $>4.35$. As A3COSMOS includes  data at various bands, we selected measurements with observed wavelengths between 850 and 890 $\mu$m, comparable to the Band 7 measurements from AS2COSMOS.

First, we matched our XID+ deblended 850 $\mu$m catalogue with the A3COSMOS blind catalogue by finding the closest match within 0.5\arcsec, resulting in 354 matches. In comparison, there are 457 matches between the super-deblended catalogue and the blind catalogue. The top-left panel of Fig. \ref{A3COSMOS} compares the deblended fluxes with the blind ALMA photometry. For the super-deblended fluxes, 87\% of the matches (with ALMA fluxes $>4$ mJy) agree with the ALMA blind photometry within a factor of two (i.e. in the shaded region). For the XID+ deblended fluxes, 90\% of the matches (with ALMA fluxes $>4$ mJy) agree with the ALMA blind photometry within a factor of two, demonstrating a higher level of flux accuracy. Compared to the XID+ deblended fluxes, more super-deblended fluxes show significant underestimation relative to ALMA. In the bottom-left panel of Fig. \ref{A3COSMOS}, we show that the probability distributions of the flux difference normalised by the $1\sigma$ flux uncertainty estimate, for sources with ALMA measured fluxes $>4$ mJy. 

We now compare with the ALMA prior photometry catalogue. We matched our XID+ deblended catalogue with the ALMA prior catalogue by finding the closest match within 0.5\arcsec, resulting in 788 matches. In comparison, there are 810 matches between the super-deblended catalogue and the ALMA prior catalogue. The right column of Fig. \ref{A3COSMOS} compares the deblended fluxes with the fluxes from the ALMA prior catalogue. Qualitatively, we see similar patterns as in the left column when comparing with the ALMA blind catalogue. For the super-deblended fluxes, 75\% of the matches (with ALMA fluxes $>4$ mJy) agree with the ALMA prior photometry within a factor of two. For the XID+ deblended fluxes, 77\% of the matches (with ALMA fluxes $>4$ mJy) agree with the ALMA prior photometry within a factor of two. Our deblended fluxes clearly extend to fainter sources than the super-deblended fluxes, which could be mostly due to the deeper 850 $\mu$m map from S2COSMOS compared to S2CLS-COSMOS. Other differences in the prior catalogues and deblending methodologies might also play a role.

\section{Example science applications}\label{science}

Our deblended far-IR and sub-mm photometry catalogue can be used in a wide range of studies related to dusty galaxy evolution. For example, one could investigate the number counts, the monochromatic luminosity functions as well as the integrated IR luminosity functions, and the contribution from dusty galaxies to the cosmic star-formation history. These long-wavelength data are also critical for studying the SMF of dusty sources and how they compare with the general SMFs, the fraction of DSFGs as a function of stellar mass and redshift, and the fraction of dust-obscured SFR as a function of stellar mass and redshift. Here we present two example science applications, which are the galaxy SFMS and the FIRC. 

\subsection{The galaxy star-formation main sequence}


\begin{figure}[ht]
    \centering
\includegraphics[width=9cm]{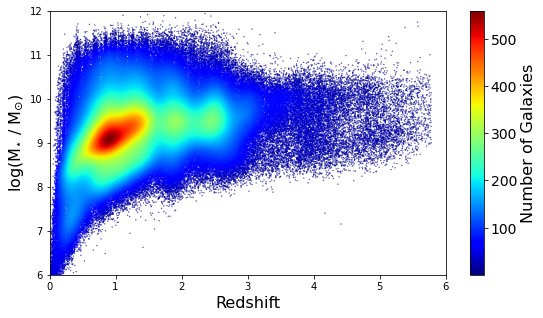}
    \caption{Stellar mass $M_{\star}$ vs redshift for all galaxies in our deblended catalogue, with colour-coding corresponding to number of sources.}
    \label{fig:MS-mass-z}
\end{figure}

\begin{figure*}[ht]
    \centering    \includegraphics[width=18.5cm, height=16cm]{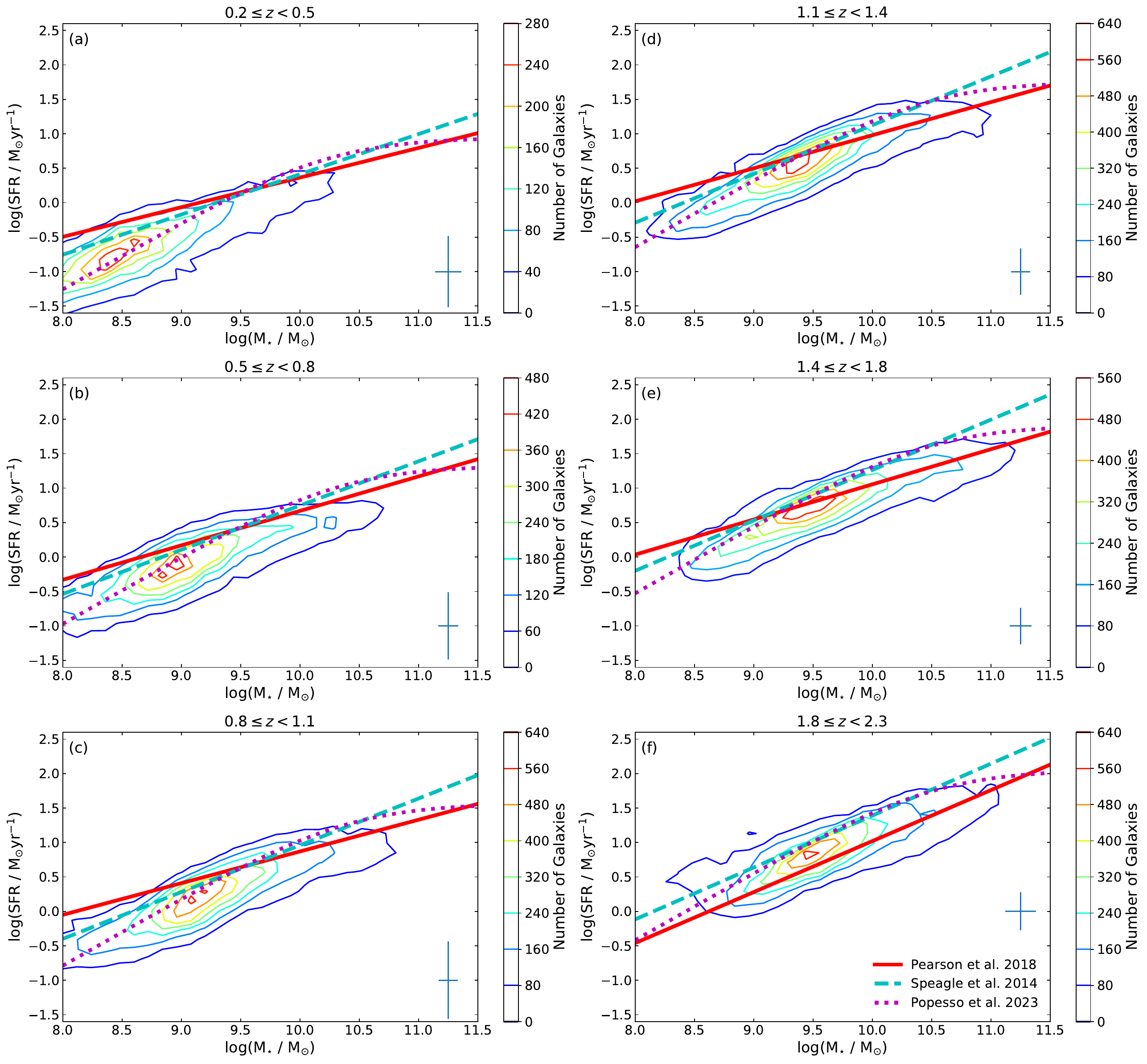}
    \caption{SFR as a function of M$_{\star}$ for (a) $0.2 \leq z < 0.5$, (b) $0.5 \leq z < 0.8$, (c) $0.8 \leq z < 1.1$, (d) $1.1 \leq z < 1.4$, (e) $1.4 \leq z < 1.8$, (f) $1.8 \leq z < 2.3$. Red lines are the results from \citet{2018A&A...615A.146P}, dashed cyan lines are the results from \citet{2014ApJS..214...15S}, and dotted magenta lines are the results from \citet{2023MNRAS.519.1526P}. Blue cross indicates the typical uncertainty in M$_{\star}$ and SFR. Colour-coding correspond to number of sources from low (dark blue) to high (red). Data are binned into the same redshift bins as in \citet{2018A&A...615A.146P}.}
    \label{fig:MS-fitting}
\end{figure*}

The SFMS is a tight correlation between stellar mass ($M_{\star}$) and SFR from the local Universe to high redshift \citep{2004MNRAS.351.1151B, 2007ApJ...660L..43N, 2014ApJS..214...15S, 2023MNRAS.519.1526P}. To estimate $M_{\star}$ and SFR, we use CIGALE and the setup of \citet{2018A&A...615A.146P}, with a delayed exponentially declining star-formation history plus an exponentially declining burst, \citet{2003MNRAS.344.1000B} stellar population, a double power law dust attenuation model, the \citet{2014ApJ...780..172D} dust emission model, and the \citet{2006MNRAS.366..767F} AGN templates.  We then selected galaxies for which the reduced $\chi^{2}$ of the best fit model is $<5$. Fig. \ref{fig:MS-mass-z} shows the distribution of the sources in our deblended catalogue in stellar mass vs. redshift. Using the derived estimates of stellar mass and SFR, we plot the correlation between the two properties in six redshift bins from $z\sim0.2$ to $z\sim2.3$ in Fig. \ref{fig:MS-fitting} and compare with the SFMS relations from \citet{2018A&A...615A.146P}, which were derived using COSMOS2015 \citep{2016ApJS..224...24L} and a less advanced deblending method for only the SPIRE data \citep{2017A&A...603A.102P, 2018A&A...615A.146P}. Thus, \citet{2018A&A...615A.146P} is an ideal comparison to check if there are any differences our updated deblending makes to the SFMS. We also make a comparison with \citet{2014ApJS..214...15S}, which compiled a large number of SFMS studies made before their publication, and \citet{2023MNRAS.519.1526P}, who also made a compilation of earlier SFMS works but used a SFMS with a high mass turnover. Our galaxies in Fig. \ref{fig:MS-fitting} are selected as star-forming using a UVJ colour-colour cut \citep{2011ApJ...735...86W, 2018A&A...615A.146P}.

Since this exercise is only meant to be a simple demonstration of what kind of scientific studies would be enabled by our data, we do not attempt a detailed analysis which would demand carefully taking into account various selection effects. With these caveats in mind and based on a simple visual inspection, we find that our new deblended data appear to give rise to a steeper SFMS than \citet{2018A&A...615A.146P} at $z < 1.8$ and more in line with \citet{2014ApJS..214...15S} and \citet{2023MNRAS.519.1526P}. This work also appears to show that the normalisations (the value of the SFMS at log(M$_{\star}$/M$_{\odot}$) = 10.5) of the SFMS are now more consistent with \citet{2014ApJS..214...15S} and \citet{2023MNRAS.519.1526P}, at least at $z < 1.4$, partially resolving the known lower normalisation of the \citet{2018A&A...615A.146P} SFMS described in \citet{2019MNRAS.490.5285P}.

\subsection{The far-IR to radio correlation}

\begin{figure}[ht]
    \centering
    \includegraphics[width=9cm]{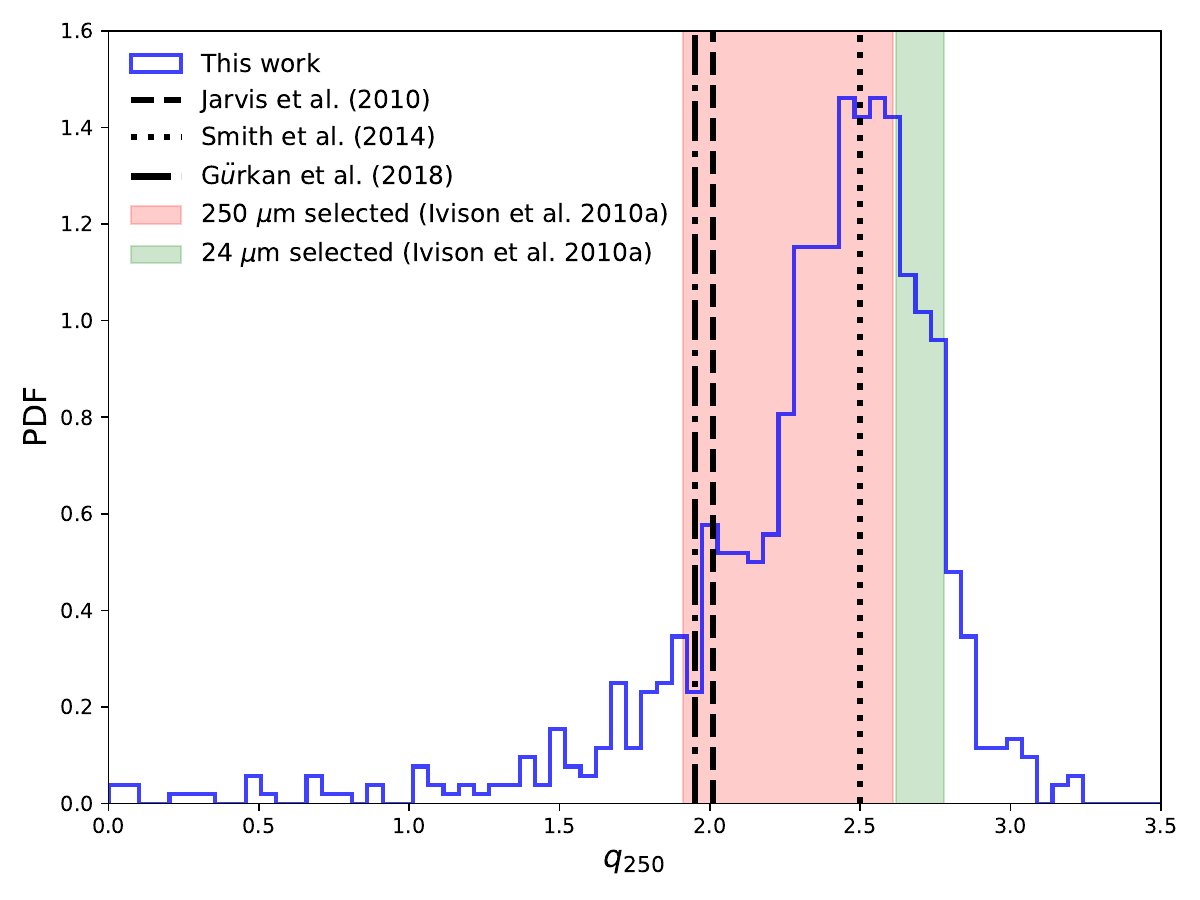}
    \caption{Distribution of the monochromatic $q_{250}$ value of the MIGHTEE sources that do not have a match in COSMOS2020. $k-$correction is not applied due to the lack of photo-$z$ estimates. The various vertical lines and the shaded regions correspond to similar measurements from previous studies.}
    \label{fig:FIRC}
\end{figure}

The correlation between the far-IR and radio emission in star-forming galaxies is one of the tightest known in astronomy. Although the FIRC has been studied extensively  at various IR and radio frequencies, the exact physical origin is not yet known and it is still under debate whether it evolves with redshift and other physical parameters such as stellar mass. As a second example, we briefly explore the FIRC using our deblended fluxes of the MIGHTEE sources at 1.3 GHz. Here we focus on the 1038 MIGHTEE sources that are not included in COSMOS2020. We emphasise again that in deblending the FIR/sub-mm maps for the additional radio sources, only positional priors are used (i.e., no information regarding their radio fluxes are used in the deblending process).

It is common to use a dimensionless parameter $q$, defined as the logarithmic ratio of the far-IR luminosity to the radio luminosity in the rest-frame, to describe the FIRC. Different studies often also use slight variations of $q$, e.g., by using either integrated or monochromatic (far-)IR luminosity. Also, the choice of the radio frequency can vary. Here we compute a monochromatic $q_{250}$ as below,
\begin{equation}
    q_{250} = \log_{10} \left(\frac{S_{250}}{S_{1.3 GHz}} \right).
\end{equation}
Ideally, one need to apply $k$-correction to derive the rest-frame  250 $\mu$m and 1.3 GHz flux densities, particularly if the sample extends a wide redshift range. However, as our sample consists of the 1038 MIGHTEE sources that are not present in COSMOS2020, we do not have photo-$z$ information for these sources.  Therefore, we simply calculate the $q_{250}$ value in the observed frame. 

Fig. \ref{fig:FIRC} shows the distribution of our measured monochromatic $q_{250}$ values for the MIGHTEE sources that do not have a match in COSMOS2020. We also compare with similar measurements from previous studies of the FIRC at 1.4 GHz. For example, \citet{2010MNRAS.402..245I} 
used multi-wavelength data across the redshift range $0<z<3$ in the extended Chandra Deep Field South field, including 250, 350 and 500 $\mu$m data from the Balloon-borne Large Aperture
Submillimetre Telescope (BLAST;  \citealt{2009Natur.458..737D}) and 1.4 GHz data from the VLA
\citep{2008ApJS..179..114M}. For their 250 $\mu$m selected sample, they measured $q_{250} = 2.26\pm0.35$, without applying $k$-correction. Using a stacking analysis on their 24 $\mu$m selected sample, they measured $q_{250} = 2.70\pm0.08$ without applying $k$-correction. \citet{2010MNRAS.409...92J} 
studied the FIRC at $z<0.5$ using the {\it Herschel} Astrophysical TeraHertz Large Area Survey (H-ATLAS; \citealt{2010PASP..122..499E})  Science Demonstration Phase
(SDP) data and radio data from the NRAO VLA Northern Sky Survey (NVSS; \citealt{1998AJ....115.1693C}) and the Faint Images of the Radio Sky at Twenty-one centimetres (FIRST; \citealt{1995ApJ...450..559B}) and found a mean value of $2.01\pm0.04$ for $q_{250}$. \citet{2014MNRAS.445.2232S} used a sample of 250 $\mu$m-selected galaxies at $z<0.5$ from H-ATLAS and cross-matched with the 1.4 GHz flux density measurements from FIRST. They reported a value around 2.5 for $q_{250}$ based on a stacking analysis. Finally, \citet{2018MNRAS.475.3010G} selected a sample of galaxies from the Sloan Digital Sky Survey (SDSS DR7; \citealt{2009ApJS..182..543A}). Matching the sample to the far-IR data from the H-ATLAS survey in the North Galactic Pole (NGP) field and the 1.4 GHz data from FIRST, they measured a best-fit value of $\sim1.95\pm0.2$ for $q_{250}$. Although it is difficult to make proper comparisons between our measured $q_{250}$ values and the previous results due to various differences (such as the probed redshift range, sample selection, the adopted radio frequency and whether $k-$correction is applied), we see a reasonably good agreement. This  demonstrates yet again the high quality of our deblended far-IR and sub-mm photometry.


\section{Conclusions}\label{conclude}

Far-IR \& sub-mm data are key to measuring the obscured star-formation activity. The star-formation properties of very dusty galaxies (for which the commonly adopted energy balance assumption could also break down) cannot be constrained reliably without  far-IR/sub-mm data. The dusty galaxy population is also essential in understanding the formation and evolution of massive galaxies, as most of the massive population are shown to be dusty at high redshifts. However, there is currently no planned IR mission until the 2030s at the earliest. In this paper, we present a new deblended far-IR and sub-mm point source catalogue derived using a probabilistic and progressive deblending approach. We have combined the strengths of previous efforts and also improved on several aspects including the construction of the prior catalogue and taking into account the correlations between bands. Crucially, the construction of the prior benefits from an improved multi-wavelength catalogue in COSMOS and from a deep learning based emulator to speed up the flux prediction step which needs to be carried out progressively. 

Based on testing using realistic simulations of the far-IR and sub-mm sky, we have demonstrated the good performance of our deblending methodology. For the {\it Spitzer}/MIPS 24 $\mu$m map and the {\it Herschel}/PACS 100 \& 160 $\mu$m maps which are dominated by instrument noise, we can deblend fluxes which are median unbiased and have approximately Gaussian uncertainties all the way down to the 1$\sigma$ instrument noise. For the {\it Herschel}/SPIRE maps which are dominated by confusion noise, we show that with our deblending methodology we can deblend fluxes down to roughly the $1\sigma$ confusion noise. There is a slight systematic flux underestimation towards faint flux levels, which reaches approximately 10\%, 15\% and 25\% at 250, 350 and 500 $\mu$m, respectively, for the faintest sources. After testing with simulations, we deblend the real data and compare with previous blindly extracted catalogues and the super-deblended catalogue from \citet{2018ApJ...864...56J}. 
As an additional test, we deblend the SCUBA-2 850 $\mu$m and test our deblended photometry using ALMA Band 7 continuum measurements. We publicly release our XID+ deblended far-IR and sub-mm point source catalogues through the {\it Herschel} Extragalactic Legacy Project \footnote{\url{https://hedam.lam.fr/HELP/dataproducts/dmu26/dmu26_XID+COSMOS2024/}}. In Table \ref{cat_final}, we provide an explanation of the columns contained in the catalogue. 

We use the COSMOS field as the first application in this work. In the future, we plan to apply our deblending methodology to other premier extragalactic survey fields. On the observational front, a lot of progress has been made over the past few years and exciting new data are expected to arrive soon which can be used to further improve the prior information used to deblend the long-wavelength data. For example, the COSMOS-Web (\citet{2023ApJ...954...31C}, PIs: Kartaltepe \& Casey, ID=1727) treasury program using JWST will cover 0.54 deg$^2$ the COSMOS field with NIRCam imaging and 0.19 deg$^2$ with MIRI imaging. The recently launched {\it Euclid} satellite will also observe the COSMOS field as one of its calibration fields. The deep optical and near-IR {\it Euclid} data, together with the rich multi-wavelength data such as deep IRAC mid-IR observations will be crucial for more accurate photometric redshift and stellar mass estimates, thereby providing better prior catalogue for future deblending work. On the other hand, new deep sub-mm observations from the ground can be used to directly validate any deblending approach. For example, the upcoming A-MKID sub-mm camera on the APEX telescope based on the latest microwave kinetic inductance detector (MKID) technologies will survey the sky at 350 and 850 $\mu$m simultaneously. Thanks to its sensitivity, large format, and field of view (15$\times$15 arcmin$^2$), A-MKID will be an unprecedented sub-mm survey machine, capable of imaging large areas down to confusion limits in a reasonable time. Benefiting hugely from the 12-metre dish size, the resulting beam size at 350 $\mu$m will be around 7\arcsec (a factor of 3.4 smaller than {\it Herschel}).


\begin{table*}\label{cat_final}
        \caption{Columns contained in our XID+ deblended far-IR and sub-mm point source catalogue.}
        \label{table:cat}
        \centering
        \begin{tabular}{l l l}
                \hline
                Name & Unit & Description\\
                \hline
                ID             & - & The COSMOS2020 ID (negative numbers for radio sources) \\
                R.A. & - & Right Ascension from COSMOS2020 (or radio positions)\\
                Dec & - & Declination  from COSMOS2020 (or radio positions)\\
                \hline
                F\_24 & mJy& 24 $\mu$m flux density (median)\\
                FErr\_24\_u  & mJy & 24 $\mu$m flux density (84th Percentile); $\sigma^+=$ FErr\_24\_u $-$ FErr\_24 \\
                FErr\_24\_l & mJy & 24 $\mu$m flux density (16th Percentile); $\sigma^-=$ FErr\_24 $-$ FErr\_24\_l\\
           FErr\_24\_$1\sigma$ & mJy & maximum of $\sigma^+$ and $\sigma^-$ \\

                Bkg\_24 & mJy/Beam & Fitted Background of 24 $\mu$m map (median)\\
                Sig\_conf\_24 & mJy/Beam & Fitted residual noise component due to confusion (median) \\
                
Sig\_tot\_24 & mJy/Beam & total error = $\sqrt{ ({\rm Sig\_conf\_24})^2 + ({\rm FErr\_24\_1\sigma)^2} }$ \\

                Rhat\_24 & - & Convergence Statistic (ideally $<1.2$)\\
                
                n\_eff\_24 & - & Number of effective samples (ideally $>40$)\\
                
                Post\_24 & mJy  & 3000 samplings from the posterior PDF of the 24 $\mu$m flux density\\

                tile\_MIPS & - & tile number\\
                \hline
                       F\_100/160 & mJy& 100/160 $\mu$m flux density (median)\\
                FErr\_100/160\_u  & mJy & 100/160 $\mu$m flux density (84th Percentile)\\
                FErr\_100/160\_l & mJy & 100/160 $\mu$m flux density (16th Percentile)\\
                
                FErr\_100/160\_$1\sigma$ & mJy & maximum of $\sigma^+$ and $\sigma^-$ \\

                Bkg\_100/160 & mJy/Beam & Fitted Background of 100/160 $\mu$m map (median)\\
                Sig\_conf\_100/160 & mJy/Beam & Fitted residual noise component due to confusion (median) \\

                Sig\_tot\_100/160 & mJy/Beam & total error = $\sqrt{ ({\rm Sig\_conf\_100/160})^2 + ({\rm FErr\_100/160\_1\sigma)^2} }$ \\

                Rhat\_100/160 & - & Convergence Statistic (ideally $<1.2$)\\
                n\_eff\_100/160 & - & Number of effective samples (ideally $>40$)\\
                Post\_100/160 & mJy  & 3000 samplings from the posterior PDF of the 100/160 $\mu$m flux density\\
                tile\_PACS & - & tile number\\
                \hline
                F\_250/350/500 & mJy& 250/350/500 $\mu$m flux density (median)\\
                FErr\_250/350/500\_u  & mJy & 250/350/500 $\mu$m flux density (84th Percentile)\\
                FErr\_250/350/500\_l & mJy & 250/350/500 $\mu$m flux density (16th Percentile)\\

                FErr\_250/350/500\_$1\sigma$ & mJy & maximum of $\sigma^+$ and $\sigma^-$ \\

Bkg\_250/350/500 & mJy/Beam & Fitted Background of 250/350/500 $\mu$m map (median)\\
                Sig\_conf\_250/350/500 & mJy/Beam & Fitted residual noise component due to confusion (median) \\
           
                Sig\_tot\_250/350/500 & mJy/Beam & total error = $\sqrt{ ({\rm Sig\_conf\_250/350/500})^2 + ({\rm FErr\_250/350/500\_1\sigma)^2} }$ \\
                Rhat\_250/350/500 & - & Convergence Statistic (ideally $<1.2$)\\
                n\_eff\_250/350/500 & - & Number of effective samples (ideally $>40$)\\
                Post\_250/350/500 & mJy  & 3000 samplings from the posterior PDF of the 250/350/500 $\mu$m flux density\\
                tile\_SPIRE & - & tile number\\
                \hline
                                
                F\_850 & mJy& 850 $\mu$m flux density (median)\\
                FErr\_850\_u  & mJy & 850 $\mu$m flux density (84th Percentile)\\
                FErr\_850\_l & mJy & 850 $\mu$m flux density (16th Percentile)\\
                
                FErr\_850\_$1\sigma$ & mJy & maximum of $\sigma^+$ and $\sigma^-$ \\

                Bkg\_850 & mJy/Beam & Fitted Background of 850 $\mu$m map (median)\\
                Sig\_conf\_850 & mJy/Beam & Fitted residual noise component due to confusion (median) \\

Sig\_tot\_850 & mJy/Beam & total error = $\sqrt{ ({\rm Sig\_conf\_850})^2 + ({\rm FErr\_850\_1\sigma)^2} }$ \\
                
                Rhat\_850 & - & Convergence Statistic (ideally $<1.2$)\\
                n\_eff\_850 & - & Number of effective samples (ideally $>40$)\\
                Post\_850 & mJy  & 3000 samplings from the posterior PDF of the 850 $\mu$m flux density\\
                tile\_SCUBA & - & tile number\\
                \hline
        \end{tabular}
  \tablefoot{
$\sigma^+$ can be calculated from the difference of the 84th percentile and the median. $\sigma^-$ can be calculated from the difference between the median and the 16th percentile. The $1\sigma$ uncertainty of the source flux density can be derived from the maximum of $\sigma^+$ and $\sigma^-$. For a final estimate of the flux uncertainty, one can use the total error derived from combining the $1\sigma$ uncertainty and the residual confusion noise in quadrature. For the MIPS and SPIRE bands, this is equivalent to scaling the $1\sigma$ uncertainty by a factor of two. For the PACS and SCUBA-2 bands, the residual confusion noise is much smaller than the instrument noise level. Consequently, the total error is very close to the $1\sigma$ uncertainty.
  }
\end{table*}




\begin{acknowledgements}
We thank the anonymous referee for the constructive comments and suggestions.
      
      We thank Ian Smail and Emanuele Daddi for providing many very helpful comments and suggestions which significantly improved the quality of the paper. 
      
      This publication is part of the project ‘Clash of the titans:
deciphering the enigmatic role of cosmic collisions’ (with project number VI.Vidi.193.113 of the research programme Vidi which is (partly) financed by the Dutch Research Council (NWO).

      Using data from the public data release of the PACS Evolutionary Probe PEP (Lutz et al. 2011).
      
      W.J.P. has been supported by the Polish National Science Center project UMO-2020/37/B/ST9/00466. 

IHW, MJJ and CLH acknowledge generous support from the Hintze Family Charitable Foundation through the Oxford Hintze Centre for Astrophysical Surveys. CLH acknowledges support from the Leverhulme Trust through an Early Career Research Fellowship.

MV acknowledges financial support from the Inter-University Institute for Data Intensive Astronomy (IDIA), a partnership of the University of Cape Town, the University of Pretoria and the University of the Western Cape, and from the South African Department of Science and Innovation's National Research Foundation under the ISARP RADIOMAP+ Joint Research Scheme (DSI-NRF Grant Number 150551) and the CPRR HIPPO Project (DSI-NRF Grant Number SRUG22031677). 

LM acknowledges support from the Italian Ministry of Foreign Affairs and International Cooperation (MAECI Grant Number ZA18GR02) and the South African Department of Science and Innovation’s National Research Foundation (DSI-NRF Grant Number 113121) under the ISARP RADIOSKY2020 Joint Research Scheme.

Based on observations collected at the European Southern Observatory under ESO programme ID 179.A-2005 and on data products produced by CALET and the Cambridge Astronomy Survey Unit on behalf of the UltraVISTA consortium.

      The MeerKAT telescope is operated by the South African Radio Astronomy Observatory, which is a facility of the National Research Foundation, an agency of the Department of Science and Innovation. We acknowledge the use of the ilifu cloud computing facility – www.ilifu.ac.za, a partnership between the University of Cape Town, the University of the Western Cape, Stellenbosch University, Sol Plaatje University and the Cape Peninsula University of Technology. The Ilifu facility is supported by contributions from the Inter-University Institute for Data Intensive Astronomy (IDIA – a partnership between the University of Cape Town, the University of Pretoria and the University of the Western Cape, the Computational Biology division at UCT and the Data Intensive Research Initiative of South Africa (DIRISA). The authors acknowledge the Centre for High Performance Computing (CHPC), South Africa, for providing computational resources to this research project.

      The James Clerk Maxwell Telescope is operated by the East Asian Observatory on behalf of The National Astronomical Observatory of Japan; Academia Sinica Institute of Astronomy and Astrophysics; the Korea Astronomy and Space Science Institute; the Operation, Maintenance and Upgrading Fund for Astronomical Telescopes and Facility Instruments, budgeted from the Ministry of Finance (MOF) of China and administrated by the Chinese Academy of Sciences (CAS), as well as the National Key R\&D Program of China (No.\ 2017YFA0402700). Additional funding support is provided by the Science and Technology Facilities Council of the United Kingdom and participating universities in the United Kingdom and Canada (ST/M007634/1, ST/M003019/1, ST/N005856/1). The James Clerk Maxwell Telescope has historically been operated by the Joint Astronomy Centre on behalf of the Science and Technology Facilities Council of the United Kingdom, the National Research Council of Canada and the Netherlands Organization for Scientific Research and data from observations undertaken during this period of operation is used in this manuscript. This research used the JCMT project IDs M16AL002, M17BL006 (S2COSMOS) and S2CLS (MJLCS01) and the facilities of the Canadian Astronomy Data Centre operated by the National Research Council of Canada with the support of the Canadian Space Agency.
\end{acknowledgements}

%
%

\bibliographystyle{aa}
\bibliography{MS-references}

\begin{thebibliography}{124}
\expandafter\ifx\csname natexlab\endcsname\relax\def\natexlab#1{#1}\fi

\bibitem[{{Abadi} {et~al.}(2016){Abadi}, {Barham}, {Chen}, {Chen}, {Davis}, {Dean}, {Devin}, {Ghemawat}, {Irving}, {Isard}, {Kudlur}, {Levenberg}, {Monga}, {Moore}, {Murray}, {Steiner}, {Tucker}, {Vasudevan}, {Warden}, {Wicke}, {Yu}, \& {Zheng}}]{2016arXiv160508695A}
{Abadi}, M., {Barham}, P., {Chen}, J., {et~al.} 2016, arXiv e-prints, arXiv:1605.08695

\bibitem[{{Abazajian} {et~al.}(2009){Abazajian}, {Adelman-McCarthy}, {Ag{\"u}eros}, {Allam}, {Allende Prieto}, {An}, {Anderson}, {Anderson}, {Annis}, {Bahcall}, {Bailer-Jones}, {Barentine}, {Bassett}, {Becker}, {Beers}, {Bell}, {Belokurov}, {Berlind}, {Berman}, {Bernardi}, {Bickerton}, {Bizyaev}, {Blakeslee}, {Blanton}, {Bochanski}, {Boroski}, {Brewington}, {Brinchmann}, {Brinkmann}, {Brunner}, {Budav{\'a}ri}, {Carey}, {Carliles}, {Carr}, {Castander}, {Cinabro}, {Connolly}, {Csabai}, {Cunha}, {Czarapata}, {Davenport}, {de Haas}, {Dilday}, {Doi}, {Eisenstein}, {Evans}, {Evans}, {Fan}, {Friedman}, {Frieman}, {Fukugita}, {G{\"a}nsicke}, {Gates}, {Gillespie}, {Gilmore}, {Gonzalez}, {Gonzalez}, {Grebel}, {Gunn}, {Gy{\"o}ry}, {Hall}, {Harding}, {Harris}, {Harvanek}, {Hawley}, {Hayes}, {Heckman}, {Hendry}, {Hennessy}, {Hindsley}, {Hoblitt}, {Hogan}, {Hogg}, {Holtzman}, {Hyde}, {Ichikawa}, {Ichikawa}, {Im}, {Ivezi{\'c}}, {Jester}, {Jiang}, {Johnson}, {Jorgensen}, {Juri{\'c}}, {Kent}, {Kessler}, {Kleinman}, {Knapp},
  {Konishi}, {Kron}, {Krzesinski}, {Kuropatkin}, {Lampeitl}, {Lebedeva}, {Lee}, {Lee}, {French Leger}, {L{\'e}pine}, {Li}, {Lima}, {Lin}, {Long}, {Loomis}, {Loveday}, {Lupton}, {Magnier}, {Malanushenko}, {Malanushenko}, {Mandelbaum}, {Margon}, {Marriner}, {Mart{\'\i}nez-Delgado}, {Matsubara}, {McGehee}, {McKay}, {Meiksin}, {Morrison}, {Mullally}, {Munn}, {Murphy}, {Nash}, {Nebot}, {Neilsen}, {Newberg}, {Newman}, {Nichol}, {Nicinski}, {Nieto-Santisteban}, {Nitta}, {Okamura}, {Oravetz}, {Ostriker}, {Owen}, {Padmanabhan}, {Pan}, {Park}, {Pauls}, {Peoples}, {Percival}, {Pier}, {Pope}, {Pourbaix}, {Price}, {Purger}, {Quinn}, {Raddick}, {Re Fiorentin}, {Richards}, {Richmond}, {Riess}, {Rix}, {Rockosi}, {Sako}, {Schlegel}, {Schneider}, {Scholz}, {Schreiber}, {Schwope}, {Seljak}, {Sesar}, {Sheldon}, {Shimasaku}, {Sibley}, {Simmons}, {Sivarani}, {Allyn Smith}, {Smith}, {Smol{\v{c}}i{\'c}}, {Snedden}, {Stebbins}, {Steinmetz}, {Stoughton}, {Strauss}, {SubbaRao}, {Suto}, {Szalay}, {Szapudi}, {Szkody}, {Tanaka},
  {Tegmark}, {Teodoro}, {Thakar}, {Tremonti}, {Tucker}, {Uomoto}, {Vanden Berk}, {Vandenberg}, {Vidrih}, {Vogeley}, {Voges}, {Vogt}, {Wadadekar}, {Watters}, {Weinberg}, {West}, {White}, {Wilhite}, {Wonders}, {Yanny}, {Yocum}, {York}, {Zehavi}, {Zibetti}, \& {Zucker}}]{2009ApJS..182..543A}
{Abazajian}, K.~N., {Adelman-McCarthy}, J.~K., {Ag{\"u}eros}, M.~A., {et~al.} 2009, \apjs, 182, 543

\bibitem[{{Aihara} {et~al.}(2019){Aihara}, {AlSayyad}, {Ando}, {Armstrong}, {Bosch}, {Egami}, {Furusawa}, {Furusawa}, {Goulding}, {Harikane}, {Hikage}, {Ho}, {Hsieh}, {Huang}, {Ikeda}, {Imanishi}, {Ito}, {Iwata}, {Jaelani}, {Kakuma}, {Kawana}, {Kikuta}, {Kobayashi}, {Koike}, {Komiyama}, {Li}, {Liang}, {Lin}, {Luo}, {Lupton}, {Lust}, {MacArthur}, {Matsuoka}, {Mineo}, {Miyatake}, {Miyazaki}, {More}, {Murata}, {Namiki}, {Nishizawa}, {Oguri}, {Okabe}, {Okamoto}, {Okura}, {Ono}, {Onodera}, {Onoue}, {Osato}, {Ouchi}, {Shibuya}, {Strauss}, {Sugiyama}, {Suto}, {Takada}, {Takagi}, {Takata}, {Takita}, {Tanaka}, {Terai}, {Toba}, {Uchiyama}, {Utsumi}, {Wang}, {Wang}, \& {Yamada}}]{2019PASJ...71..114A}
{Aihara}, H., {AlSayyad}, Y., {Ando}, M., {et~al.} 2019, \pasj, 71, 114

\bibitem[{{Algera} {et~al.}(2023){Algera}, {Inami}, {Oesch}, {Sommovigo}, {Bouwens}, {Topping}, {Schouws}, {Stefanon}, {Stark}, {Aravena}, {Barrufet}, {da Cunha}, {Dayal}, {Endsley}, {Ferrara}, {Fudamoto}, {Gonzalez}, {Graziani}, {Hodge}, {Hygate}, {de Looze}, {Nanayakkara}, {Schneider}, \& {van der Werf}}]{2023MNRAS.518.6142A}
{Algera}, H. S.~B., {Inami}, H., {Oesch}, P.~A., {et~al.} 2023, \mnras, 518, 6142

\bibitem[{{An} {et~al.}(2019){An}, {Simpson}, {Smail}, {Swinbank}, {Ma}, {Liu}, {Lang}, {Schinnerer}, {Karim}, {Magnelli}, {Leslie}, {Bertoldi}, {Chen}, {Geach}, {Matsuda}, {Stach}, {Wardlow}, {Gullberg}, {Ivison}, {Ao}, {Coogan}, {Thomson}, {Chapman}, {Wang}, {Wang}, {Yang}, {Asquith}, {Bourne}, {Coppin}, {Hine}, {Ho}, {Hwang}, {Kato}, {Lacaille}, {Lewis}, {Oteo}, {Scholtz}, {Sawicki}, \& {Smith}}]{2019ApJ...886...48A}
{An}, F.~X., {Simpson}, J.~M., {Smail}, I., {et~al.} 2019, \apj, 886, 48

\bibitem[{{An} {et~al.}(2018){An}, {Stach}, {Smail}, {Swinbank}, {Almaini}, {Simpson}, {Hartley}, {Maltby}, {Ivison}, {Arumugam}, {Wardlow}, {Cooke}, {Gullberg}, {Thomson}, {Chen}, {Simpson}, {Geach}, {Scott}, {Dunlop}, {Farrah}, {van der Werf}, {Blain}, {Conselice}, {Micha{\l}owski}, {Chapman}, \& {Coppin}}]{2018ApJ...862..101A}
{An}, F.~X., {Stach}, S.~M., {Smail}, I., {et~al.} 2018, \apj, 862, 101

\bibitem[{{Arnouts} {et~al.}(2002){Arnouts}, {Moscardini}, {Vanzella}, {Colombi}, {Cristiani}, {Fontana}, {Giallongo}, {Matarrese}, \& {Saracco}}]{2002MNRAS.329..355A}
{Arnouts}, S., {Moscardini}, L., {Vanzella}, E., {et~al.} 2002, \mnras, 329, 355

\bibitem[{{Becker} {et~al.}(1995){Becker}, {White}, \& {Helfand}}]{1995ApJ...450..559B}
{Becker}, R.~H., {White}, R.~L., \& {Helfand}, D.~J. 1995, \apj, 450, 559

\bibitem[{{Berta} {et~al.}(2011){Berta}, {Magnelli}, {Nordon}, {Lutz}, {Wuyts}, {Altieri}, {Andreani}, {Aussel}, {Casta{\~n}eda}, {Cepa}, {Cimatti}, {Daddi}, {Elbaz}, {F{\"o}rster Schreiber}, {Genzel}, {Le Floc'h}, {Maiolino}, {P{\'e}rez-Fournon}, {Poglitsch}, {Popesso}, {Pozzi}, {Riguccini}, {Rodighiero}, {Sanchez-Portal}, {Sturm}, {Tacconi}, \& {Valtchanov}}]{2011A&A...532A..49B}
{Berta}, S., {Magnelli}, B., {Nordon}, R., {et~al.} 2011, \aap, 532, A49

\bibitem[{{B{\'e}thermin} {et~al.}(2015){B{\'e}thermin}, {Daddi}, {Magdis}, {Lagos}, {Sargent}, {Albrecht}, {Aussel}, {Bertoldi}, {Buat}, {Galametz}, {Heinis}, {Ilbert}, {Karim}, {Koekemoer}, {Lacey}, {Le Floc'h}, {Navarrete}, {Pannella}, {Schreiber}, {Smol{\v{c}}i{\'c}}, {Symeonidis}, \& {Viero}}]{2015A&A...573A.113B}
{B{\'e}thermin}, M., {Daddi}, E., {Magdis}, G., {et~al.} 2015, \aap, 573, A113

\bibitem[{{B{\'e}thermin} {et~al.}(2010){B{\'e}thermin}, {Dole}, {Cousin}, \& {Bavouzet}}]{2010A&A...516A..43B}
{B{\'e}thermin}, M., {Dole}, H., {Cousin}, M., \& {Bavouzet}, N. 2010, \aap, 516, A43

\bibitem[{{B{\'e}thermin} {et~al.}(2022){B{\'e}thermin}, {Van Cuyck}, {Beelen}, \& {Gkogkou}}]{2022ascl.soft04016B}
{B{\'e}thermin}, M., {Van Cuyck}, M., {Beelen}, A., \& {Gkogkou}, A. 2022, {pySIDES: Simulated Infrared Dusty Extragalactic Sky in Python}, Astrophysics Source Code Library, record ascl:2204.016

\bibitem[{{B{\'e}thermin} {et~al.}(2017){B{\'e}thermin}, {Wu}, {Lagache}, {Davidzon}, {Ponthieu}, {Cousin}, {Wang}, {Dor{\'e}}, {Daddi}, \& {Lapi}}]{2017A&A...607A..89B}
{B{\'e}thermin}, M., {Wu}, H.-Y., {Lagache}, G., {et~al.} 2017, \aap, 607, A89

\bibitem[{{Bourne} {et~al.}(2017){Bourne}, {Dunlop}, {Merlin}, {Parsa}, {Schreiber}, {Castellano}, {Conselice}, {Coppin}, {Farrah}, {Fontana}, {Geach}, {Halpern}, {Knudsen}, {Micha{\l}owski}, {Mortlock}, {Santini}, {Scott}, {Shu}, {Simpson}, {Simpson}, {Smith}, \& {van der Werf}}]{2017MNRAS.467.1360B}
{Bourne}, N., {Dunlop}, J.~S., {Merlin}, E., {et~al.} 2017, \mnras, 467, 1360

\bibitem[{Braak {et~al.}(2010)Braak, Boer, Totir, Winkler, Smith, \& Bink}]{braak2010identity}
Braak, C. J.~t., Boer, M.~P., Totir, L.~R., {et~al.} 2010, Genetics, 185, 1045

\bibitem[{{Brammer} {et~al.}(2008){Brammer}, {van Dokkum}, \& {Coppi}}]{2008ApJ...686.1503B}
{Brammer}, G.~B., {van Dokkum}, P.~G., \& {Coppi}, P. 2008, \apj, 686, 1503

\bibitem[{{Brinchmann} {et~al.}(2004){Brinchmann}, {Charlot}, {White}, {Tremonti}, {Kauffmann}, {Heckman}, \& {Brinkmann}}]{2004MNRAS.351.1151B}
{Brinchmann}, J., {Charlot}, S., {White}, S.~D.~M., {et~al.} 2004, \mnras, 351, 1151

\bibitem[{{Bruzual} \& {Charlot}(2003)}]{2003MNRAS.344.1000B}
{Bruzual}, G. \& {Charlot}, S. 2003, \mnras, 344, 1000

\bibitem[{{Bussmann} {et~al.}(2015){Bussmann}, {Riechers}, {Fialkov}, {Scudder}, {Hayward}, {Cowley}, {Bock}, {Calanog}, {Chapman}, {Cooray}, {De Bernardis}, {Farrah}, {Fu}, {Gavazzi}, {Hopwood}, {Ivison}, {Jarvis}, {Lacey}, {Loeb}, {Oliver}, {P{\'e}rez-Fournon}, {Rigopoulou}, {Roseboom}, {Scott}, {Smith}, {Vieira}, {Wang}, \& {Wardlow}}]{2015ApJ...812...43B}
{Bussmann}, R.~S., {Riechers}, D., {Fialkov}, A., {et~al.} 2015, \apj, 812, 43

\bibitem[{{Calzetti} {et~al.}(2000){Calzetti}, {Armus}, {Bohlin}, {Kinney}, {Koornneef}, \& {Storchi-Bergmann}}]{2000ApJ...533..682C}
{Calzetti}, D., {Armus}, L., {Bohlin}, R.~C., {et~al.} 2000, \apj, 533, 682

\bibitem[{{Casey}(2012)}]{2012MNRAS.425.3094C}
{Casey}, C.~M. 2012, \mnras, 425, 3094

\bibitem[{{Casey} {et~al.}(2023){Casey}, {Kartaltepe}, {Drakos}, {Franco}, {Harish}, {Paquereau}, {Ilbert}, {Rose}, {Cox}, {Nightingale}, {Robertson}, {Silverman}, {Koekemoer}, {Massey}, {McCracken}, {Rhodes}, {Akins}, {Allen}, {Amvrosiadis}, {Arango-Toro}, {Bagley}, {Bongiorno}, {Capak}, {Champagne}, {Chartab}, {Ch{\'a}vez Ortiz}, {Chworowsky}, {Cooke}, {Cooper}, {Darvish}, {Ding}, {Faisst}, {Finkelstein}, {Fujimoto}, {Gentile}, {Gillman}, {Gould}, {Gozaliasl}, {Hayward}, {He}, {Hemmati}, {Hirschmann}, {Jahnke}, {Jin}, {Khostovan}, {Kokorev}, {Lambrides}, {Laigle}, {Larson}, {Leung}, {Liu}, {Liaudat}, {Long}, {Magdis}, {Mahler}, {Mainieri}, {Manning}, {Maraston}, {Martin}, {McCleary}, {McKinney}, {McPartland}, {Mobasher}, {Pattnaik}, {Renzini}, {Rich}, {Sanders}, {Sattari}, {Scognamiglio}, {Scoville}, {Sheth}, {Shuntov}, {Sparre}, {Suzuki}, {Talia}, {Toft}, {Trakhtenbrot}, {Urry}, {Valentino}, {Vanderhoof}, {Vardoulaki}, {Weaver}, {Whitaker}, {Wilkins}, {Yang}, \& {Zavala}}]{2023ApJ...954...31C}
{Casey}, C.~M., {Kartaltepe}, J.~S., {Drakos}, N.~E., {et~al.} 2023, \apj, 954, 31

\bibitem[{{Casey} {et~al.}(2014){Casey}, {Narayanan}, \& {Cooray}}]{2014PhR...541...45C}
{Casey}, C.~M., {Narayanan}, D., \& {Cooray}, A. 2014, \physrep, 541, 45

\bibitem[{{Chabrier}(2003)}]{2003PASP..115..763C}
{Chabrier}, G. 2003, \pasp, 115, 763

\bibitem[{{Chapin} {et~al.}(2011){Chapin}, {Chapman}, {Coppin}, {Devlin}, {Dunlop}, {Greve}, {Halpern}, {Hasselfield}, {Hughes}, {Ivison}, {Marsden}, {Moncelsi}, {Netterfield}, {Pascale}, {Scott}, {Smail}, {Viero}, {Walter}, {Weiss}, \& {van der Werf}}]{2011MNRAS.411..505C}
{Chapin}, E.~L., {Chapman}, S.~C., {Coppin}, K.~E., {et~al.} 2011, \mnras, 411, 505

\bibitem[{{Charlot} \& {Fall}(2000)}]{2000ApJ...539..718C}
{Charlot}, S. \& {Fall}, S.~M. 2000, \apj, 539, 718

\bibitem[{{Condon} {et~al.}(1998){Condon}, {Cotton}, {Greisen}, {Yin}, {Perley}, {Taylor}, \& {Broderick}}]{1998AJ....115.1693C}
{Condon}, J.~J., {Cotton}, W.~D., {Greisen}, E.~W., {et~al.} 1998, \aj, 115, 1693

\bibitem[{{Dale} {et~al.}(2014){Dale}, {Helou}, {Magdis}, {Armus}, {D{\'\i}az-Santos}, \& {Shi}}]{2014ApJ...784...83D}
{Dale}, D.~A., {Helou}, G., {Magdis}, G.~E., {et~al.} 2014, \apj, 784, 83

\bibitem[{{Delhaize} {et~al.}(2017){Delhaize}, {Smol{\v{c}}i{\'c}}, {Delvecchio}, {Novak}, {Sargent}, {Baran}, {Magnelli}, {Zamorani}, {Schinnerer}, {Murphy}, {Aravena}, {Berta}, {Bondi}, {Capak}, {Carilli}, {Ciliegi}, {Civano}, {Ilbert}, {Karim}, {Laigle}, {Le F{\`e}vre}, {Marchesi}, {McCracken}, {Salvato}, {Seymour}, \& {Tasca}}]{2017A&A...602A...4D}
{Delhaize}, J., {Smol{\v{c}}i{\'c}}, V., {Delvecchio}, I., {et~al.} 2017, \aap, 602, A4

\bibitem[{{Devlin} {et~al.}(2009){Devlin}, {Ade}, {Aretxaga}, {Bock}, {Chapin}, {Griffin}, {Gundersen}, {Halpern}, {Hargrave}, {Hughes}, {Klein}, {Marsden}, {Martin}, {Mauskopf}, {Moncelsi}, {Netterfield}, {Ngo}, {Olmi}, {Pascale}, {Patanchon}, {Rex}, {Scott}, {Semisch}, {Thomas}, {Truch}, {Tucker}, {Tucker}, {Viero}, \& {Wiebe}}]{2009Natur.458..737D}
{Devlin}, M.~J., {Ade}, P. A.~R., {Aretxaga}, I., {et~al.} 2009, \nat, 458, 737

\bibitem[{{Dole} {et~al.}(2004){Dole}, {Rieke}, {Lagache}, {Puget}, {Alonso-Herrero}, {Bai}, {Blaylock}, {Egami}, {Engelbracht}, {Gordon}, {Hines}, {Kelly}, {Le Floc'h}, {Misselt}, {Morrison}, {Muzerolle}, {Papovich}, {P{\'e}rez-Gonz{\'a}lez}, {Rieke}, {Rigby}, {Neugebauer}, {Stansberry}, {Su}, {Young}, {Beichman}, \& {Richards}}]{2004ApJS..154...93D}
{Dole}, H., {Rieke}, G.~H., {Lagache}, G., {et~al.} 2004, \apjs, 154, 93

\bibitem[{{Draine} {et~al.}(2014){Draine}, {Aniano}, {Krause}, {Groves}, {Sandstrom}, {Braun}, {Leroy}, {Klaas}, {Linz}, {Rix}, {Schinnerer}, {Schmiedeke}, \& {Walter}}]{2014ApJ...780..172D}
{Draine}, B.~T., {Aniano}, G., {Krause}, O., {et~al.} 2014, \apj, 780, 172

\bibitem[{{Driver} {et~al.}(2008){Driver}, {Popescu}, {Tuffs}, {Graham}, {Liske}, \& {Baldry}}]{2008ApJ...678L.101D}
{Driver}, S.~P., {Popescu}, C.~C., {Tuffs}, R.~J., {et~al.} 2008, \apjl, 678, L101

\bibitem[{{Dudzevi{\v{c}}i{\={u}}t{\.{e}}} {et~al.}(2020){Dudzevi{\v{c}}i{\={u}}t{\.{e}}}, {Smail}, {Swinbank}, {Stach}, {Almaini}, {da Cunha}, {An}, {Arumugam}, {Birkin}, {Blain}, {Chapman}, {Chen}, {Conselice}, {Coppin}, {Dunlop}, {Farrah}, {Geach}, {Gullberg}, {Hartley}, {Hodge}, {Ivison}, {Maltby}, {Scott}, {Simpson}, {Simpson}, {Thomson}, {Walter}, {Wardlow}, {Weiss}, \& {van der Werf}}]{2020MNRAS.494.3828D}
{Dudzevi{\v{c}}i{\={u}}t{\.{e}}}, U., {Smail}, I., {Swinbank}, A.~M., {et~al.} 2020, \mnras, 494, 3828

\bibitem[{{Eales} {et~al.}(2010){Eales}, {Dunne}, {Clements}, {Cooray}, {De Zotti}, {Dye}, {Ivison}, {Jarvis}, {Lagache}, {Maddox}, {Negrello}, {Serjeant}, {Thompson}, {Van Kampen}, {Amblard}, {Andreani}, {Baes}, {Beelen}, {Bendo}, {Benford}, {Bertoldi}, {Bock}, {Bonfield}, {Boselli}, {Bridge}, {Buat}, {Burgarella}, {Carlberg}, {Cava}, {Chanial}, {Charlot}, {Christopher}, {Coles}, {Cortese}, {Dariush}, {da Cunha}, {Dalton}, {Danese}, {Dannerbauer}, {Driver}, {Dunlop}, {Fan}, {Farrah}, {Frayer}, {Frenk}, {Geach}, {Gardner}, {Gomez}, {Gonz{\'a}lez-Nuevo}, {Gonz{\'a}lez-Solares}, {Griffin}, {Hardcastle}, {Hatziminaoglou}, {Herranz}, {Hughes}, {Ibar}, {Jeong}, {Lacey}, {Lapi}, {Lawrence}, {Lee}, {Leeuw}, {Liske}, {L{\'o}pez-Caniego}, {M{\"u}ller}, {Nandra}, {Panuzzo}, {Papageorgiou}, {Patanchon}, {Peacock}, {Pearson}, {Phillipps}, {Pohlen}, {Popescu}, {Rawlings}, {Rigby}, {Rigopoulou}, {Robotham}, {Rodighiero}, {Sansom}, {Schulz}, {Scott}, {Smith}, {Sibthorpe}, {Smail}, {Stevens}, {Sutherland}, {Takeuchi},
  {Tedds}, {Temi}, {Tuffs}, {Trichas}, {Vaccari}, {Valtchanov}, {van der Werf}, {Verma}, {Vieria}, {Vlahakis}, \& {White}}]{2010PASP..122..499E}
{Eales}, S., {Dunne}, L., {Clements}, D., {et~al.} 2010, \pasp, 122, 499

\bibitem[{{Elbaz} {et~al.}(2011){Elbaz}, {Dickinson}, {Hwang}, {D{\'\i}az-Santos}, {Magdis}, {Magnelli}, {Le Borgne}, {Galliano}, {Pannella}, {Chanial}, {Armus}, {Charmandaris}, {Daddi}, {Aussel}, {Popesso}, {Kartaltepe}, {Altieri}, {Valtchanov}, {Coia}, {Dannerbauer}, {Dasyra}, {Leiton}, {Mazzarella}, {Alexander}, {Buat}, {Burgarella}, {Chary}, {Gilli}, {Ivison}, {Juneau}, {Le Floc'h}, {Lutz}, {Morrison}, {Mullaney}, {Murphy}, {Pope}, {Scott}, {Brodwin}, {Calzetti}, {Cesarsky}, {Charlot}, {Dole}, {Eisenhardt}, {Ferguson}, {F{\"o}rster Schreiber}, {Frayer}, {Giavalisco}, {Huynh}, {Koekemoer}, {Papovich}, {Reddy}, {Surace}, {Teplitz}, {Yun}, \& {Wilson}}]{2011A&A...533A.119E}
{Elbaz}, D., {Dickinson}, M., {Hwang}, H.~S., {et~al.} 2011, \aap, 533, A119

\bibitem[{{Euclid Collaboration} {et~al.}(2023){Euclid Collaboration}, {Bisigello}, {Conselice}, {Baes}, {Bolzonella}, {Brescia}, {Cavuoti}, {Cucciati}, {Humphrey}, {Hunt}, {Maraston}, {Pozzetti}, {Tortora}, {van Mierlo}, {Aghanim}, {Auricchio}, {Baldi}, {Bender}, {Bodendorf}, {Bonino}, {Branchini}, {Brinchmann}, {Camera}, {Capobianco}, {Carbone}, {Carretero}, {Castander}, {Castellano}, {Cimatti}, {Congedo}, {Conversi}, {Copin}, {Corcione}, {Courbin}, {Cropper}, {Da Silva}, {Degaudenzi}, {Douspis}, {Dubath}, {Duncan}, {Dupac}, {Dusini}, {Farrens}, {Ferriol}, {Frailis}, {Franceschi}, {Franzetti}, {Fumana}, {Garilli}, {Gillard}, {Gillis}, {Giocoli}, {Grazian}, {Grupp}, {Guzzo}, {Haugan}, {Holmes}, {Hormuth}, {Hornstrup}, {Jahnke}, {K{\"u}mmel}, {Kermiche}, {Kiessling}, {Kilbinger}, {Kohley}, {Kunz}, {Kurki-Suonio}, {Ligori}, {Lilje}, {Lloro}, {Maiorano}, {Mansutti}, {Marggraf}, {Markovic}, {Marulli}, {Massey}, {Maurogordato}, {Medinaceli}, {Meneghetti}, {Merlin}, {Meylan}, {Moresco}, {Moscardini}, {Munari},
  {Niemi}, {Padilla}, {Paltani}, {Pasian}, {Pedersen}, {Pettorino}, {Polenta}, {Poncet}, {Popa}, {Raison}, {Renzi}, {Rhodes}, {Riccio}, {Rix}, {Romelli}, {Roncarelli}, {Rosset}, {Rossetti}, {Saglia}, {Sapone}, {Sartoris}, {Schneider}, {Scodeggio}, {Secroun}, {Seidel}, {Sirignano}, {Sirri}, {Stanco}, {Tallada-Cresp{\'\i}}, {Tavagnacco}, {Taylor}, {Tereno}, {Toledo-Moreo}, {Torradeflot}, {Tutusaus}, {Valentijn}, {Valenziano}, {Vassallo}, {Wang}, {Zacchei}, {Zamorani}, {Zoubian}, {Andreon}, {Bardelli}, {Boucaud}, {Colodro-Conde}, {Di Ferdinando}, {Graci{\'a}-Carpio}, {Lindholm}, {Maino}, {Mei}, {Scottez}, {Sureau}, {Tenti}, {Zucca}, {Borlaff}, {Ballardini}, {Biviano}, {Bozzo}, {Burigana}, {Cabanac}, {Cappi}, {Carvalho}, {Casas}, {Castignani}, {Cooray}, {Coupon}, {Courtois}, {Cuby}, {Davini}, {De Lucia}, {Desprez}, {Dole}, {Escartin}, {Escoffier}, {Farina}, {Fotopoulou}, {Ganga}, {Garcia-Bellido}, {George}, {Giacomini}, {Gozaliasl}, {Hildebrandt}, {Hook}, {Huertas-Company}, {Kansal}, {Keihanen}, {Kirkpatrick},
  {Loureiro}, {Mac{\'\i}as-P{\'e}rez}, {Magliocchetti}, {Mainetti}, {Marcin}, {Martinelli}, {Martinet}, {Metcalf}, {Monaco}, {Morgante}, {Nadathur}, {Nucita}, {Patrizii}, {Peel}, {Potter}, {Pourtsidou}, {P{\"o}ntinen}, {Reimberg}, {S{\'a}nchez}, {Sakr}, {Schirmer}, {Sefusatti}, {Sereno}, {Stadel}, {Teyssier}, {Valieri}, {Valiviita}, \& {Viel}}]{2023MNRAS.520.3529E}
{Euclid Collaboration}, {Bisigello}, L., {Conselice}, C.~J., {et~al.} 2023, \mnras, 520, 3529

\bibitem[{{Euclid Collaboration} {et~al.}(2022){Euclid Collaboration}, {Moneti}, {McCracken}, {Shuntov}, {Kauffmann}, {Capak}, {Davidzon}, {Ilbert}, {Scarlata}, {Toft}, {Weaver}, {Chary}, {Cuby}, {Faisst}, {Masters}, {McPartland}, {Mobasher}, {Sanders}, {Scaramella}, {Stern}, {Szapudi}, {Teplitz}, {Zalesky}, {Amara}, {Auricchio}, {Bodendorf}, {Bonino}, {Branchini}, {Brau-Nogue}, {Brescia}, {Brinchmann}, {Capobianco}, {Carbone}, {Carretero}, {Castander}, {Castellano}, {Cavuoti}, {Cimatti}, {Cledassou}, {Congedo}, {Conselice}, {Conversi}, {Copin}, {Corcione}, {Costille}, {Cropper}, {Da Silva}, {Degaudenzi}, {Douspis}, {Dubath}, {Duncan}, {Dupac}, {Dusini}, {Farrens}, {Ferriol}, {Fosalba}, {Frailis}, {Franceschi}, {Fumana}, {Garilli}, {Gillis}, {Giocoli}, {Granett}, {Grazian}, {Grupp}, {Haugan}, {Hoekstra}, {Holmes}, {Hormuth}, {Hudelot}, {Jahnke}, {Kermiche}, {Kiessling}, {Kilbinger}, {Kitching}, {Kohley}, {K{\"u}mmel}, {Kunz}, {Kurki-Suonio}, {Ligori}, {Lilje}, {Lloro}, {Maiorano}, {Mansutti}, {Marggraf},
  {Markovic}, {Marulli}, {Massey}, {Maurogordato}, {Meneghetti}, {Merlin}, {Meylan}, {Moresco}, {Moscardini}, {Munari}, {Niemi}, {Padilla}, {Paltani}, {Pasian}, {Pedersen}, {Pires}, {Poncet}, {Popa}, {Pozzetti}, {Raison}, {Rebolo}, {Rhodes}, {Rix}, {Roncarelli}, {Rossetti}, {Saglia}, {Schneider}, {Secroun}, {Seidel}, {Serrano}, {Sirignano}, {Sirri}, {Stanco}, {Tallada-Cresp{\'\i}}, {Taylor}, {Tereno}, {Toledo-Moreo}, {Torradeflot}, {Wang}, {Welikala}, {Weller}, {Zamorani}, {Zoubian}, {Andreon}, {Bardelli}, {Camera}, {Graci{\'a}-Carpio}, {Medinaceli}, {Mei}, {Polenta}, {Romelli}, {Sureau}, {Tenti}, {Vassallo}, {Zacchei}, {Zucca}, {Baccigalupi}, {Balaguera-Antol{\'\i}nez}, {Bernardeau}, {Biviano}, {Bolzonella}, {Bozzo}, {Burigana}, {Cabanac}, {Cappi}, {Carvalho}, {Casas}, {Castignani}, {Colodro-Conde}, {Coupon}, {Courtois}, {Di Ferdinando}, {Farina}, {Finelli}, {Flose-Reimberg}, {Fotopoulou}, {Galeotta}, {Ganga}, {Garcia-Bellido}, {Gaztanaga}, {Gozaliasl}, {Hook}, {Joachimi}, {Kansal}, {Keihanen},
  {Kirkpatrick}, {Lindholm}, {Mainetti}, {Maino}, {Maoli}, {Martinelli}, {Martinet}, {Maturi}, {Metcalf}, {Morgante}, {Morisset}, {Nucita}, {Patrizii}, {Potter}, {Renzi}, {Riccio}, {S{\'a}nchez}, {Sapone}, {Schirmer}, {Schultheis}, {Scottez}, {Sefusatti}, {Teyssier}, {Tubio}, {Tutusaus}, {Valiviita}, {Viel}, \& {Hildebrandt}}]{2022A&A...658A.126E}
{Euclid Collaboration}, {Moneti}, A., {McCracken}, H.~J., {et~al.} 2022, \aap, 658, A126

\bibitem[{{Fixsen} {et~al.}(1998){Fixsen}, {Dwek}, {Mather}, {Bennett}, \& {Shafer}}]{1998ApJ...508..123F}
{Fixsen}, D.~J., {Dwek}, E., {Mather}, J.~C., {Bennett}, C.~L., \& {Shafer}, R.~A. 1998, \apj, 508, 123

\bibitem[{{Fritz} {et~al.}(2006){Fritz}, {Franceschini}, \& {Hatziminaoglou}}]{2006MNRAS.366..767F}
{Fritz}, J., {Franceschini}, A., \& {Hatziminaoglou}, E. 2006, \mnras, 366, 767

\bibitem[{{Gao} {et~al.}(2021){Gao}, {Wang}, {Efstathiou}, {Ma{\l}ek}, {Best}, {Bonato}, {Farrah}, {Kondapally}, {McCheyne}, \& {R{\"o}ttgering}}]{2021A&A...654A.117G}
{Gao}, F., {Wang}, L., {Efstathiou}, A., {et~al.} 2021, \aap, 654, A117

\bibitem[{{Gardner} {et~al.}(2006){Gardner}, {Mather}, {Clampin}, {Doyon}, {Greenhouse}, {Hammel}, {Hutchings}, {Jakobsen}, {Lilly}, {Long}, {Lunine}, {McCaughrean}, {Mountain}, {Nella}, {Rieke}, {Rieke}, {Rix}, {Smith}, {Sonneborn}, {Stiavelli}, {Stockman}, {Windhorst}, \& {Wright}}]{2006SSRv..123..485G}
{Gardner}, J.~P., {Mather}, J.~C., {Clampin}, M., {et~al.} 2006, \ssr, 123, 485

\bibitem[{{Geach} {et~al.}(2017){Geach}, {Dunlop}, {Halpern}, {Smail}, {van der Werf}, {Alexander}, {Almaini}, {Aretxaga}, {Arumugam}, {Asboth}, {Banerji}, {Beanlands}, {Best}, {Blain}, {Birkinshaw}, {Chapin}, {Chapman}, {Chen}, {Chrysostomou}, {Clarke}, {Clements}, {Conselice}, {Coppin}, {Cowley}, {Danielson}, {Eales}, {Edge}, {Farrah}, {Gibb}, {Harrison}, {Hine}, {Hughes}, {Ivison}, {Jarvis}, {Jenness}, {Jones}, {Karim}, {Koprowski}, {Knudsen}, {Lacey}, {Mackenzie}, {Marsden}, {McAlpine}, {McMahon}, {Meijerink}, {Micha{\l}owski}, {Oliver}, {Page}, {Peacock}, {Rigopoulou}, {Robson}, {Roseboom}, {Rotermund}, {Scott}, {Serjeant}, {Simpson}, {Simpson}, {Smith}, {Spaans}, {Stanley}, {Stevens}, {Swinbank}, {Targett}, {Thomson}, {Valiante}, {Wake}, {Webb}, {Willott}, {Zavala}, \& {Zemcov}}]{2017MNRAS.465.1789G}
{Geach}, J.~E., {Dunlop}, J.~S., {Halpern}, M., {et~al.} 2017, \mnras, 465, 1789

\bibitem[{{Gkogkou} {et~al.}(2023){Gkogkou}, {B{\'e}thermin}, {Lagache}, {Van Cuyck}, {Jullo}, {Aravena}, {Beelen}, {Benoit}, {Bounmy}, {Calvo}, {Catalano}, {Cora}, {Croton}, {de la Torre}, {Fasano}, {Ferrara}, {Goupy}, {Hoarau}, {Hu}, {Ishiyama}, {Knudsen}, {Lambert}, {Mac{\'\i}as-P{\'e}rez}, {Marpaud}, {Mellema}, {Monfardini}, {Pallottini}, {Ponthieu}, {Prada}, {Roehlly}, {Vallini}, \& {Walter}}]{2023A&A...670A..16G}
{Gkogkou}, A., {B{\'e}thermin}, M., {Lagache}, G., {et~al.} 2023, \aap, 670, A16

\bibitem[{{Gruppioni} {et~al.}(2020){Gruppioni}, {B{\'e}thermin}, {Loiacono}, {Le F{\`e}vre}, {Capak}, {Cassata}, {Faisst}, {Schaerer}, {Silverman}, {Yan}, {Bardelli}, {Boquien}, {Carraro}, {Cimatti}, {Dessauges-Zavadsky}, {Ginolfi}, {Fujimoto}, {Hathi}, {Jones}, {Khusanova}, {Koekemoer}, {Lagache}, {Lemaux}, {Oesch}, {Pozzi}, {Riechers}, {Rodighiero}, {Romano}, {Talia}, {Vallini}, {Vergani}, {Zamorani}, \& {Zucca}}]{2020A&A...643A...8G}
{Gruppioni}, C., {B{\'e}thermin}, M., {Loiacono}, F., {et~al.} 2020, \aap, 643, A8

\bibitem[{{G{\"u}rkan} {et~al.}(2018){G{\"u}rkan}, {Hardcastle}, {Smith}, {Best}, {Bourne}, {Calistro-Rivera}, {Heald}, {Jarvis}, {Prandoni}, {R{\"o}ttgering}, {Sabater}, {Shimwell}, {Tasse}, \& {Williams}}]{2018MNRAS.475.3010G}
{G{\"u}rkan}, G., {Hardcastle}, M.~J., {Smith}, D.~J.~B., {et~al.} 2018, \mnras, 475, 3010

\bibitem[{{Hatziminaoglou} {et~al.}(2018){Hatziminaoglou}, {Farrah}, {Humphreys}, {Manrique}, {P{\'e}rez-Fournon}, {Pitchford}, {Salvador-Sol{\'e}}, \& {Wang}}]{2018MNRAS.480.4974H}
{Hatziminaoglou}, E., {Farrah}, D., {Humphreys}, E., {et~al.} 2018, \mnras, 480, 4974

\bibitem[{{Hauser} \& {Dwek}(2001)}]{2001ARA&A..39..249H}
{Hauser}, M.~G. \& {Dwek}, E. 2001, \araa, 39, 249

\bibitem[{{Heywood} {et~al.}(2022){Heywood}, {Jarvis}, {Hale}, {Whittam}, {Bester}, {Hugo}, {Kenyon}, {Prescott}, {Smirnov}, {Tasse}, {Afonso}, {Best}, {Collier}, {Deane}, {Frank}, {Hardcastle}, {Knowles}, {Maddox}, {Murphy}, {Prandoni}, {Randriamampandry}, {Santos}, {Sekhar}, {Tabatabaei}, {Taylor}, \& {Thorat}}]{2022MNRAS.509.2150H}
{Heywood}, I., {Jarvis}, M.~J., {Hale}, C.~L., {et~al.} 2022, \mnras, 509, 2150

\bibitem[{{Hodge} {et~al.}(2013){Hodge}, {Karim}, {Smail}, {Swinbank}, {Walter}, {Biggs}, {Ivison}, {Weiss}, {Alexander}, {Bertoldi}, {Brandt}, {Chapman}, {Coppin}, {Cox}, {Danielson}, {Dannerbauer}, {De Breuck}, {Decarli}, {Edge}, {Greve}, {Knudsen}, {Menten}, {Rix}, {Schinnerer}, {Simpson}, {Wardlow}, \& {van der Werf}}]{2013ApJ...768...91H}
{Hodge}, J.~A., {Karim}, A., {Smail}, I., {et~al.} 2013, \apj, 768, 91

\bibitem[{{Holland} {et~al.}(2013){Holland}, {Bintley}, {Chapin}, {Chrysostomou}, {Davis}, {Dempsey}, {Duncan}, {Fich}, {Friberg}, {Halpern}, {Irwin}, {Jenness}, {Kelly}, {MacIntosh}, {Robson}, {Scott}, {Ade}, {Atad-Ettedgui}, {Berry}, {Craig}, {Gao}, {Gibb}, {Hilton}, {Hollister}, {Kycia}, {Lunney}, {McGregor}, {Montgomery}, {Parkes}, {Tilanus}, {Ullom}, {Walther}, {Walton}, {Woodcraft}, {Amiri}, {Atkinson}, {Burger}, {Chuter}, {Coulson}, {Doriese}, {Dunare}, {Economou}, {Niemack}, {Parsons}, {Reintsema}, {Sibthorpe}, {Smail}, {Sudiwala}, \& {Thomas}}]{2013MNRAS.430.2513H}
{Holland}, W.~S., {Bintley}, D., {Chapin}, E.~L., {et~al.} 2013, \mnras, 430, 2513

\bibitem[{{Hurley} {et~al.}(2017){Hurley}, {Oliver}, {Betancourt}, {Clarke}, {Cowley}, {Duivenvoorden}, {Farrah}, {Griffin}, {Lacey}, {Le Floc'h}, {Papadopoulos}, {Sargent}, {Scudder}, {Vaccari}, {Valtchanov}, \& {Wang}}]{2017MNRAS.464..885H}
{Hurley}, P.~D., {Oliver}, S., {Betancourt}, M., {et~al.} 2017, \mnras, 464, 885

\bibitem[{{Ilbert} {et~al.}(2006){Ilbert}, {Arnouts}, {McCracken}, {Bolzonella}, {Bertin}, {Le F{\`e}vre}, {Mellier}, {Zamorani}, {Pell{\`o}}, {Iovino}, {Tresse}, {Le Brun}, {Bottini}, {Garilli}, {Maccagni}, {Picat}, {Scaramella}, {Scodeggio}, {Vettolani}, {Zanichelli}, {Adami}, {Bardelli}, {Cappi}, {Charlot}, {Ciliegi}, {Contini}, {Cucciati}, {Foucaud}, {Franzetti}, {Gavignaud}, {Guzzo}, {Marano}, {Marinoni}, {Mazure}, {Meneux}, {Merighi}, {Paltani}, {Pollo}, {Pozzetti}, {Radovich}, {Zucca}, {Bondi}, {Bongiorno}, {Busarello}, {de La Torre}, {Gregorini}, {Lamareille}, {Mathez}, {Merluzzi}, {Ripepi}, {Rizzo}, \& {Vergani}}]{2006A&A...457..841I}
{Ilbert}, O., {Arnouts}, S., {McCracken}, H.~J., {et~al.} 2006, \aap, 457, 841

\bibitem[{{Ivison} {et~al.}(2010){Ivison}, {Alexander}, {Biggs}, {Brandt}, {Chapin}, {Coppin}, {Devlin}, {Dickinson}, {Dunlop}, {Dye}, {Eales}, {Frayer}, {Halpern}, {Hughes}, {Ibar}, {Kov{\'a}cs}, {Marsden}, {Moncelsi}, {Netterfield}, {Pascale}, {Patanchon}, {Rafferty}, {Rex}, {Schinnerer}, {Scott}, {Semisch}, {Smail}, {Swinbank}, {Truch}, {Tucker}, {Viero}, {Walter}, {Wei{\ss}}, {Wiebe}, \& {Xue}}]{2010MNRAS.402..245I}
{Ivison}, R.~J., {Alexander}, D.~M., {Biggs}, A.~D., {et~al.} 2010, \mnras, 402, 245

\bibitem[{{Jarvis} {et~al.}(2016){Jarvis}, {Taylor}, {Agudo}, {Allison}, {Deane}, {Frank}, {Gupta}, {Heywood}, {Maddox}, {McAlpine}, {Santos}, {Scaife}, {Vaccari}, {Zwart}, {Adams}, {Bacon}, {Baker}, {Bassett}, {Best}, {Beswick}, {Blyth}, {Brown}, {Bruggen}, {Cluver}, {Colafrancesco}, {Cotter}, {Cress}, {Dav{\'e}}, {Ferrari}, {Hardcastle}, {Hale}, {Harrison}, {Hatfield}, {Klockner}, {Kolwa}, {Malefahlo}, {Marubini}, {Mauch}, {Moodley}, {Morganti}, {Norris}, {Peters}, {Prandoni}, {Prescott}, {Oliver}, {Oozeer}, {Rottgering}, {Seymour}, {Simpson}, {Smirnov}, \& {Smith}}]{2016mks..confE...6J}
{Jarvis}, M., {Taylor}, R., {Agudo}, I., {et~al.} 2016, in MeerKAT Science: On the Pathway to the SKA, 6

\bibitem[{{Jarvis} {et~al.}(2010){Jarvis}, {Smith}, {Bonfield}, {Hardcastle}, {Falder}, {Stevens}, {Ivison}, {Auld}, {Baes}, {Baldry}, {Bamford}, {Bourne}, {Buttiglione}, {Cava}, {Cooray}, {Dariush}, {de Zotti}, {Dunlop}, {Dunne}, {Dye}, {Eales}, {Fritz}, {Hill}, {Hopwood}, {Hughes}, {Ibar}, {Jones}, {Kelvin}, {Lawrence}, {Leeuw}, {Loveday}, {Maddox}, {Micha{\l}owski}, {Negrello}, {Norberg}, {Pohlen}, {Prescott}, {Rigby}, {Robotham}, {Rodighiero}, {Scott}, {Sharp}, {Temi}, {Thompson}, {van der Werf}, {van Kampen}, {Vlahakis}, \& {White}}]{2010MNRAS.409...92J}
{Jarvis}, M.~J., {Smith}, D.~J.~B., {Bonfield}, D.~G., {et~al.} 2010, \mnras, 409, 92

\bibitem[{{Jin} {et~al.}(2018){Jin}, {Daddi}, {Liu}, {Smol{\v{c}}i{\'c}}, {Schinnerer}, {Calabr{\`o}}, {Gu}, {Delhaize}, {Delvecchio}, {Gao}, {Salvato}, {Puglisi}, {Dickinson}, {Bertoldi}, {Sargent}, {Novak}, {Magdis}, {Aretxaga}, {Wilson}, \& {Capak}}]{2018ApJ...864...56J}
{Jin}, S., {Daddi}, E., {Liu}, D., {et~al.} 2018, \apj, 864, 56

\bibitem[{{Jonas} \& {MeerKAT Team}(2016)}]{2016mks..confE...1J}
{Jonas}, J. \& {MeerKAT Team}. 2016, in MeerKAT Science: On the Pathway to the SKA, 1

\bibitem[{Kingma \& Ba(2014)}]{kingma2014adam}
Kingma, D.~P. \& Ba, J. 2014, arXiv preprint arXiv:1412.6980

\bibitem[{{Komatsu} {et~al.}(2011){Komatsu}, {Smith}, {Dunkley}, {Bennett}, {Gold}, {Hinshaw}, {Jarosik}, {Larson}, {Nolta}, {Page}, {Spergel}, {Halpern}, {Hill}, {Kogut}, {Limon}, {Meyer}, {Odegard}, {Tucker}, {Weiland}, {Wollack}, \& {Wright}}]{2011ApJS..192...18K}
{Komatsu}, E., {Smith}, K.~M., {Dunkley}, J., {et~al.} 2011, \apjs, 192, 18

\bibitem[{{Laigle} {et~al.}(2016){Laigle}, {McCracken}, {Ilbert}, {Hsieh}, {Davidzon}, {Capak}, {Hasinger}, {Silverman}, {Pichon}, {Coupon}, {Aussel}, {Le Borgne}, {Caputi}, {Cassata}, {Chang}, {Civano}, {Dunlop}, {Fynbo}, {Kartaltepe}, {Koekemoer}, {Le F{\`e}vre}, {Le Floc'h}, {Leauthaud}, {Lilly}, {Lin}, {Marchesi}, {Milvang-Jensen}, {Salvato}, {Sanders}, {Scoville}, {Smolcic}, {Stockmann}, {Taniguchi}, {Tasca}, {Toft}, {Vaccari}, \& {Zabl}}]{2016ApJS..224...24L}
{Laigle}, C., {McCracken}, H.~J., {Ilbert}, O., {et~al.} 2016, \apjs, 224, 24

\bibitem[{{Lang} {et~al.}(2016){Lang}, {Hogg}, \& {Mykytyn}}]{2016ascl.soft04008L}
{Lang}, D., {Hogg}, D.~W., \& {Mykytyn}, D. 2016, {The Tractor: Probabilistic astronomical source detection and measurement}, Astrophysics Source Code Library, record ascl:1604.008

\bibitem[{{Larson} {et~al.}(2011){Larson}, {Dunkley}, {Hinshaw}, {Komatsu}, {Nolta}, {Bennett}, {Gold}, {Halpern}, {Hill}, {Jarosik}, {Kogut}, {Limon}, {Meyer}, {Odegard}, {Page}, {Smith}, {Spergel}, {Tucker}, {Weiland}, {Wollack}, \& {Wright}}]{2011ApJS..192...16L}
{Larson}, D., {Dunkley}, J., {Hinshaw}, G., {et~al.} 2011, \apjs, 192, 16

\bibitem[{{Laureijs} {et~al.}(2011){Laureijs}, {Amiaux}, {Arduini}, {Augu{\`e}res}, {Brinchmann}, {Cole}, {Cropper}, {Dabin}, {Duvet}, {Ealet}, {Garilli}, {Gondoin}, {Guzzo}, {Hoar}, {Hoekstra}, {Holmes}, {Kitching}, {Maciaszek}, {Mellier}, {Pasian}, {Percival}, {Rhodes}, {Saavedra Criado}, {Sauvage}, {Scaramella}, {Valenziano}, {Warren}, {Bender}, {Castander}, {Cimatti}, {Le F{\`e}vre}, {Kurki-Suonio}, {Levi}, {Lilje}, {Meylan}, {Nichol}, {Pedersen}, {Popa}, {Rebolo Lopez}, {Rix}, {Rottgering}, {Zeilinger}, {Grupp}, {Hudelot}, {Massey}, {Meneghetti}, {Miller}, {Paltani}, {Paulin-Henriksson}, {Pires}, {Saxton}, {Schrabback}, {Seidel}, {Walsh}, {Aghanim}, {Amendola}, {Bartlett}, {Baccigalupi}, {Beaulieu}, {Benabed}, {Cuby}, {Elbaz}, {Fosalba}, {Gavazzi}, {Helmi}, {Hook}, {Irwin}, {Kneib}, {Kunz}, {Mannucci}, {Moscardini}, {Tao}, {Teyssier}, {Weller}, {Zamorani}, {Zapatero Osorio}, {Boulade}, {Foumond}, {Di Giorgio}, {Guttridge}, {James}, {Kemp}, {Martignac}, {Spencer}, {Walton}, {Bl{\"u}mchen}, {Bonoli},
  {Bortoletto}, {Cerna}, {Corcione}, {Fabron}, {Jahnke}, {Ligori}, {Madrid}, {Martin}, {Morgante}, {Pamplona}, {Prieto}, {Riva}, {Toledo}, {Trifoglio}, {Zerbi}, {Abdalla}, {Douspis}, {Grenet}, {Borgani}, {Bouwens}, {Courbin}, {Delouis}, {Dubath}, {Fontana}, {Frailis}, {Grazian}, {Koppenh{\"o}fer}, {Mansutti}, {Melchior}, {Mignoli}, {Mohr}, {Neissner}, {Noddle}, {Poncet}, {Scodeggio}, {Serrano}, {Shane}, {Starck}, {Surace}, {Taylor}, {Verdoes-Kleijn}, {Vuerli}, {Williams}, {Zacchei}, {Altieri}, {Escudero Sanz}, {Kohley}, {Oosterbroek}, {Astier}, {Bacon}, {Bardelli}, {Baugh}, {Bellagamba}, {Benoist}, {Bianchi}, {Biviano}, {Branchini}, {Carbone}, {Cardone}, {Clements}, {Colombi}, {Conselice}, {Cresci}, {Deacon}, {Dunlop}, {Fedeli}, {Fontanot}, {Franzetti}, {Giocoli}, {Garcia-Bellido}, {Gow}, {Heavens}, {Hewett}, {Heymans}, {Holland}, {Huang}, {Ilbert}, {Joachimi}, {Jennins}, {Kerins}, {Kiessling}, {Kirk}, {Kotak}, {Krause}, {Lahav}, {van Leeuwen}, {Lesgourgues}, {Lombardi}, {Magliocchetti}, {Maguire},
  {Majerotto}, {Maoli}, {Marulli}, {Maurogordato}, {McCracken}, {McLure}, {Melchiorri}, {Merson}, {Moresco}, {Nonino}, {Norberg}, {Peacock}, {Pello}, {Penny}, {Pettorino}, {Di Porto}, {Pozzetti}, {Quercellini}, {Radovich}, {Rassat}, {Roche}, {Ronayette}, {Rossetti}, {Sartoris}, {Schneider}, {Semboloni}, {Serjeant}, {Simpson}, {Skordis}, {Smadja}, {Smartt}, {Spano}, {Spiro}, {Sullivan}, {Tilquin}, {Trotta}, {Verde}, {Wang}, {Williger}, {Zhao}, {Zoubian}, \& {Zucca}}]{2011arXiv1110.3193L}
{Laureijs}, R., {Amiaux}, J., {Arduini}, S., {et~al.} 2011, arXiv e-prints, arXiv:1110.3193

\bibitem[{{Le Floc'h} {et~al.}(2009){Le Floc'h}, {Aussel}, {Ilbert}, {Riguccini}, {Frayer}, {Salvato}, {Arnouts}, {Surace}, {Feruglio}, {Rodighiero}, {Capak}, {Kartaltepe}, {Heinis}, {Sheth}, {Yan}, {McCracken}, {Thompson}, {Sanders}, {Scoville}, \& {Koekemoer}}]{2009ApJ...703..222L}
{Le Floc'h}, E., {Aussel}, H., {Ilbert}, O., {et~al.} 2009, \apj, 703, 222

\bibitem[{{Lee} {et~al.}(2013){Lee}, {Sanders}, {Casey}, {Scoville}, {Hung}, {Le Floc'h}, {Ilbert}, {Aussel}, {Capak}, {Kartaltepe}, {Roseboom}, {Salvato}, {Aravena}, {Berta}, {Bock}, {Oliver}, {Riguccini}, \& {Symeonidis}}]{2013ApJ...778..131L}
{Lee}, N., {Sanders}, D.~B., {Casey}, C.~M., {et~al.} 2013, \apj, 778, 131

\bibitem[{{Liu} {et~al.}(2018){Liu}, {Daddi}, {Dickinson}, {Owen}, {Pannella}, {Sargent}, {B{\'e}thermin}, {Magdis}, {Gao}, {Shu}, {Wang}, {Jin}, \& {Inami}}]{2018ApJ...853..172L}
{Liu}, D., {Daddi}, E., {Dickinson}, M., {et~al.} 2018, \apj, 853, 172

\bibitem[{{Liu} {et~al.}(2019){Liu}, {Lang}, {Magnelli}, {Schinnerer}, {Leslie}, {Fudamoto}, {Bondi}, {Groves}, {Jim{\'e}nez-Andrade}, {Harrington}, {Karim}, {Oesch}, {Sargent}, {Vardoulaki}, {B{\v{a}}descu}, {Moser}, {Bertoldi}, {Battisti}, {da Cunha}, {Zavala}, {Vaccari}, {Davidzon}, {Riechers}, \& {Aravena}}]{2019ApJS..244...40L}
{Liu}, D., {Lang}, P., {Magnelli}, B., {et~al.} 2019, \apjs, 244, 40

\bibitem[{{Long} {et~al.}(2023){Long}, {Casey}, {del P. Lagos}, {Lambrides}, {Zavala}, {Champagne}, {Cooper}, \& {Cooray}}]{2023ApJ...953...11L}
{Long}, A.~S., {Casey}, C.~M., {del P. Lagos}, C., {et~al.} 2023, \apj, 953, 11

\bibitem[{{Lutz} {et~al.}(2011){Lutz}, {Poglitsch}, {Altieri}, {Andreani}, {Aussel}, {Berta}, {Bongiovanni}, {Brisbin}, {Cava}, {Cepa}, {Cimatti}, {Daddi}, {Dominguez-Sanchez}, {Elbaz}, {F{\"o}rster Schreiber}, {Genzel}, {Grazian}, {Gruppioni}, {Harwit}, {Le Floc'h}, {Magdis}, {Magnelli}, {Maiolino}, {Nordon}, {P{\'e}rez Garc{\'\i}a}, {Popesso}, {Pozzi}, {Riguccini}, {Rodighiero}, {Saintonge}, {Sanchez Portal}, {Santini}, {Shao}, {Sturm}, {Tacconi}, {Valtchanov}, {Wetzstein}, \& {Wieprecht}}]{2011A&A...532A..90L}
{Lutz}, D., {Poglitsch}, A., {Altieri}, B., {et~al.} 2011, \aap, 532, A90

\bibitem[{{Madau} \& {Dickinson}(2014)}]{2014ARA&A..52..415M}
{Madau}, P. \& {Dickinson}, M. 2014, \araa, 52, 415

\bibitem[{{Magnelli} {et~al.}(2010){Magnelli}, {Lutz}, {Berta}, {Altieri}, {Andreani}, {Aussel}, {Casta{\~n}eda}, {Cava}, {Cepa}, {Cimatti}, {Daddi}, {Dannerbauer}, {Dominguez}, {Elbaz}, {F{\"o}rster Schreiber}, {Genzel}, {Grazian}, {Gruppioni}, {Magdis}, {Maiolino}, {Nordon}, {P{\'e}rez Fournon}, {P{\'e}rez Garc{\'\i}a}, {Poglitsch}, {Popesso}, {Pozzi}, {Riguccini}, {Rodighiero}, {Saintonge}, {Santini}, {Sanchez-Portal}, {Shao}, {Sturm}, {Tacconi}, {Valtchanov}, {Wieprecht}, \& {Wiezorrek}}]{2010A&A...518L..28M}
{Magnelli}, B., {Lutz}, D., {Berta}, S., {et~al.} 2010, \aap, 518, L28

\bibitem[{{Magnelli} {et~al.}(2012){Magnelli}, {Lutz}, {Santini}, {Saintonge}, {Berta}, {Albrecht}, {Altieri}, {Andreani}, {Aussel}, {Bertoldi}, {B{\'e}thermin}, {Bongiovanni}, {Capak}, {Chapman}, {Cepa}, {Cimatti}, {Cooray}, {Daddi}, {Danielson}, {Dannerbauer}, {Dunlop}, {Elbaz}, {Farrah}, {F{\"o}rster Schreiber}, {Genzel}, {Hwang}, {Ibar}, {Ivison}, {Le Floc'h}, {Magdis}, {Maiolino}, {Nordon}, {Oliver}, {P{\'e}rez Garc{\'\i}a}, {Poglitsch}, {Popesso}, {Pozzi}, {Riguccini}, {Rodighiero}, {Rosario}, {Roseboom}, {Salvato}, {Sanchez-Portal}, {Scott}, {Smail}, {Sturm}, {Swinbank}, {Tacconi}, {Valtchanov}, {Wang}, \& {Wuyts}}]{2012A&A...539A.155M}
{Magnelli}, B., {Lutz}, D., {Santini}, P., {et~al.} 2012, \aap, 539, A155

\bibitem[{{Magnelli} {et~al.}(2013){Magnelli}, {Popesso}, {Berta}, {Pozzi}, {Elbaz}, {Lutz}, {Dickinson}, {Altieri}, {Andreani}, {Aussel}, {B{\'e}thermin}, {Bongiovanni}, {Cepa}, {Charmandaris}, {Chary}, {Cimatti}, {Daddi}, {F{\"o}rster Schreiber}, {Genzel}, {Gruppioni}, {Harwit}, {Hwang}, {Ivison}, {Magdis}, {Maiolino}, {Murphy}, {Nordon}, {Pannella}, {P{\'e}rez Garc{\'\i}a}, {Poglitsch}, {Rosario}, {Sanchez-Portal}, {Santini}, {Scott}, {Sturm}, {Tacconi}, \& {Valtchanov}}]{2013A&A...553A.132M}
{Magnelli}, B., {Popesso}, P., {Berta}, S., {et~al.} 2013, \aap, 553, A132

\bibitem[{{Martis} {et~al.}(2016){Martis}, {Marchesini}, {Brammer}, {Muzzin}, {Labb{\'e}}, {Momcheva}, {Skelton}, {Stefanon}, {van Dokkum}, \& {Whitaker}}]{2016ApJ...827L..25M}
{Martis}, N.~S., {Marchesini}, D., {Brammer}, G.~B., {et~al.} 2016, \apjl, 827, L25

\bibitem[{{McCracken} {et~al.}(2012){McCracken}, {Milvang-Jensen}, {Dunlop}, {Franx}, {Fynbo}, {Le F{\`e}vre}, {Holt}, {Caputi}, {Goranova}, {Buitrago}, {Emerson}, {Freudling}, {Hudelot}, {L{\'o}pez-Sanjuan}, {Magnard}, {Mellier}, {M{\o}ller}, {Nilsson}, {Sutherland}, {Tasca}, \& {Zabl}}]{2012A&A...544A.156M}
{McCracken}, H.~J., {Milvang-Jensen}, B., {Dunlop}, J., {et~al.} 2012, \aap, 544, A156

\bibitem[{{Miller} {et~al.}(2008){Miller}, {Fomalont}, {Kellermann}, {Mainieri}, {Norman}, {Padovani}, {Rosati}, \& {Tozzi}}]{2008ApJS..179..114M}
{Miller}, N.~A., {Fomalont}, E.~B., {Kellermann}, K.~I., {et~al.} 2008, \apjs, 179, 114

\bibitem[{{Moneti} {et~al.}(2023){Moneti}, {McCracken}, {Hudelot}, {Rouberol}, {Herent}, {Mellier}, {Dunlop}, {Le Fevre}, {Franx}, {Fynbo}, {Bowler}, {Caputi}, {Kauffmann}, {Milvang-Jensen}, {Gonzalez-Fernandez}, {Gonzalez-Solares}, {Irwin}, {Lewis}, {Blake}, {Cross}, {Read}, \& {Sutorius}}]{2023yCat.2373....0M}
{Moneti}, A., {McCracken}, H.~J., {Hudelot}, W., {et~al.} 2023, VizieR Online Data Catalog, II/373

\bibitem[{Nair \& Hinton(2010)}]{nair2010rectified}
Nair, V. \& Hinton, G.~E. 2010, in Proceedings of the 27th international conference on machine learning (ICML-10), 807--814

\bibitem[{{Nguyen} {et~al.}(2010){Nguyen}, {Schulz}, {Levenson}, {Amblard}, {Arumugam}, {Aussel}, {Babbedge}, {Blain}, {Bock}, {Boselli}, {Buat}, {Castro-Rodriguez}, {Cava}, {Chanial}, {Chapin}, {Clements}, {Conley}, {Conversi}, {Cooray}, {Dowell}, {Dwek}, {Eales}, {Elbaz}, {Fox}, {Franceschini}, {Gear}, {Glenn}, {Griffin}, {Halpern}, {Hatziminaoglou}, {Ibar}, {Isaak}, {Ivison}, {Lagache}, {Lu}, {Madden}, {Maffei}, {Mainetti}, {Marchetti}, {Marsden}, {Marshall}, {O'Halloran}, {Oliver}, {Omont}, {Page}, {Panuzzo}, {Papageorgiou}, {Pearson}, {Perez Fournon}, {Pohlen}, {Rangwala}, {Rigopoulou}, {Rizzo}, {Roseboom}, {Rowan-Robinson}, {Scott}, {Seymour}, {Shupe}, {Smith}, {Stevens}, {Symeonidis}, {Trichas}, {Tugwell}, {Vaccari}, {Valtchanov}, {Vigroux}, {Wang}, {Ward}, {Wiebe}, {Wright}, {Xu}, \& {Zemcov}}]{2010A&A...518L...5N}
{Nguyen}, H.~T., {Schulz}, B., {Levenson}, L., {et~al.} 2010, \aap, 518, L5

\bibitem[{{Noeske} {et~al.}(2007){Noeske}, {Weiner}, {Faber}, {Papovich}, {Koo}, {Somerville}, {Bundy}, {Conselice}, {Newman}, {Schiminovich}, {Le Floc'h}, {Coil}, {Rieke}, {Lotz}, {Primack}, {Barmby}, {Cooper}, {Davis}, {Ellis}, {Fazio}, {Guhathakurta}, {Huang}, {Kassin}, {Martin}, {Phillips}, {Rich}, {Small}, {Willmer}, \& {Wilson}}]{2007ApJ...660L..43N}
{Noeske}, K.~G., {Weiner}, B.~J., {Faber}, S.~M., {et~al.} 2007, \apjl, 660, L43

\bibitem[{{Noll} {et~al.}(2009){Noll}, {Burgarella}, {Giovannoli}, {Buat}, {Marcillac}, \& {Mu{\~n}oz-Mateos}}]{2009A&A...507.1793N}
{Noll}, S., {Burgarella}, D., {Giovannoli}, E., {et~al.} 2009, \aap, 507, 1793

\bibitem[{{Oliver} {et~al.}(2012){Oliver}, {Bock}, {Altieri}, {Amblard}, {Arumugam}, {Aussel}, {Babbedge}, {Beelen}, {B{\'e}thermin}, {Blain}, {Boselli}, {Bridge}, {Brisbin}, {Buat}, {Burgarella}, {Castro-Rodr{\'\i}guez}, {Cava}, {Chanial}, {Cirasuolo}, {Clements}, {Conley}, {Conversi}, {Cooray}, {Dowell}, {Dubois}, {Dwek}, {Dye}, {Eales}, {Elbaz}, {Farrah}, {Feltre}, {Ferrero}, {Fiolet}, {Fox}, {Franceschini}, {Gear}, {Giovannoli}, {Glenn}, {Gong}, {Gonz{\'a}lez Solares}, {Griffin}, {Halpern}, {Harwit}, {Hatziminaoglou}, {Heinis}, {Hurley}, {Hwang}, {Hyde}, {Ibar}, {Ilbert}, {Isaak}, {Ivison}, {Lagache}, {Le Floc'h}, {Levenson}, {Faro}, {Lu}, {Madden}, {Maffei}, {Magdis}, {Mainetti}, {Marchetti}, {Marsden}, {Marshall}, {Mortier}, {Nguyen}, {O'Halloran}, {Omont}, {Page}, {Panuzzo}, {Papageorgiou}, {Patel}, {Pearson}, {P{\'e}rez-Fournon}, {Pohlen}, {Rawlings}, {Raymond}, {Rigopoulou}, {Riguccini}, {Rizzo}, {Rodighiero}, {Roseboom}, {Rowan-Robinson}, {S{\'a}nchez Portal}, {Schulz}, {Scott}, {Seymour}, {Shupe},
  {Smith}, {Stevens}, {Symeonidis}, {Trichas}, {Tugwell}, {Vaccari}, {Valtchanov}, {Vieira}, {Viero}, {Vigroux}, {Wang}, {Ward}, {Wardlow}, {Wright}, {Xu}, \& {Zemcov}}]{2012MNRAS.424.1614O}
{Oliver}, S.~J., {Bock}, J., {Altieri}, B., {et~al.} 2012, \mnras, 424, 1614

\bibitem[{{Oliver} {et~al.}(2010){Oliver}, {Wang}, {Smith}, {Altieri}, {Amblard}, {Arumugam}, {Auld}, {Aussel}, {Babbedge}, {Blain}, {Bock}, {Boselli}, {Buat}, {Burgarella}, {Castro-Rodr{\'\i}guez}, {Cava}, {Chanial}, {Clements}, {Conley}, {Conversi}, {Cooray}, {Dowell}, {Dwek}, {Eales}, {Elbaz}, {Fox}, {Franceschini}, {Gear}, {Glenn}, {Griffin}, {Halpern}, {Hatziminaoglou}, {Ibar}, {Isaak}, {Ivison}, {Lagache}, {Levenson}, {Lu}, {Madden}, {Maffei}, {Mainetti}, {Marchetti}, {Mitchell-Wynne}, {Mortier}, {Nguyen}, {O'Halloran}, {Omont}, {Page}, {Panuzzo}, {Papageorgiou}, {Pearson}, {P{\'e}rez-Fournon}, {Pohlen}, {Rawlings}, {Raymond}, {Rigopoulou}, {Rizzo}, {Roseboom}, {Rowan-Robinson}, {S{\'a}nchez Portal}, {Savage}, {Schulz}, {Scott}, {Seymour}, {Shupe}, {Stevens}, {Symeonidis}, {Trichas}, {Tugwell}, {Vaccari}, {Valiante}, {Valtchanov}, {Vieira}, {Vigroux}, {Ward}, {Wright}, {Xu}, \& {Zemcov}}]{2010A&A...518L..21O}
{Oliver}, S.~J., {Wang}, L., {Smith}, A.~J., {et~al.} 2010, \aap, 518, L21

\bibitem[{{Pannella} {et~al.}(2009){Pannella}, {Carilli}, {Daddi}, {McCracken}, {Owen}, {Renzini}, {Strazzullo}, {Civano}, {Koekemoer}, {Schinnerer}, {Scoville}, {Smol{\v{c}}i{\'c}}, {Taniguchi}, {Aussel}, {Kneib}, {Ilbert}, {Mellier}, {Salvato}, {Thompson}, \& {Willott}}]{2009ApJ...698L.116P}
{Pannella}, M., {Carilli}, C.~L., {Daddi}, E., {et~al.} 2009, \apjl, 698, L116

\bibitem[{{Pearson} {et~al.}(2018){Pearson}, {Wang}, {Hurley}, {Ma{\l}ek}, {Buat}, {Burgarella}, {Farrah}, {Oliver}, {Smith}, \& {van der Tak}}]{2018A&A...615A.146P}
{Pearson}, W.~J., {Wang}, L., {Hurley}, P.~D., {et~al.} 2018, \aap, 615, A146

\bibitem[{{Pearson} {et~al.}(2017){Pearson}, {Wang}, {van der Tak}, {Hurley}, {Burgarella}, \& {Oliver}}]{2017A&A...603A.102P}
{Pearson}, W.~J., {Wang}, L., {van der Tak}, F.~F.~S., {et~al.} 2017, \aap, 603, A102

\bibitem[{{Popesso} {et~al.}(2023){Popesso}, {Concas}, {Cresci}, {Belli}, {Rodighiero}, {Inami}, {Dickinson}, {Ilbert}, {Pannella}, \& {Elbaz}}]{2023MNRAS.519.1526P}
{Popesso}, P., {Concas}, A., {Cresci}, G., {et~al.} 2023, \mnras, 519, 1526

\bibitem[{{Popesso} {et~al.}(2019){Popesso}, {Morselli}, {Concas}, {Schreiber}, {Rodighiero}, {Cresci}, {Belli}, {Ilbert}, {Erfanianfar}, {Mancini}, {Inami}, {Dickinson}, {Pannella}, \& {Elbaz}}]{2019MNRAS.490.5285P}
{Popesso}, P., {Morselli}, L., {Concas}, A., {et~al.} 2019, \mnras, 490, 5285

\bibitem[{{Puget} {et~al.}(1996){Puget}, {Abergel}, {Bernard}, {Boulanger}, {Burton}, {Desert}, \& {Hartmann}}]{1996A&A...308L...5P}
{Puget}, J.~L., {Abergel}, A., {Bernard}, J.~P., {et~al.} 1996, \aap, 308, L5

\bibitem[{{Rieke} {et~al.}(2009){Rieke}, {Alonso-Herrero}, {Weiner}, {P{\'e}rez-Gonz{\'a}lez}, {Blaylock}, {Donley}, \& {Marcillac}}]{2009ApJ...692..556R}
{Rieke}, G.~H., {Alonso-Herrero}, A., {Weiner}, B.~J., {et~al.} 2009, \apj, 692, 556

\bibitem[{{Rodighiero} {et~al.}(2006){Rodighiero}, {Lari}, {Pozzi}, {Gruppioni}, {Fadda}, {Franceschini}, {Lonsdale}, {Surace}, {Shupe}, \& {Fang}}]{2006MNRAS.371.1891R}
{Rodighiero}, G., {Lari}, C., {Pozzi}, F., {et~al.} 2006, \mnras, 371, 1891

\bibitem[{{Roseboom} {et~al.}(2012){Roseboom}, {Ivison}, {Greve}, {Amblard}, {Arumugam}, {Auld}, {Aussel}, {Bethermin}, {Blain}, {Bock}, {Boselli}, {Brisbin}, {Buat}, {Burgarella}, {Castro-Rodr{\'\i}guez}, {Cava}, {Chanial}, {Chapin}, {Chapman}, {Clements}, {Conley}, {Conversi}, {Cooray}, {Dowell}, {Dunlop}, {Dwek}, {Eales}, {Elbaz}, {Farrah}, {Franceschini}, {Glenn}, {Griffin}, {Halpern}, {Hatziminaoglou}, {Ibar}, {Isaak}, {Lagache}, {Levenson}, {Lu}, {Madden}, {Maffei}, {Mainetti}, {Marchetti}, {Marsden}, {Morrison}, {Mortier}, {Nguyen}, {O'Halloran}, {Oliver}, {Omont}, {Page}, {Panuzzo}, {Papageorgiou}, {Pearson}, {P{\'e}rez-Fournon}, {Pohlen}, {Rawlings}, {Raymond}, {Rigopoulou}, {Rizzo}, {Rodighiero}, {Rowan-Robinson}, {Schulz}, {Scott}, {Seymour}, {Shupe}, {Smith}, {Stevens}, {Symeonidis}, {Trichas}, {Tugwell}, {Vaccari}, {Valtchanov}, {Vieira}, {Viero}, {Vigroux}, {Wardlow}, {Wang}, {Wright}, {Xu}, \& {Zemcov}}]{2012MNRAS.419.2758R}
{Roseboom}, I.~G., {Ivison}, R.~J., {Greve}, T.~R., {et~al.} 2012, \mnras, 419, 2758

\bibitem[{{Roseboom} {et~al.}(2010){Roseboom}, {Oliver}, {Kunz}, {Altieri}, {Amblard}, {Arumugam}, {Auld}, {Aussel}, {Babbedge}, {B{\'e}thermin}, {Blain}, {Bock}, {Boselli}, {Brisbin}, {Buat}, {Burgarella}, {Castro-Rodr{\'\i}guez}, {Cava}, {Chanial}, {Chapin}, {Clements}, {Conley}, {Conversi}, {Cooray}, {Dowell}, {Dwek}, {Dye}, {Eales}, {Elbaz}, {Farrah}, {Fox}, {Franceschini}, {Gear}, {Glenn}, {Solares}, {Griffin}, {Halpern}, {Harwit}, {Hatziminaoglou}, {Huang}, {Ibar}, {Isaak}, {Ivison}, {Lagache}, {Levenson}, {Lu}, {Madden}, {Maffei}, {Mainetti}, {Marchetti}, {Marsden}, {Mortier}, {Nguyen}, {O'Halloran}, {Omont}, {Page}, {Panuzzo}, {Papageorgiou}, {Patel}, {Pearson}, {P{\'e}rez-Fournon}, {Pohlen}, {Rawlings}, {Raymond}, {Rigopoulou}, {Rizzo}, {Rowan-Robinson}, {Portal}, {Schulz}, {Scott}, {Seymour}, {Shupe}, {Smith}, {Stevens}, {Symeonidis}, {Trichas}, {Tugwell}, {Vaccari}, {Valtchanov}, {Vieira}, {Vigroux}, {Wang}, {Ward}, {Wright}, {Xu}, \& {Zemcov}}]{2010MNRAS.409...48R}
{Roseboom}, I.~G., {Oliver}, S.~J., {Kunz}, M., {et~al.} 2010, \mnras, 409, 48

\bibitem[{{Sanders} {et~al.}(2007){Sanders}, {Salvato}, {Aussel}, {Ilbert}, {Scoville}, {Surace}, {Frayer}, {Sheth}, {Helou}, {Brooke}, {Bhattacharya}, {Yan}, {Kartaltepe}, {Barnes}, {Blain}, {Calzetti}, {Capak}, {Carilli}, {Carollo}, {Comastri}, {Daddi}, {Ellis}, {Elvis}, {Fall}, {Franceschini}, {Giavalisco}, {Hasinger}, {Impey}, {Koekemoer}, {Le F{\`e}vre}, {Lilly}, {Liu}, {McCracken}, {Mobasher}, {Renzini}, {Rich}, {Schinnerer}, {Shopbell}, {Taniguchi}, {Thompson}, {Urry}, \& {Williams}}]{2007ApJS..172...86S}
{Sanders}, D.~B., {Salvato}, M., {Aussel}, H., {et~al.} 2007, \apjs, 172, 86

\bibitem[{{Sawicki} {et~al.}(2019){Sawicki}, {Arnouts}, {Huang}, {Coupon}, {Golob}, {Gwyn}, {Foucaud}, {Moutard}, {Iwata}, {Liu}, {Chen}, {Desprez}, {Harikane}, {Ono}, {Strauss}, {Tanaka}, {Thibert}, {Balogh}, {Bundy}, {Chapman}, {Gunn}, {Hsieh}, {Ilbert}, {Jing}, {LeF{\`e}vre}, {Li}, {Matsuda}, {Miyazaki}, {Nagao}, {Nishizawa}, {Ouchi}, {Shimasaku}, {Silverman}, {de la Torre}, {Tresse}, {Wang}, {Willott}, {Yamada}, {Yang}, \& {Yee}}]{2019MNRAS.489.5202S}
{Sawicki}, M., {Arnouts}, S., {Huang}, J., {et~al.} 2019, \mnras, 489, 5202

\bibitem[{{Schinnerer} {et~al.}(2010){Schinnerer}, {Sargent}, {Bondi}, {Smol{\v{c}}i{\'c}}, {Datta}, {Carilli}, {Bertoldi}, {Blain}, {Ciliegi}, {Koekemoer}, \& {Scoville}}]{2010ApJS..188..384S}
{Schinnerer}, E., {Sargent}, M.~T., {Bondi}, M., {et~al.} 2010, \apjs, 188, 384

\bibitem[{{Scoville} {et~al.}(2007){Scoville}, {Aussel}, {Brusa}, {Capak}, {Carollo}, {Elvis}, {Giavalisco}, {Guzzo}, {Hasinger}, {Impey}, {Kneib}, {LeFevre}, {Lilly}, {Mobasher}, {Renzini}, {Rich}, {Sanders}, {Schinnerer}, {Schminovich}, {Shopbell}, {Taniguchi}, \& {Tyson}}]{2007ApJS..172....1S}
{Scoville}, N., {Aussel}, H., {Brusa}, M., {et~al.} 2007, \apjs, 172, 1

\bibitem[{{Scudder} {et~al.}(2016){Scudder}, {Oliver}, {Hurley}, {Griffin}, {Sargent}, {Scott}, {Wang}, \& {Wardlow}}]{2016MNRAS.460.1119S}
{Scudder}, J.~M., {Oliver}, S., {Hurley}, P.~D., {et~al.} 2016, \mnras, 460, 1119

\bibitem[{{Serra} {et~al.}(2011){Serra}, {Amblard}, {Temi}, {Burgarella}, {Giovannoli}, {Buat}, {Noll}, \& {Im}}]{2011ApJ...740...22S}
{Serra}, P., {Amblard}, A., {Temi}, P., {et~al.} 2011, \apj, 740, 22

\bibitem[{{Shirley} {et~al.}(2021){Shirley}, {Duncan}, {Campos Varillas}, {Hurley}, {Ma{\l}ek}, {Roehlly}, {Smith}, {Aussel}, {Bakx}, {Buat}, {Burgarella}, {Christopher}, {Duivenvoorden}, {Eales}, {Efstathiou}, {Gonz{\'a}lez Solares}, {Griffin}, {Jarvis}, {Faro}, {Marchetti}, {McCheyne}, {Papadopoulos}, {Penner}, {Pons}, {Prescott}, {Rigby}, {Rottgering}, {Saxena}, {Scudder}, {Vaccari}, {Wang}, \& {Oliver}}]{2021MNRAS.507..129S}
{Shirley}, R., {Duncan}, K., {Campos Varillas}, M.~C., {et~al.} 2021, \mnras, 507, 129

\bibitem[{{Simpson} {et~al.}(2020){Simpson}, {Smail}, {Dudzevi{\v{c}}i{\={u}}t{\.{e}}}, {Matsuda}, {Hsieh}, {Wang}, {Swinbank}, {Stach}, {An}, {Birkin}, {Ao}, {Bunker}, {Chapman}, {Chen}, {Coppin}, {Ikarashi}, {Ivison}, {Mitsuhashi}, {Saito}, {Umehata}, {Wang}, \& {Zhao}}]{2020MNRAS.495.3409S}
{Simpson}, J.~M., {Smail}, I., {Dudzevi{\v{c}}i{\={u}}t{\.{e}}}, U., {et~al.} 2020, \mnras, 495, 3409

\bibitem[{{Simpson} {et~al.}(2019){Simpson}, {Smail}, {Swinbank}, {Chapman}, {Chen}, {Geach}, {Matsuda}, {Wang}, {Wang}, {Yang}, {Ao}, {Asquith}, {Bourne}, {Coogan}, {Coppin}, {Gullberg}, {Hine}, {Ho}, {Hwang}, {Ivison}, {Kato}, {Lacaille}, {Lewis}, {Liu}, {Micha{\l}owski}, {Oteo}, {Sawicki}, {Scholtz}, {Smith}, {Thomson}, \& {Wardlow}}]{2019ApJ...880...43S}
{Simpson}, J.~M., {Smail}, I., {Swinbank}, A.~M., {et~al.} 2019, \apj, 880, 43

\bibitem[{{Smith} {et~al.}(2012){Smith}, {Wang}, {Oliver}, {Auld}, {Bock}, {Brisbin}, {Burgarella}, {Chanial}, {Chapin}, {Clements}, {Conversi}, {Cooray}, {Dowell}, {Eales}, {Farrah}, {Franceschini}, {Glenn}, {Griffin}, {Ivison}, {Mortier}, {Page}, {Papageorgiou}, {Pearson}, {P{\'e}rez-Fournon}, {Pohlen}, {Rawlings}, {Raymond}, {Rodighiero}, {Roseboom}, {Rowan-Robinson}, {Savage}, {Scott}, {Seymour}, {Symeonidis}, {Tugwell}, {Vaccari}, {Valtchanov}, {Vigroux}, {Ward}, {Wright}, \& {Zemcov}}]{2012MNRAS.419..377S}
{Smith}, A.~J., {Wang}, L., {Oliver}, S.~J., {et~al.} 2012, \mnras, 419, 377

\bibitem[{{Smith} {et~al.}(2014){Smith}, {Jarvis}, {Hardcastle}, {Vaccari}, {Bourne}, {Dunne}, {Ibar}, {Maddox}, {Prescott}, {Vlahakis}, {Eales}, {Maddox}, {Smith}, {Valiante}, \& {de Zotti}}]{2014MNRAS.445.2232S}
{Smith}, D.~J.~B., {Jarvis}, M.~J., {Hardcastle}, M.~J., {et~al.} 2014, \mnras, 445, 2232

\bibitem[{{Smol{\v{c}}i{\'c}} {et~al.}(2017){Smol{\v{c}}i{\'c}}, {Novak}, {Bondi}, {Ciliegi}, {Mooley}, {Schinnerer}, {Zamorani}, {Navarrete}, {Bourke}, {Karim}, {Vardoulaki}, {Leslie}, {Delhaize}, {Carilli}, {Myers}, {Baran}, {Delvecchio}, {Miettinen}, {Banfield}, {Balokovi{\'c}}, {Bertoldi}, {Capak}, {Frail}, {Hallinan}, {Hao}, {Herrera Ruiz}, {Horesh}, {Ilbert}, {Intema}, {Jeli{\'c}}, {Kl{\"o}ckner}, {Krpan}, {Kulkarni}, {McCracken}, {Laigle}, {Middleberg}, {Murphy}, {Sargent}, {Scoville}, \& {Sheth}}]{2017A&A...602A...1S}
{Smol{\v{c}}i{\'c}}, V., {Novak}, M., {Bondi}, M., {et~al.} 2017, \aap, 602, A1

\bibitem[{{Speagle} {et~al.}(2014){Speagle}, {Steinhardt}, {Capak}, \& {Silverman}}]{2014ApJS..214...15S}
{Speagle}, J.~S., {Steinhardt}, C.~L., {Capak}, P.~L., \& {Silverman}, J.~D. 2014, \apjs, 214, 15

\bibitem[{Srivastava {et~al.}(2014)Srivastava, Hinton, Krizhevsky, Sutskever, \& Salakhutdinov}]{srivastava2014dropout}
Srivastava, N., Hinton, G., Krizhevsky, A., Sutskever, I., \& Salakhutdinov, R. 2014, The journal of machine learning research, 15, 1929

\bibitem[{{Stalevski} {et~al.}(2012){Stalevski}, {Fritz}, {Baes}, {Nakos}, \& {Popovi{\'c}}}]{2012MNRAS.420.2756S}
{Stalevski}, M., {Fritz}, J., {Baes}, M., {Nakos}, T., \& {Popovi{\'c}}, L.~{\v{C}}. 2012, \mnras, 420, 2756

\bibitem[{Team(2015)}]{stan2015stan}
Team, S.~D. 2015, Stan: A C++ library for probability and sampling, version 2.8. 0

\bibitem[{Team(2018)}]{stan2018pystan}
Team, S.~D. 2018, Version 2.19. 1.1

\bibitem[{Tibshirani(1996)}]{tibshirani1996regression}
Tibshirani, R. 1996, Journal of the Royal Statistical Society Series B: Statistical Methodology, 58, 267

\bibitem[{{Wang} {et~al.}(2021){Wang}, {Gao}, {Best}, {Duncan}, {Hardcastle}, {Kondapally}, {Ma{\l}ek}, {McCheyne}, {Sabater}, {Shimwell}, {Tasse}, {Bonato}, {Bondi}, {Cochrane}, {Farrah}, {G{\"u}rkan}, {Haskell}, {Pearson}, {Prandoni}, {R{\"o}ttgering}, {Smith}, {Vaccari}, \& {Williams}}]{2021A&A...648A...8W}
{Wang}, L., {Gao}, F., {Best}, P.~N., {et~al.} 2021, \aap, 648, A8

\bibitem[{{Wang} {et~al.}(2016){Wang}, {Norberg}, {Gunawardhana}, {Heinis}, {Baldry}, {Bland-Hawthorn}, {Bourne}, {Brough}, {Brown}, {Cluver}, {Cooray}, {da Cunha}, {Driver}, {Dunne}, {Dye}, {Eales}, {Grootes}, {Holwerda}, {Hopkins}, {Ibar}, {Ivison}, {Lacey}, {Lara-Lopez}, {Loveday}, {Maddox}, {Micha{\l}owski}, {Oteo}, {Owers}, {Popescu}, {Smith}, {Taylor}, {Tuffs}, \& {van der Werf}}]{2016MNRAS.461.1898W}
{Wang}, L., {Norberg}, P., {Gunawardhana}, M.~L.~P., {et~al.} 2016, \mnras, 461, 1898

\bibitem[{{Wang} {et~al.}(2019){Wang}, {Pearson}, {Cowley}, {Trayford}, {B{\'e}thermin}, {Gruppioni}, {Hurley}, \& {Micha{\l}owski}}]{2019A&A...624A..98W}
{Wang}, L., {Pearson}, W.~J., {Cowley}, W., {et~al.} 2019, \aap, 624, A98

\bibitem[{{Wang} {et~al.}(2014){Wang}, {Viero}, {Clarke}, {Bock}, {Buat}, {Conley}, {Farrah}, {Guo}, {Heinis}, {Magdis}, {Marchetti}, {Marsden}, {Norberg}, {Oliver}, {Page}, {Roehlly}, {Roseboom}, {Schulz}, {Smith}, {Vaccari}, \& {Zemcov}}]{2014MNRAS.444.2870W}
{Wang}, L., {Viero}, M., {Clarke}, C., {et~al.} 2014, \mnras, 444, 2870

\bibitem[{{Weaver} {et~al.}(2023){Weaver}, {Davidzon}, {Toft}, {Ilbert}, {McCracken}, {Gould}, {Jespersen}, {Steinhardt}, {Lagos}, {Capak}, {Casey}, {Chartab}, {Faisst}, {Hayward}, {Kartaltepe}, {Kauffmann}, {Koekemoer}, {Kokorev}, {Laigle}, {Liu}, {Long}, {Magdis}, {McPartland}, {Milvang-Jensen}, {Mobasher}, {Moneti}, {Peng}, {Sanders}, {Shuntov}, {Sneppen}, {Valentino}, {Zalesky}, \& {Zamorani}}]{2023A&A...677A.184W}
{Weaver}, J.~R., {Davidzon}, I., {Toft}, S., {et~al.} 2023, \aap, 677, A184

\bibitem[{{Weaver} {et~al.}(2022){Weaver}, {Kauffmann}, {Ilbert}, {McCracken}, {Moneti}, {Toft}, {Brammer}, {Shuntov}, {Davidzon}, {Hsieh}, {Laigle}, {Anastasiou}, {Jespersen}, {Vinther}, {Capak}, {Casey}, {McPartland}, {Milvang-Jensen}, {Mobasher}, {Sanders}, {Zalesky}, {Arnouts}, {Aussel}, {Dunlop}, {Faisst}, {Franx}, {Furtak}, {Fynbo}, {Gould}, {Greve}, {Gwyn}, {Kartaltepe}, {Kashino}, {Koekemoer}, {Kokorev}, {Le F{\`e}vre}, {Lilly}, {Masters}, {Magdis}, {Mehta}, {Peng}, {Riechers}, {Salvato}, {Sawicki}, {Scarlata}, {Scoville}, {Shirley}, {Silverman}, {Sneppen}, {Smolc̆i{\'c}}, {Steinhardt}, {Stern}, {Tanaka}, {Taniguchi}, {Teplitz}, {Vaccari}, {Wang}, \& {Zamorani}}]{2022ApJS..258...11W}
{Weaver}, J.~R., {Kauffmann}, O.~B., {Ilbert}, O., {et~al.} 2022, \apjs, 258, 11

\bibitem[{{Whitaker} {et~al.}(2011){Whitaker}, {Labb{\'e}}, {van Dokkum}, {Brammer}, {Kriek}, {Marchesini}, {Quadri}, {Franx}, {Muzzin}, {Williams}, {Bezanson}, {Illingworth}, {Lee}, {Lundgren}, {Nelson}, {Rudnick}, {Tal}, \& {Wake}}]{2011ApJ...735...86W}
{Whitaker}, K.~E., {Labb{\'e}}, I., {van Dokkum}, P.~G., {et~al.} 2011, \apj, 735, 86

\bibitem[{{Whitaker} {et~al.}(2017){Whitaker}, {Pope}, {Cybulski}, {Casey}, {Popping}, \& {Yun}}]{2017ApJ...850..208W}
{Whitaker}, K.~E., {Pope}, A., {Cybulski}, R., {et~al.} 2017, \apj, 850, 208

\bibitem[{{Whitaker} {et~al.}(2012){Whitaker}, {van Dokkum}, {Brammer}, \& {Franx}}]{2012ApJ...754L..29W}
{Whitaker}, K.~E., {van Dokkum}, P.~G., {Brammer}, G., \& {Franx}, M. 2012, \apjl, 754, L29

\bibitem[{{Wuyts} {et~al.}(2011){Wuyts}, {F{\"o}rster Schreiber}, {van der Wel}, {Magnelli}, {Guo}, {Genzel}, {Lutz}, {Aussel}, {Barro}, {Berta}, {Cava}, {Graci{\'a}-Carpio}, {Hathi}, {Huang}, {Kocevski}, {Koekemoer}, {Lee}, {Le Floc'h}, {McGrath}, {Nordon}, {Popesso}, {Pozzi}, {Riguccini}, {Rodighiero}, {Saintonge}, \& {Tacconi}}]{2011ApJ...742...96W}
{Wuyts}, S., {F{\"o}rster Schreiber}, N.~M., {van der Wel}, A., {et~al.} 2011, \apj, 742, 96

\bibitem[{{Zavala} {et~al.}(2021){Zavala}, {Casey}, {Manning}, {Aravena}, {Bethermin}, {Caputi}, {Clements}, {Cunha}, {Drew}, {Finkelstein}, {Fujimoto}, {Hayward}, {Hodge}, {Kartaltepe}, {Knudsen}, {Koekemoer}, {Long}, {Magdis}, {Man}, {Popping}, {Sanders}, {Scoville}, {Sheth}, {Staguhn}, {Toft}, {Treister}, {Vieira}, \& {Yun}}]{2021ApJ...909..165Z}
{Zavala}, J.~A., {Casey}, C.~M., {Manning}, S.~M., {et~al.} 2021, \apj, 909, 165

\bibitem[{Zou(2006)}]{zou2006adaptive}
Zou, H. 2006, Journal of the American statistical association, 101, 1418

\end{thebibliography}

\begin{appendix} 

\section{CIGALE parameters}\label{appendix_cigale}

Table \ref{tab:parameter} shows the choices of the SED model components and parameters in the 16 CIGALE runs, with all possible combinations of two star-formation history models, two dust attenuation models, two dust emission models, and two AGN models.

\begin{table*}
\caption{Choices of the SED model components and parameters in the 16 CIGALE runs.}
    \centering
    \begin{tabular}{lll}
    \hline
    \multicolumn{3}{c}{\textbf{star-formation history}}\\
    \hline
    delayed $\tau$+starburst&e-folding time of the main population&500,1000,3000,5000,8000 Myrs\\
    &age the main population&500, 1000-13000 (step 1000) Myrs\\
    &e-folding time of the late starburst population&9000,13000 Myrs\\
    &age of the late starburst population&1,30,100,500 Myrs\\
   &mass fraction of the late starburst population&0.01,0.05,0.1,0.2\\
    \hline
    double exponential&e-folding time of the main population&1000-9000 (step 2000) Myrs\\
    &age the main population&500, 1000-13000 (step 1000) Myrs\\
    &e-folding time of the late starburst population&10000 Myrs\\
    &age of the late starburst population&1,30,100,500 Myrs\\
   &mass fraction of the late starburst population&0.01,0.05,0.1,0.2\\
    \hline
     \multicolumn{3}{c}{\textbf{dust attenuation}}\\
    \hline
    modified Charlot and Fall 2000&V-band attenuation in the interstellar medium&0.1,0.5-3.5 (step 0.5)\\
    &Av$_{ISM} $/ Av$_{BC}$+Av$_{ISM} $&0.25,0.5,0.75\\
&power-law slope of the attenuation in the birth clouds&-0.7\\
&power-law slope of the attenuation in the ISM&-0.7\\
\hline
modified Calzetti 2000&color excess&0.01,0.05,0.1,0.2,0.3,0.4,0.5\\
&reduction factor&0.25,0.5,0.75\\
&Slope of the power law modifying the attenuation curve& -0.5,-0.4,-0.3,-0.2,-0.1,0\\
&Extinction law&Milky Way\\
\hline
\multicolumn{3}{c}{\textbf{dust emission}}\\
\hline
Draine et al. 2014&Mass fraction of PAH&0.47\\
&minimum radiation field&5,15,25\\
&power-law slope $\alpha$ in dM/dU  $\propto U^\alpha$ &2\\
&Fraction illuminated from min to max radiation field&0.02\\
\hline
Dale et al. 2014&alpha slope&1.5-4 (step 0.5)\\
\hline
\multicolumn{3}{c}{\textbf{AGN templates}}\\
\hline
Fritz et al. 2006 & Full opening angle of the dust torus (degree) & 100\\
&viewing angle (degree)&0.001,50.1,89.990 \\
&AGN fraction&0,0.1,0.3,0.5,0.7\\
&optical depth at 9.7 $\mu$m&0.3,1,6\\
&linked to the radial dust distribution in the torus&-0.5\\
&linked to the angular dust distribution in the torus&0\\
\hline
SKIRTOR& optical depth at 9.7 $\mu$m&7\\
&torus density radial parameter $p$&1\\
&torus density angular parameter $q$ &1\\
&half opening angle (degree)&40 \\
&viewing angle (degree) &30,70  \\
&AGN fraction&0,0.1,0.3,0.5,0.7\\
&extinction law of polar dust&SMC\\
&E(B-V) of polar dust&0,0.1,0.2,0.3,0.4\\
&temperature of polar dust&100\\
&emissivity of polar dust&1.6\\
\hline
    \end{tabular}   
    \label{tab:parameter}
\tablefoot{The 16 CIGALE runs involve all possible combinations of two star-formation history models (delayed $\tau$ + starburst or double exponential), two dust attenuation models (modified \citet{2000ApJ...539..718C} or modified \citet{2000ApJ...533..682C} attenuation law), two dust emission models (IR SED templates from the \citet{2014ApJ...780..172D} model or the \citet{2014ApJ...784...83D} model), and two AGN models (smooth AGN templates from \citet{2006MNRAS.366..767F} or clumpy AGN templates from the SKIRTOR model in \citet{2012MNRAS.420.2756S}).
    }
\end{table*}

\section{Scaling the flux uncertainty}\label{appendix_scaling}

To make the computational time manageable, XID+ ignores the fact that residual confusion noise is correlated between pixels. Instead, it is assumed to be constant in each tile and is modelled as Gaussian fluctuations around a global background, ${\rm N(B, \Sigma^{residual}_{confusion})}$, as described in Eq. \ref{eq_xid}. If we add this residual confusion noise ${\rm \Sigma^{residual}_{confusion}}$ in quadrature to the $1\sigma$ flux uncertainty estimate to form a ``total" noise, then the ``total" flux uncertainties behave as expected in the sense that the distribution of the flux differences normalised by the "total" flux uncertainties follows the standard Gaussian. 

\begin{figure}
    \centering
\includegraphics[width=.49\textwidth]{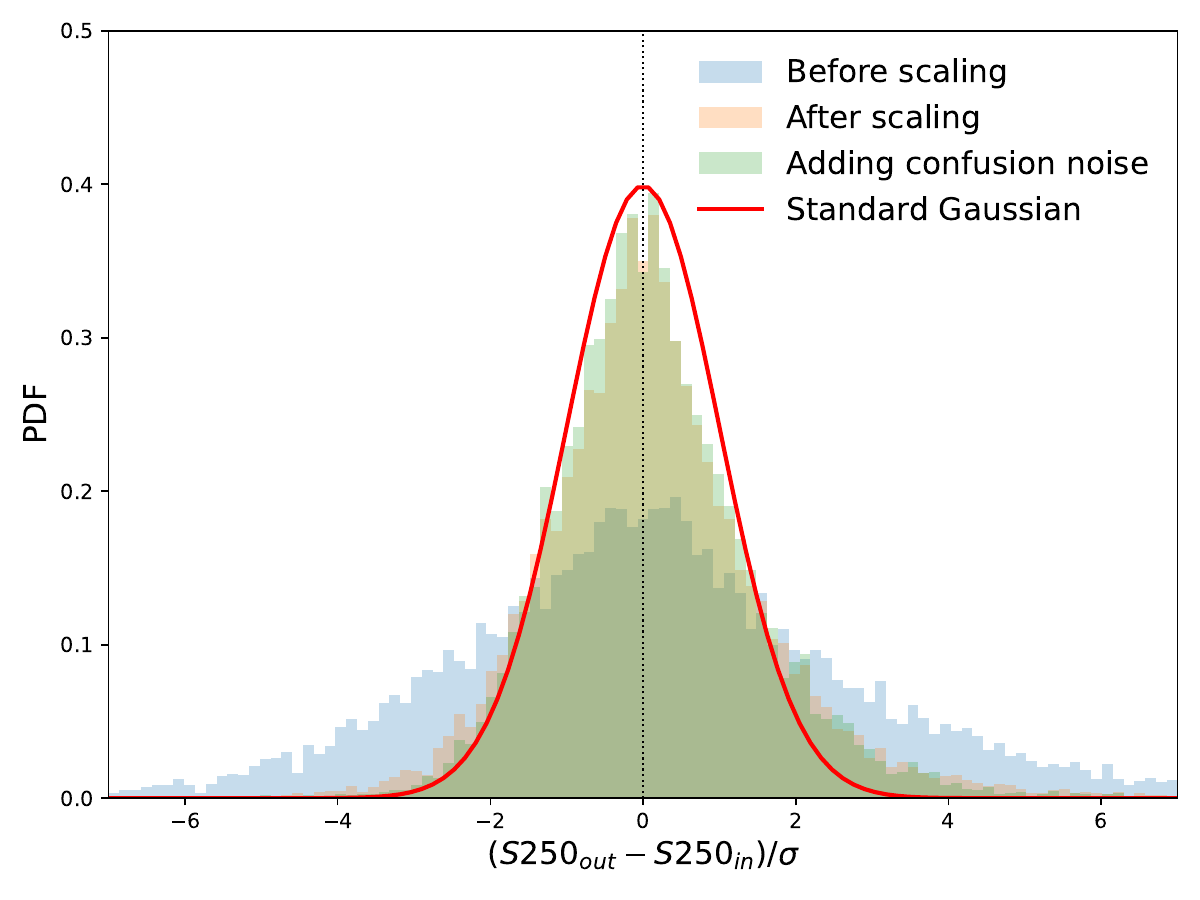}
    \caption{Distribution of flux difference normalised by the $1\sigma$ flux uncertainty estimate, before and after scaling $1\sigma$ by a factor of two.  We also show the distribution after adding in the residual confusion noise in quadrature to the $1\sigma$ flux uncertainty.}
    \label{fig:spire_norm_v2}
\end{figure}

\begin{figure}[ht]
    \centering
    \includegraphics[width=9cm]{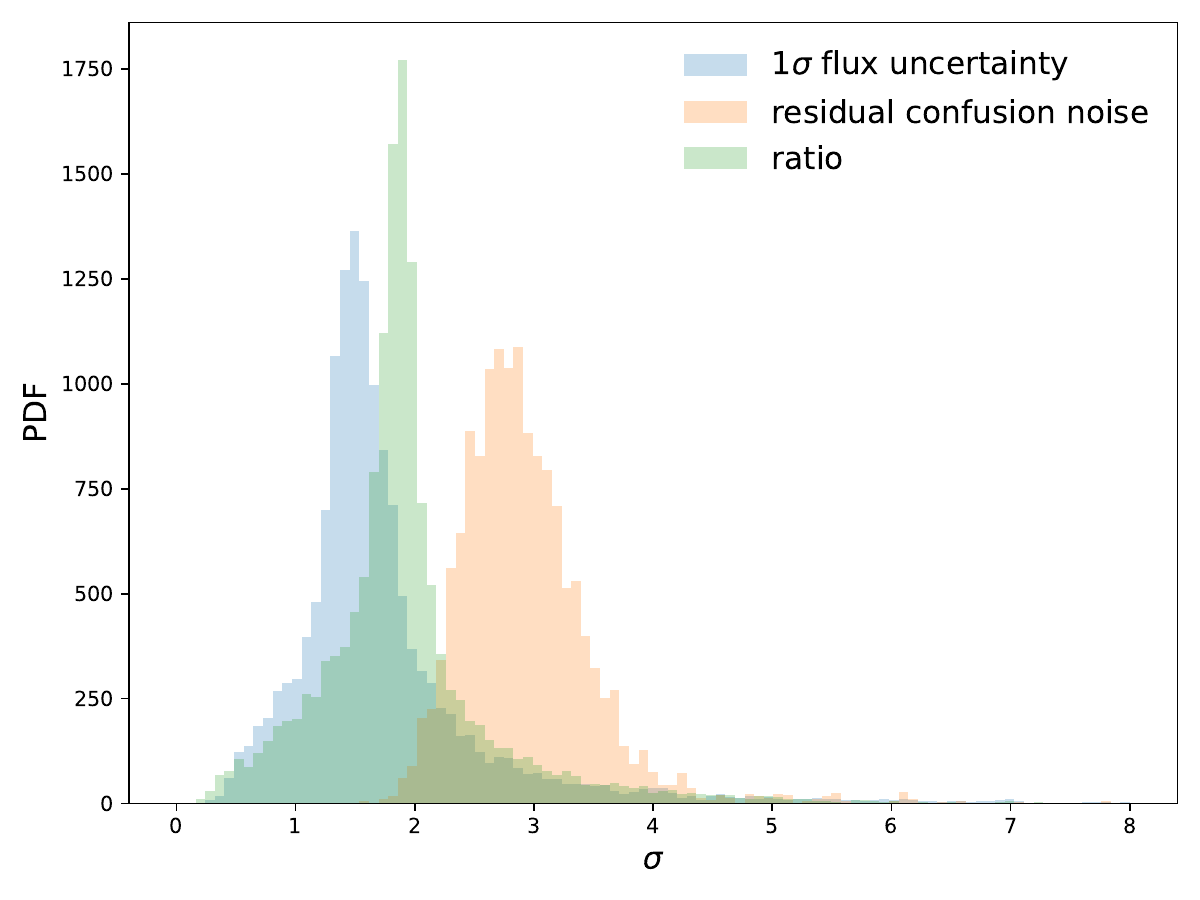}
    \caption{Distribution of the $1\sigma$ flux uncertainty at 250 $\mu$m (derived from the maximum of $\sigma^+$ and $\sigma^-$) is shown by the blue histogram. The brown histogram shows that distribution of the residual confusion noise. The green histogram shows the ratio of the two.}
    \label{fig:sigma}
\end{figure}

In the MIPS and SPIRE bands, scaling the $1\sigma$ flux uncertainty by a factor of two also makes the distribution of the normalised flux difference to follow the standard Gaussian. In other words, the effect of scaling by a factor of two is roughly the same as combining the $1\sigma$ flux uncertainty with the residual confusion noise in quadrature, as shown in Fig. \ref{fig:spire_norm_v2}. To understand why they coincide, consider using XID+ to fit a map with $n_1 \times n_2 = M$ pixels. XID+ assumes that the map $\rm{\bf d}$ (a flattened vector of $M$ pixels) is formed from the contribution of the known sources and two independent noise terms (instrument noise and residual confusion noise). The contribution of the known sources $\Sigma_{i=1}^{S} \rm{\bf{P}} \rm{f_i}$, where $\rm{\bf{P}}$ is the point response function (PRF) and $\rm{f_i}$ is the flux of a given source, can be written in a linear form, 
\begin{equation}
\rm{\bf d} = \rm{\bf A f}. 
\end{equation}
Here $\rm{\bf f}$ is a vector of $S$ elements (corresponding to the number of known sources) and $\rm{\bf A}$ is a $M \times S$ pointing matrix. Therefore, the maximum-likelihood solution can be found as
\begin{equation}
\rm{\bf f} = (\rm{\bf A}^T \rm{\bf N_d}^{-1} \rm{\bf A})^{-1} \rm{\bf A}^{T} \rm{\bf N_d}^{-1} \rm{\bf d},
\end{equation}
where $\rm{\bf N_d}$ is a diagonal matrix with $\rm{\bf N_{d, ii}} = \sigma^2_{inst, ii} + \sigma^2_{conf}$.
If we make an assumption that the sources do not overlap with each other, then the Fisher information matrix can be simplified to a $S\times S$ diagonal matrix with each entry equal to 
\begin{equation}
\Sigma {\rm PRF}_i^2 / <\sigma^2_{inst, i} + \sigma^2_{conf}>.
\label{eq_Fisher}
\end{equation}
By the Cram\'{e}r-Rao lower bound, the inverse of Eq. \ref{eq_Fisher} represents the lower limit of the variance of the estimated flux density. If we make a further simplifying assumption that the instrument noise does not vary much across the pixels, then the minimum of the variance would be 
\begin{equation}
\rm{var}_{min} = <\sigma^2_{inst} + \sigma^2_{conf}> / \Sigma PRF_i^2.
\end{equation}
Take the deblended 250 $\mu$m fluxes as an example. From Fig. \ref{fig:sigma}, the residual confusion noise is $\sim3$ mJy, which is a factor of 1.8 higher than the instrument noise (see Table 3). Given $\Sigma {\rm PRF}_i^2 = 5.2$ for the 250 $\mu$m beam map, we can derive the minimum $1\sigma$ flux uncertainty is $\sim2$ times smaller than the residual confusion noise, as  
\begin{equation}
\rm{var}^{250}_{min} = <(\sigma_{conf}/1.8)^2 + \sigma^2_{conf}> / 5.2 = 0.25\sigma^2_{conf}.
\end{equation}
Therefore combining the $1\sigma$ uncertainty with the residual noise in quadrature has roughly the same effect as scaling the $1\sigma$ uncertainty by a factor of two,
\begin{equation}
\sigma_{total} = \sqrt{(1\sigma)^2 + \sigma^2_{conf}} = \sqrt{5}\sigma.
\end{equation}
Performing these calculations for the MIPS band shows the same conclusion. For the PACS and SCUBA-2 bands, because the instrument noise is much greater than the confusion noise, the $1\sigma$ flux uncertainty is dominated by instrument noise and therefore does not need to be scaled up as for the other bands. In other words, the total error after including the residual confusion noise is almost the same as the $1\sigma$ flux uncertainty.

\end{appendix}

\end{document}